\documentclass[]{aastex631}

\newcommand{\cln}{\textsc{58}~}
\newcommand{\clnlensing}{\textsc{24}~}
\newcommand{\clndiagram}{\textsc{54}~}
\newcommand{\paperi}{\citetalias{Fu2022}}

\defcitealias{Fu2022}{Paper~I}

\def\rr1#1{#1}

\graphicspath{{./}}

\begin{document}

\title{LoVoCCS. II. Weak Lensing Mass Distributions, Red-Sequence Galaxy Distributions, and Their Alignment with the Brightest Cluster Galaxy in \cln Nearby X-ray-Luminous Galaxy Clusters}

\correspondingauthor{Shenming Fu}
\email{shenming.fu.astro@gmail.com}

\author[0000-0001-5422-1958]{Shenming~Fu}
\affiliation{NSF's National Optical-Infrared Astronomy Research Laboratory, 950 North Cherry Avenue, Tucson, AZ 85719, USA}

\author[0000-0003-0751-7312]{Ian~Dell'Antonio}
\affiliation{Department of Physics, Brown University, 182 Hope Street, Box 1843, Providence, RI 02912, USA}

\author[0000-0003-0936-7223]{Zacharias~Escalante}
\affiliation{Department of Physics, Brown University, 182 Hope Street, Box 1843, Providence, RI 02912, USA}

\author[0000-0003-3397-6838]{Jessica~Nelson}
\affiliation{Department of Physics, Brown University, 182 Hope Street, Box 1843, Providence, RI 02912, USA}

\author[0000-0003-2314-5336]{Anthony~Englert}
\affiliation{Department of Physics, Brown University, 182 Hope Street, Box 1843, Providence, RI 02912, USA}

\author[0000-0003-4774-4288]{S\o ren~Helhoski}
\affiliation{Department of Physics, Brown University, 182 Hope Street, Box 1843, Providence, RI 02912, USA}

\author[0000-0002-7342-3229]{Rahul~Shinde}
\affiliation{Department of Physics, Brown University, 182 Hope Street, Box 1843, Providence, RI 02912, USA}

\author{Julia~Brockland}
\affiliation{Department of Physics, Brown University, 182 Hope Street, Box 1843, Providence, RI 02912, USA}

\author{Philip~LaDuca}
\affiliation{Department of Physics, Brown University, 182 Hope Street, Box 1843, Providence, RI 02912, USA}

\author{Christelyn~Larkin}
\affiliation{Department of Physics, Brown University, 182 Hope Street, Box 1843, Providence, RI 02912, USA}

\author{Lucca~Paris}
\affiliation{Department of Physics, Brown University, 182 Hope Street, Box 1843, Providence, RI 02912, USA}

\author{Shane~Weiner}
\affiliation{Department of Physics, Brown University, 182 Hope Street, Box 1843, Providence, RI 02912, USA}

\author[0000-0003-4811-7913]{William~K.~Black}
\affiliation{Department of Physics and Astronomy, Brigham Young University, N283 ESC, Provo, UT 84602, USA}

\author[0000-0001-7583-0621]{Ranga-Ram~Chary}
\affiliation{Infrared Processing and Analysis Center, MC 314-6, Caltech, 1200 E. California Boulevard, Pasadena, CA 91125, USA}

\author[0000-0003-2416-1557]{Douglas~Clowe}
\affiliation{Department of Physics and Astronomy, Ohio University, 1 Ohio University, Athens, OH 45701, USA}

\author[0000-0003-1371-6019]{M.~C.~Cooper}
\affiliation{Department of Physics and Astronomy, University of California, Irvine, Irvine CA, USA}

\author[0000-0002-2808-0853]{Megan~Donahue}
\affiliation{Department of Physics and Astronomy, Michigan State University, East Lansing, MI 48824, USA}

\author[0000-0002-4876-956X]{August~Evrard}
\affiliation{Department of Astronomy, University of Michigan, Ann Arbor, MI 48109, USA}
\affiliation{Department of Physics, University of Michigan, Ann Arbor, MI 48109, USA}

\author[0000-0002-3032-1783]{Mark~Lacy}
\affiliation{National Radio Astronomy Observatory, 520 Edgemont Road, Charlottesville, VA 22903, USA}

\author[0000-0003-3234-7247]{Tod~Lauer}
\affiliation{NSF's National Optical-Infrared Astronomy Research Laboratory, 950 North Cherry Avenue, Tucson, AZ 85719, USA}

\author[0000-0002-0561-7937]{Binyang~Liu}
\affiliation{Purple Mountain Observatory, Chinese Academy of Sciences, Nanjing 210023, China}

\author[0000-0002-9883-7460]{Jacqueline~McCleary}
\affiliation{Department of Physics, Northeastern University, 110 Forsyth St, Boston, MA 02115}

\author[0000-0003-1225-7084]{Massimo~Meneghetti}
\affiliation{Osservatorio di Astrofisica e Scienza dello Spazio di Bologna, Istituto Nazionale di Astrofisica Via Gobetti 93/3, I-40129, Bologna, Italy}
\affiliation{National Institute for Nuclear Physics, viale Berti Pichat 6/2, I-40127 Bologna, Italy}

\author[0000-0001-7964-9766]{Hironao~Miyatake}
\affiliation{Kobayashi-Maskawa Institute for the Origin of Particles and the Universe (KMI), Nagoya University, Nagoya, 464-8602, Japan}
\affiliation{Division of Physics and Astrophysical Science, Graduate School of Science, Nagoya University, Nagoya 464-8602, Japan}
\affiliation{Kavli Institute for the Physics and Mathematics of the Universe (WPI), The University of Tokyo Institutes for Advanced Study (UTIAS), The University of Tokyo, Chiba 277-8583, Japan}

\author[0000-0001-7847-0393]{Mireia~Montes}
\affiliation{Instituto de Astrof\'isica de Canarias, c/ V\'ia L\'actea s/n, E-38205 - La Laguna, Tenerife, Spain}
\affiliation{Departamento de Astrof\'isica, Universidad de La Laguna, E-38205 - La Laguna, Tenerife, Spain}

\author[0000-0002-5554-8896]{Priyamvada Natarajan}
\affiliation{Department of Astronomy, Yale University, 266 Whitney Avenue, New Haven, CT 06511, USA}
\affiliation{Department of Physics, Yale University, 217 Prospect Street, New Haven, CT 06520, USA}

\author[0000-0002-0144-387X]{Michelle~Ntampaka}
\affiliation{Space Telescope Science Institute, Baltimore, MD 21218, USA}
\affiliation{Department of Physics and Astronomy, Johns Hopkins University, Baltimore, MD 21218, USA}

\author[0000-0002-7957-8993]{Elena~Pierpaoli}
\affiliation{University of Southern California, Los Angeles, CA 90089, USA}

\author[0000-0002-9365-7989]{Marc~Postman}
\affiliation{Space Telescope Science Institute, 3700 San Martin Drive, Baltimore, MD 21218, USA}

\author[0000-0002-9254-144X]{Jubee~Sohn}
\affiliation{Astronomy Program, Department of Physics and Astronomy, Seoul National University, 1 Gwanak-ro, Gwanak-gu, Seoul 08826, Republic of Korea}
\affiliation{SNU Astronomy Research Center, Seoul National University, 1 Gwanak-ro, Gwanak-gu, Seoul 08826, Republic of Korea}

\author[0000-0001-9658-1396]{David~Turner}
\affiliation{Department of Physics and Astronomy, Michigan State University, East Lansing, MI 48824, USA}

\author[0000-0002-7196-4822]{Keiichi~Umetsu}
\affiliation{Academia Sinica Institute of Astronomy and Astrophysics (ASIAA), No. 1, Section 4, Roosevelt Road, Taipei 10617, Taiwan}

\author[0000-0001-6161-8988]{Yousuke~Utsumi}
\affiliation{Kavli Institute for Particle Astrophysics and Cosmology (KIPAC), SLAC National Accelerator Laboratory, Stanford University, 2575 Sand Hill Road, Menlo Park, CA 94025, USA}

\author[0000-0002-6572-7089]{Gillian~Wilson}
\affiliation{Department of Physics, University of California Merced, 5200 Lake Road, Merced, CA 95343, USA}




\begin{abstract}

The Local Volume Complete Cluster Survey (LoVoCCS) is an on-going program to observe nearly a hundred low-redshift X-ray-luminous galaxy clusters (redshifts $0.03<z<0.12$ and X-ray luminosities in the 0.1-2.4~keV band $L_{{\rm X500c}}>10^{44}$~erg/s) with the Dark Energy Camera (DECam), capturing data in $u,g,r,i,z$ bands with a $5\sigma$ point source depth of approximately 25-26th AB magnitudes. Here, we map the aperture masses  in \cln galaxy cluster fields using weak gravitational lensing.   These clusters span a variety of dynamical states, from nearly relaxed to merging systems, and approximately half of them have not been subject to detailed weak lensing analysis before. In each cluster field, we analyze the alignment between the 2D mass distribution described by the aperture mass map, the 2D red-sequence (RS) galaxy distribution, and the brightest cluster galaxy (BCG). We find that the orientations of the BCG and \rr1{the} RS distribution are strongly aligned throughout the interiors  of the clusters:  the median misalignment angle is 19 deg within 2 Mpc. We also observe the alignment between the orientations of the RS distribution and the overall cluster mass distribution (by a median difference of 32 deg within 1 Mpc), although this is constrained by galaxy shape noise and the limitations of our cluster sample size. These types of alignment suggest long-term dynamical evolution within the clusters over cosmic timescales.

\end{abstract}



\section{Introduction} \label{sec:intro}

While clusters of galaxies are the largest gravitationally bound structures in the Universe, connections and interactions still exist between close pairs of clusters, traced by inter-cluster filaments that connect them. Although the direct detection of filaments has been achieved in a small number of clusters~\citep{Dietrich2012,Jauzac+2012}, it is generally difficult to detect them without stacking data~\citep{Zhang2013}.
Low-redshift massive clusters provide opportunities to inspect and study filaments individually in high resolution because of their proximity~\citep{Haines2018}. 

The different components of a cluster -- the central galaxy (CG) or the brightest cluster galaxy (BCG), the member galaxy distribution, the gas, and the dark matter -- may come to display similar orientations as the cluster evolves. 
Such alignment in low- and medium-mass clusters  has been analyzed statistically to study cluster triaxiality~\citep[][]{Shin2018}. 
The alignment is stronger in massive clusters and has been detected in individual clusters at medium redshift ($0.1<z<0.9$) with ground-based telescopes~\citep{Oguri2010,Herbonnet19} or space telescopes~\citep{Donahue2016,Umetsu2018}, but in relatively small samples. 
In addition, cluster pairs may show alignment between their matter distributions due to the large-scale tidal field. Low-redshift (low-z; $z\lesssim0.1$) massive clusters would reveal those intra- or inter-cluster features individually and in more detail, which can be observed by even ground-based telescopes. Low-redshift clusters allow one to resolve halo substructures and facilitate studies of e.g., the correlation between cluster substructure and galaxy population/evolution~\citep{McCleary2015}.

Nearby clusters are also useful for cosmological studies by providing a low-redshift anchor for the scaling relations between the cluster mass and the observables at the late Universe. The evolution of halo mass function is sensitive to the cosmological parameters $\Omega_{\rm m}$ and $\sigma_8$ at the high mass end, where galaxy clusters reside, and thus clusters provide a key probe of dark energy~\citep{Albrecht2006,Allen2011, DESC2018}. 
Building mass functions requires accurate mass measurements of clusters with unbiased selection. Most clusters in the Universe are not massive enough to produce significant gravitational lensing signals that overcome measurement noise, and their individual masses are difficult to be directly measured. However, their masses can be statistically inferred from scaling relations between the mass and the observables, e.g., member galaxy richness and gas properties (X-ray emission or the Sunyaev-Zel'dovich/SZ effect).  Weak lensing (WL) on the other hand, yields unbiased mass estimates over a large ensemble of clusters and thus permits calibration of those scaling relations~\citep{Mcclintock2019,Miyatake2019,Umetsu2020}. 
To construct a scaling relation, the cluster sample needs to be complete and span a wide range of dynamical states to avoid biased cosmological inference~\citep{Kettula2015}.

In the past, it was difficult to study clusters at very low redshift via WL, because those clusters require high-quality PSF modeling over a large sky area and very deep observations to reduce shape noise as their lensing signal ($\propto D_{ls} D_l$) is low compared to clusters at medium redshift. However, nowadays nearby clusters can be well captured by recently developed large aperture and large Field-of-View (FoV) instruments, e.g., the Dark Energy Camera (DECam; FoV $2.2\deg$) mounted on the 4-meter Blanco Telescope at the Cerro Tololo Inter-American Observatory (CTIO), and the future LSST Camera (FoV $3.5\deg$) mounted on the 8.4-meter Simonyi Survey Telescope at the Rubin Observatory. The large aperture enables fast and deep observations of hundreds of thousands of galaxies at the same time. Meanwhile, recent tools strengthen the PSF modeling over a large FoV which improves the shape measurements for lensing analysis~\citep{Bosch2018}. As a result, accurate lensing measurements of nearby clusters have become easier to achieve than in earlier years. DECam has already proved its excellent performance in the recently completed Dark Energy Survey~\citep[DES;][]{DES2016}, and as a comparison, the 10-year LSST~\citep{Ivezic2019} will be $\sim3$ magnitudes deeper over a $\sim3$ times larger sky footprint. In addition to optical observations, X-ray observations (e.g., Chandra, XMM/Newton, eROSITA) and SZ observations (e.g., Planck) provide vast amount of information about the hot gas in nearby clusters. 

We proposed the Local Volume Complete Cluster Survey (LoVoCCS; Proposal ID: 2019A-0308; PI: Ian Dell'Antonio) to observe  a volume-complete sample of 107 nearby X-ray luminous clusters ($0.03 < z < 0.12$; [0.1–2.4 keV]~$L_{\rm X500c}>10^{44}$ erg/s; $M_{\rm 500c}\gtrsim2\times10^{14}M_\odot$, $M_{\rm 200c}\gtrsim3\times10^{14}M_\odot$) in low Galactic extinction fields using DECam $u,g,r,i,z$-bands. Here, the subscript 500c (200c) means the mass within a radius where the average overdensity is 500 (200) times the critical density at that redshift; the mass $M_{\rm 500c}$ is given by~\citet{Piffaretti2011} and we convert it to $M_{\rm 200c}$ using typical NFW halos~\citep{Navarro1997,Child2018}. 
The survey started in 2019 and is expected to conclude by 2024. The deep co-added images of each cluster field have reached LSST Year $1-2$ depth ($\sim1$ mag deeper than DES; Appendix \ref{sec:app_depth}) more than two years prior to the LSST Y1 data release. Additionally, $\gtrsim10$ northern LoVoCCS clusters will not be fully covered by LSST (especially the ones outside the North Ecliptic Spur and at high Declination). These co-added images can also serve as a zeroth-year template for LSST transient studies. 

\begin{figure*}[htbp!]
    \centering
    \includegraphics[width=0.75\textwidth]{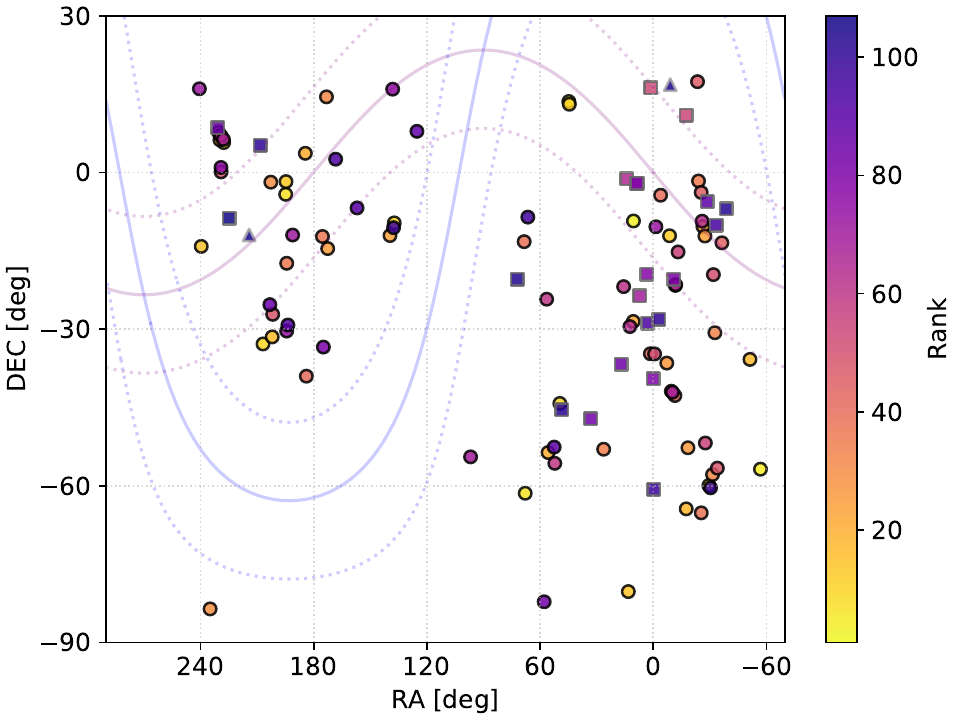}
    \caption{
    Celestial distribution of LoVoCCS clusters with their ranks and observation completeness (107 clusters in total).
    The black-edge circles denote clusters that have finished observation (83 clusters). 
    The gray-edge squares denote clusters that were partially observed (22 clusters).
    The light-gray-edge triangles denote clusters that have not been observed (2 clusters).
    The marker face color shows the cluster rank. A higher rank corresponds to a higher X-ray luminosity (a brighter color in the figure); high rank clusters are generally more complete in observation. 
    The purple and blue solid curves represent the ecliptic plane and the Milky Way galactic plane respectively, and the dashed curves show $\pm15\deg$ from the planes.
    }
    \label{fig:clusters}
\end{figure*}

We observed the cluster sample in uniform depth but prioritize clusters that have higher X-ray luminosities. Although delayed by the pandemic and inclement weather, we have finished the observation of 83 clusters (complete to the top 52 clusters in terms of X-ray luminosity) and obtained $87\%$ of the survey data (including archival data), as of the submission of this paper (Figure~\ref{fig:clusters}). In addition, we have been awarded further DECam time to survey a sample of Planck SZ clusters near the LoVoCCS X-ray luminosity cut to increase the SZ completeness (Proposal ID: 2022A-658443; PI: Ian Dell'Antonio), and Hyper Suprime-Cam (HSC) time to observe northern clusters selected by the same criteria aiming 
to improve the all-sky completeness (Proposal ID: S20A-127, S23B-105, S24A-81; PI: Hironao Miyatake). 

We presented the survey background, pipeline, and early science results in our first LoVoCCS paper~\citep[hereafter~\paperi]{Fu2022}.
In this second paper, we study a sample of \cln clusters from LoVoCCS (Section \ref{sec:dataset}), with a focus on the mass distributions and galaxy distributions in these clusters. About half of the cluster sample ($\gtrsim \clnlensing$ clusters) have not been studied in detail via WL previously, and the rest provide comparisons of our study with previously reported results.  In this work, we analyze the performance of the photometric redshift measurement and improve the quality check procedure in our pipeline, and we extend the pipeline by adding steps to construct the distribution of red-sequence (RS) galaxies and to study cluster triaxiality by measuring the orientations and positions of the BCG, the RS galaxy distribution, and the lensing mass distribution on the plane of sky (Section \ref{sec:method}). 
In Section \ref{sec:results}, we present the optical image overlaid with the mass map and RS distribution contours of each cluster (or cluster pair/group), and we compare the orientations and locations of those cluster components (BCG, galaxy distribution, and mass distribution) in individual clusters as a function of cluster evolution stage. 
We then discuss our results in Section \ref{sec:discussion} and summarize the paper in Section \ref{sec:summary}.

Throughout this paper, we use the same cosmological parameters as~\paperi: flat $\Lambda$CDM $H_0=71$ km/s/Mpc, $\Omega_{\rm m}=0.2648$; 
switching between commonly used values slightly changes the angular diameter distances and luminosity distances in this work (by $\lesssim2\%$).

\section{Dataset}\label{sec:dataset}

Our sample of \cln clusters covers a large variety of X-ray luminosities (and therefore masses)\footnote{X-ray luminosity in the 0.1–2.4 keV band: $1.1\times10^{44}$ erg/s $<L_{\rm X500c}<8.7\times10^{44}$ erg/s; mass based on the X-ray emission:   $2.0\times10^{14}M_\odot \lesssim M_{\rm 500c}\lesssim 7.3\times10^{14}M_\odot$~\citep{Piffaretti2011}. \url{https://heasarc.gsfc.nasa.gov/W3Browse/rosat/mcxc.html}}, redshifts ($0.04<z<0.12$), and dynamical states (perturbed or nearly relaxed). 
We rank all LoVoCCS clusters by X-ray luminosity~\citep{Piffaretti2011} from 1 to 107.
The 13 highest ranked LoVoCCS clusters are included in this work, and the lowest rank in the sample is 99 (A3825).  
These \cln clusters are exposure-time complete in LoVoCCS and have high-quality processing results. We give a list of public datasets used for co-addition in Appendix~\ref{sec:app_archival}. Each cluster is covered by $\gtrsim100$ exposures. As mentioned earlier, we summarize the $5\sigma$ co-added depth of these cluster fields in Appendix~\ref{sec:app_depth}; the median depth values are $25-26$th PSF magnitudes for point objects in $u,g,r,i,z$ bands, and the CModel magnitudes for extended objects are $\sim0.3$ mag shallower. The point and extended objects are separated by the LSP parameter \texttt{extendedness} in $r$ band, which is the band for our shape measurements for lensing analysis (we require seeing $\lesssim1''$), and we have used the LSP flags \texttt{base\_PsfFlux\_flag} and \texttt{modelfit\_CModel\_flag} to select objects that have high-quality photometric measurements. We include archival $Y$-band data in the processing for photometric redshift measurements. 

\section{Methods}\label{sec:method}

The LoVoCCS pipeline has been presented in \paperi. Here we briefly summarize the pipeline (Section~\ref{sec:pipeline}), and emphasize the steps that have been taken to improve it in the remainder of this section. We describe the performance of our photometric redshift measurements (Section~\ref{ref:photo_z}) and mass map construction (Section~\ref{sec:mass_map}); the new steps taken in performing the red-sequence galaxy analysis (Section~\ref{sec:RS}) and cluster triaxiality analysis (Section~\ref{sec:triaxiality}); and the improvements in our data quality checks (Section~\ref{sec:quality_check}). 

\subsection{Pipeline Steps}\label{sec:pipeline}

We require a framework that allows us to consistently analyze the full diversity of virialized and merging clusters in LoVoCCS. For this, we use the LSST Science Pipelines (LSP) software\footnote{\url{https://pipelines.lsst.io}} \rr1{version 19.0.0} to first process the raw data and then use our own pipeline to analyze the LSP data products of each cluster uniformly. The LSP software enables us to use the latest algorithms designed for the future LSST data. The output images and catalogs are therefore be compatible with the LSST data products, which simplifies and enables future synergies. The LSP software has the ability to process the data from different instruments, including DECam, Subaru/HSC and Canada–France–Hawaii Telescope (CFHT)/MegaCam, in a similar manner. The details of LSP have been described by~\citet{Bosch2018}.

In each cluster field, we first download the public DECam data from the NOIRLab Astro Data Archive.\footnote{\url{https://astroarchive.noirlab.edu}} We use the LSP to detrend the raw DECam exposures at individual CCDs. The astrometry and photometry of each CCD are calibrated against external stellar catalogs -- Gaia~\citep{Gaia2016,Gaia2018,Gaia-DB} for astrometric calibration, and Pan-STARRS1~\citep[PS1;][]{Chambers2016,PS1-DB}, SkyMapper~\citep{Wolf2018,Onken2019,smdr1,smdr2}, and Sloan Digital Sky Survey~\citep[SDSS;][]{Alam2015} for photometric calibration. 
We then select high-quality CCD exposures by the seeing measured on the processed images: median ellipticity $<0.13$, FWHM $<1\farcs16$ in $r$ band (for lensing analysis), and ellipticity $<0.33$, FWHM $<1\farcs74$ in other bands. We improve the astrometry and photometry of those exposures by calibrating their catalogs jointly. 
Next, the exposures are mapped onto 4k$\times$4k-pixel patches ($0.3\times0.3$ deg, $0\farcs263$ per pixel),  and adjacent patches have an additional 100 pixel overlap.  The exposures are then stacked on each patch and in each band. 
After stacking, the LSP detects, de-blends, measures objects on the co-added images, and performs forced photometry among different bands on each object. We use the shape information measured on the co-added $r$-band image for lensing analysis, and we use the forced photometry to obtain the color information of each object. 

After the LSP steps, we correct for the Galactic extinction using the dust reddening map~\citep{Schlegel1998,Schlafly2011,IRSA-DB}, and further calibrate the magnitude zero points in the coadd catalog (measured from the co-added images) using the model color-terms and the color-color stellar locus.  
We then measure photometric redshifts, build lensing mass maps and shear profiles, derive lensing masses, and check the data product quality. In this work, we also add steps to construct red-sequence galaxy distributions and cluster triaxiality metrics. 

We adjust the pipeline in some special cases. The default angular cut on the CCD selection is $1.5\deg$ towards the cluster center, which results in $12\times12$ patches generally. In the field of A401 and A3558, we use larger angular cuts ($1.8\deg$ and $2\deg$ respectively) to include the nearby clusters that they are interacting with. 
For the A401/A399 field in the northern sky, since there is no SDSS coverage for the $u$-band photometric calibration, we build a synthetic reference catalog for the LSP based on the empirical  color transformations between PS1 (deep observations in $g$ and $r$) and DECam ($u$) in other cluster fields, and we further calibrate the $u$-band photometry in the coadd catalog using the model stellar locus of $g-r$ vs. $u-g$ after the extinction correction; we find that this approach produces reasonable photometric  redshifts compared to the spectroscopic redshifts.

\subsection{Photometric Redshifts}\label{ref:photo_z}

In this work, we continue using the Bayesian Photometric Redshift~\citep[BPZ;][]{Benitez2000,Coe2006} algorithm to estimate galaxy photometric redshifts (photo-zs), and we compare the results with archival public spectroscopic redshifts (spec-zs) obtained from NASA/IPAC Extragalactic Database~\citep[NED;][]{Helou1991,NED-DB}. 

\begin{figure*}[htbp!]
    \centering
    
    \plottwo{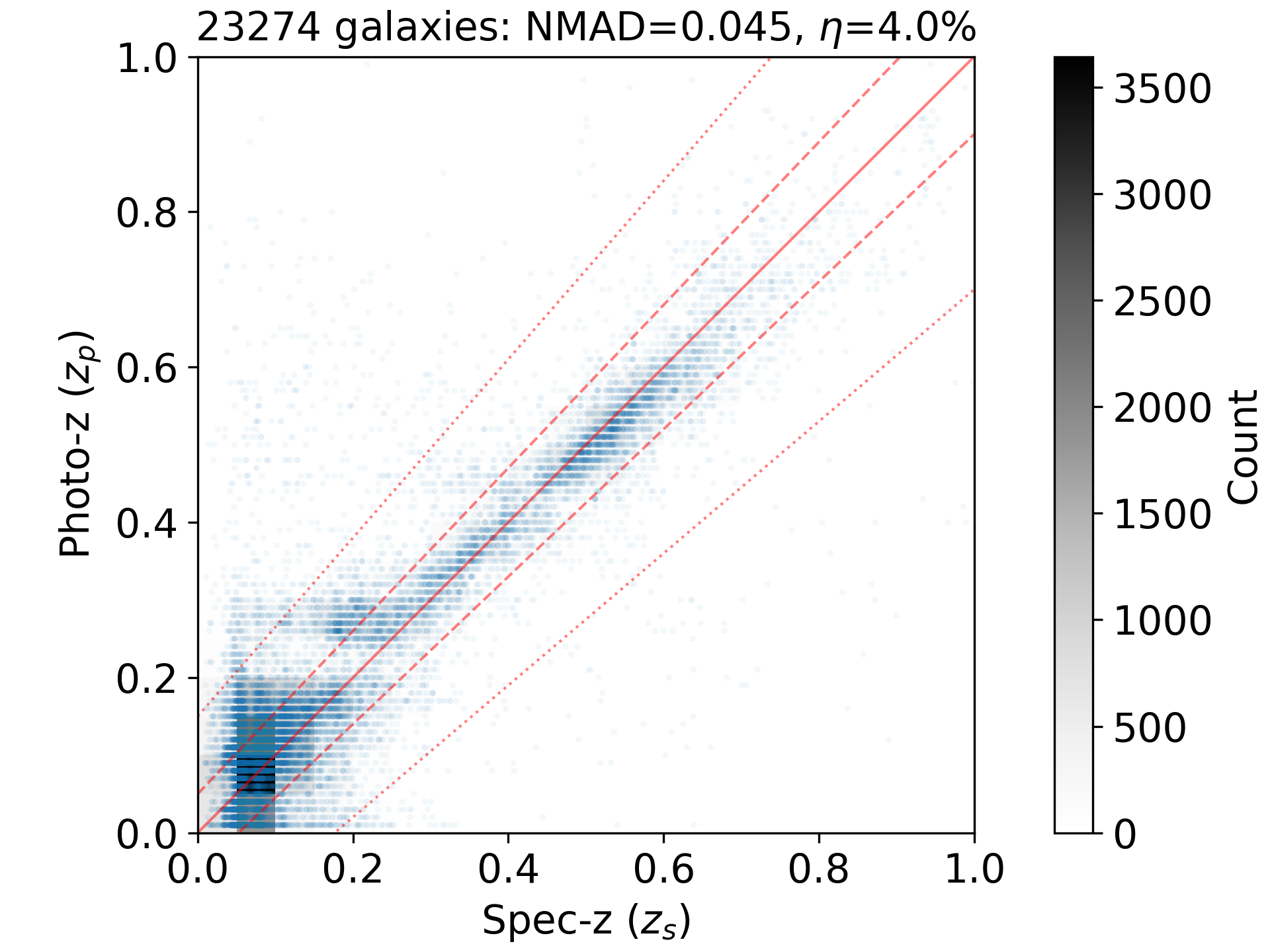}{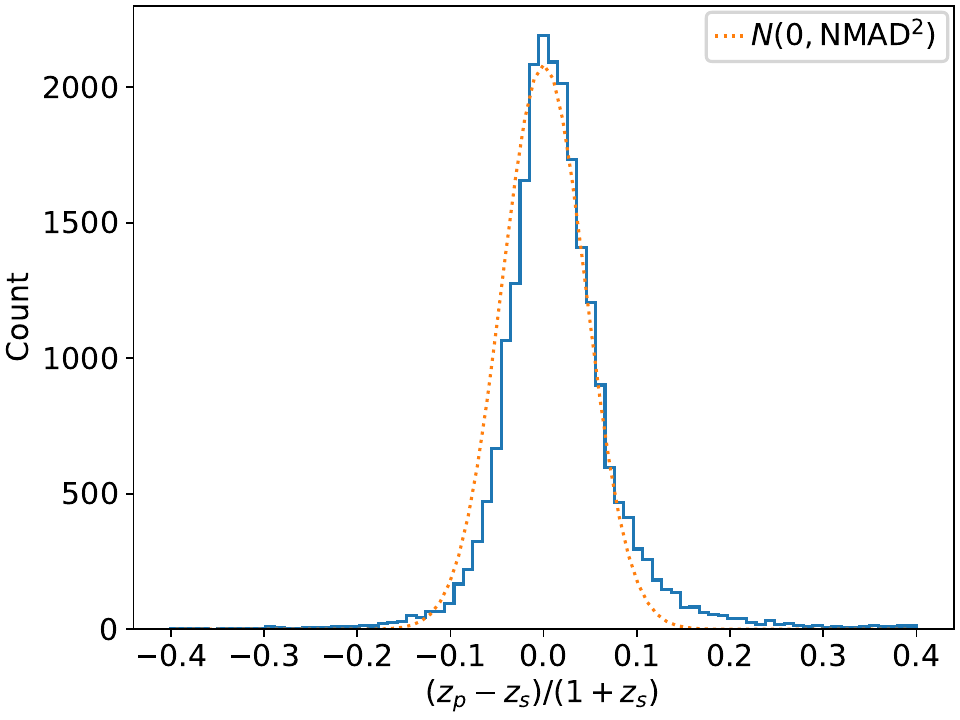}
    \caption{Comparison between the photo-zs ($z_{\rm p}$) and spec-zs ($z_{\rm s}$) in all cluster fields studied in this paper with metrics of the photo-z error $\Delta z={(z_{\rm p}-z_{\rm s})}/{(1+z_{\rm s})}$. 
    \textit{Left}: The dotted, dashed, and solid lines represent biases of $|\Delta z|=0.15, 0.05, 0$, respectively. We use a step of 0.01 in BPZ. The  normalized-median absolute deviation ($1.48\times\texttt{median}\{|\Delta z|\}$; NMAD) shows an estimation of the photo-z uncertainty, and the percentage of galaxies that have $|\Delta z|>0.15$ gives the outlier rate $\eta$.
    \textit{Right}: The blue solid curve shows the distribution of $\Delta z$, which is nearly a Gaussian $N(0, \texttt{NMAD}^2)$, the orange dotted curve; the small bias is mainly caused by the low-z galaxies (mostly cluster members) at $z_{\rm s}\lesssim0.2$.}
    \label{fig:photo_z}
\end{figure*}

As mentioned in \paperi, the photo-z outliers at low (true) redshift are mainly caused by the lack of archival $u$-band data on the periphery of cluster fields, blended objects, or special galaxy types/colors. However, the insufficiency of $u$-band data is rarely the case in the cluster central areas we observe for lensing analysis. After visual inspection, we find that the galaxy type assigned by BPZ is generally correct (we use a two-step interpolation between consecutive templates), and we also find that the spirals in clusters are more likely to have photo-z's biased high because they are redder than field galaxies due to quenching. The outliers also include cluster dwarf galaxies, which are more common in galaxies with blue BPZ galaxy types. Figure~\ref{fig:photo_z} shows an example of our photo-z quality: a comparison between the spec-zs and photo-zs of a clean sample of galaxies selected by $z_{\rm s}<1.5$, $\texttt{odds}>0.95$, $\chi^2_{\texttt{mod}}<1$, and covered by $u$ band; the photo-z uncertainty and the outlier rate are both at a few percent. Here $\texttt{odds}$ and $\chi^2_{\texttt{mod}}$ give the BPZ probability concentration and closeness of fit, respectively. Note the archival spec-z datasets mostly focus on cluster member galaxies for dynamical analysis. 

In this work, we use a photo-z cut at the cluster redshift plus 0.1 to remove foreground galaxies, instead of a fixed cut at 0.15 used in \paperi. 
Therefore, the photo-z cut is different in different cluster fields. This change improves the lensing map S/N of our higher-redshift clusters by $\lesssim10\%$. 
This change in mass map could be caused not only by the reduction of foreground galaxy contamination, but also by the high sensitivity to lensing signal of background galaxies at low redshift. Adding higher photo-z cuts can remove both foreground and background galaxies because of the photo-z noise -- we seek to find a \textit{balance} between the lensing signal and shape noise. 
In \paperi, under the original photo-z cut, we estimated that photo-z outliers could dilute the lensing signal by a few percent. As more foreground galaxies are removed under the new photo-z cut, we estimate the bias of mass maps in this work introduced by photo-z errors is even smaller ($\sim$ percent level). 
We note that in Figure~\ref{fig:photo_z} there is a cloud of galaxies at $z_{\rm s}\sim0.2$ with $z_{\rm p}\sim0.25$ (these are likely dusty/quenched), however these galaxies are mainly behind our cluster sample ($z\lesssim0.1$), and thus we do not expect their presence to affect our mass maps significantly. 
As in \paperi, we also remove sources at photo-z higher than 1.4 because their shape measurements can be noisier. 

\rr1{
In Figure~\ref{fig:photo_z}, the galaxies have magnitudes $i\lesssim 24$, which is generally brighter compared to the whole sample for lensing analysis. For the fainter lensing sources, their photo-z errors could be larger than the galaxies we use for the spec-z versus photo-z comparison. However, the photo-z bias does not occur in a specific sky area but occurs everywhere. Thus, we expect that the photo-z bias of faint objects will not materially affect the 2D structure reconstruction.  
Also, the clusters studied in this work are at low redshift, and most background sources are at high redshift. For low-z clusters, the lensing signal is not sensitive to the exact redshift of the high-z sources -- the reduced shear (more specifically the lensing distance ratio $D_{LS}/D_{S}$) changes slowly with the (true) redshift~\citep[e.g.,][]{Applegate2014}. 
In addition, faint objects are unlikely to be foreground galaxies of these low-z clusters, which means the foreground contamination caused by faint objects is small. 
}

To further filter out foreground galaxies, we examined the inclusion of  archival photometry in other wavelength bands. 
Given the depth, footprint, and seeing of archival data, we consider adding external near-infrared (NIR) photometry to remove the member/low-z galaxies (instead of increasing the wavelength coverage of background sources). Those foreground galaxies are relatively larger and brighter than the background sources, and thus they are unlikely affected by the seeing and depth of those external catalogs.  
We match the LS+WISE forced photometry catalog from the DESI Legacy Imaging Surveys~\citep[LS;][]{Dey2019} and the VISTA Hemisphere Survey catalog~\citep[VHS;][]{VHS-DB,McMahon2013}\footnote{We obtained the LS+WISE and VHS catalogs  from the NOIRLab Astro Data Lab. \url{https://datalab.noirlab.edu/index.php}} with the LoVoCCS catalog and the NED spec-z catalog. 
We find that in the color-color space of e.g., $z-w_1$ vs. $w_1-w_2$, $Y-J$ vs. $J-K_{\rm s}$, no simple cut can fully separate those foreground galaxies ($z_{\rm s}\lesssim0.1$) from the others. Also, when comparing the mass maps with and without cuts on NIR colors, we find no significant difference. 
Therefore, in the following analysis we only use the photometry from LoVoCCS. 
Another approach to the removal of foreground galaxies is to use a machine learning (ML) photo-z algorithm which takes the \textit{morphology} and \textit{special colors} of cluster members into account -- we will explore this in the future.  

In fact, as it turns out, the mass map morphology is not very sensitive to the photo-z selection as member galaxy shapes are randomly distributed in general; they do exhibit intrinsic alignment (IA) caused by tidal fields, but we expect this to be a secondary effect compared to the cluster lensing signal. Similarly, we assume that the IA between background sources is a secondary effect as well and thus do not consider it when constructing mass maps. 
Additionally, we make cuts on the photo-z quality metrics ($\texttt{odds}>0.95$ and $\chi^2_{\texttt{mod}}<4$) before making mass maps to ensure the photo-z quality and to be consistent with the mass fitting step. The photo-z quality cuts reduce the median galaxy number density from $\sim 22$ arcmin$^{-2}$ (shape and photometry cuts have been applied) to $\sim 8$ arcmin$^{-2}$, and the photo-z value cut further reduces the source density to $\sim7$ arcmin$^{-2}$. 
Our test shows that including more sources using wider photo-z quality cuts 
does not significantly improve the S/N of mass maps, and the reason could be that the sources outside our quality cuts have noisier shape measurements. 

\subsection{Aperture Mass S/N Map}\label{sec:mass_map}

As in~\paperi, we make quality cuts and conduct shear calibration on the HSM shapes~\citep{Hirata2003,Mandelbaum2005,Mandelbaum2018b} on the co-added catalog before analyzing the lensing signal. We then use the aperture mass statistics~\citep{Schneider1996} to build lensing mass S/N maps via the Schirmer filter~\citep{Schirmer2004,Hetterscheidt2005,vonderLinden2014}. 
The aperture mass signal is a convolution between the convergence map (normalized surface mass density) and a circular (radially symmetric) weight function, and it is equivalent to a convolution between the tangential shear and a circular filter function (the Schirmer filter in this case). We approximate the tangential shear by the mean of per-galaxy reduced shear estimates derived from galaxy shapes, and we compute the mean within $100\times100$ pixel bins to reduce shape noise. The noise in the aperture mass mainly comes from the shape dispersion of source galaxies. 
Therefore, both the signal and noise parts of the mass S/N map scale with the source galaxy shapes. The possible remaining biases (usually multiplicative) in the galaxy shapes caused by e.g., blending, photo-z bias, shape measurement, can be mostly factored out in the signal-to-noise ratio, and we thus consider them as secondary effects. 

The S/N mass map shows how the underlying mass is concentrated. The Schirmer filter produces the highest S/N at an aperture center when the characteristic scale of the halo structure around that center is $\sim1/10$ of the aperture radius. Thus, we vary the aperture radius starting from 3k pixels with steps of 1k pixels to determine the mass map that best  captures the cluster-scale structure. In general, we choose the aperture that maximizes the peak S/N in the mass map; if the corresponding scale size is too small or large, we consider a local maximum where the aperture corresponds to the size of a normal cluster. 
Also, the morphology, in particular, the orientation of the projected halo  
should not be strongly affected by the convolution during the mass map construction, given the symmetry of the weight function. 
Using mock lensing catalogs, we verified that the mass map algorithm recovered elliptical halo orientations  (Appendix~\ref{sec:app_mass}). 

Most of our mass maps are built based on the catalogs of central $6\times6$ patches (24k$\times$24k pixels; $1.8\times1.8$ deg) divided by the LSP. For the cluster A3558 we consider the catalog of the full $15\times15$ patches to study the nearby structure in the Shapley supercluster, but we only present the central mass map to ensure the quality. The flat sky approximation is still valid in these sky regions. In Section~\ref{sec:maps}, we present only the (curl-free) E-mode maps derived from the tangential shear towards aperture centers. We find that reassuringly, the corresponding B-mode maps derived from the cross  component show no clear patterns especially near the cluster centers. 

The large angular sizes of the low-z clusters in  LoVoCCS occasionally encompasses background clusters in the field, which may contaminate the lensing signal of the cluster of interest. If we select only the source galaxies that lie in between the foreground cluster and the background cluster, it would severely penalize the S/N. Hence, we retain background clusters when constructing the mass maps, rather than using cuts on source galaxy redshifts, to obtain sufficient S/N. Moreover, these background clusters usually span an angular region much smaller than the foreground one, and thus we expect that they only affect cluster substructure detection in general. However, if the background cluster is very close to the foreground cluster center, we remove the background source galaxies closer to the background cluster (e.g., in A3128), because the background cluster may strongly affect the mass peak position and S/N  of the cluster of interest. 
In the future, we plan to use external cluster catalogs or run cluster finders (based on optical concentration, red sequence, or X-ray/SZ) to locate background clusters, which is important for accurately determining the foreground cluster mass. 

As mentioned earlier, we use LSP flags to select high-quality objects for analysis. We find that those flags generally remove objects affected by saturated stars and satellite trails, and the cluster central morphology does not change greatly after new masks are added around bright stars; we give more examples in Section~\ref{sec:maps} when we study individual clusters.


\subsection{Red-Sequence Galaxy Distribution}\label{sec:RS}

Galaxy clusters contain a large population of red elliptical/early-type galaxies formed by galaxy merging and quenching. In the color--magnitude diagram (CMD), they exhibit a characteristic ``red sequence'' (RS), which results from the unique features in their spectral energy distributions (SED), especially the 4000\AA~break caused by old stars and metal absorption lines. The RS galaxies tend to be redder and brighter near the cluster center because of their dynamical (infalling/merging) processes and quenching history. These RS galaxies trace the overall matter distributions in clusters, and they are closely related to luminous red galaxies (LRGs). 
LRGs are common tracers of the large-scale structure (LSS) and delineate the connections between clusters. 

In this work, we include a new pipeline step that builds a red-sequence galaxy distribution in each cluster field based on its color--magnitude diagram. To best locate the RS in CMD, one can use the colors of spectroscopically confirmed cluster member galaxies. This method works well in those LoVoCCS clusters with adequate spec-z data; though the archival spec-zs in the LoVoCCS cluster fields are limited, most of them are member galaxy spec-zs. However, those spec-zs are mainly from bright galaxies, and in some clusters, the number of member spec-zs is not sufficient ($\lesssim200$) to allow reliable data fitting in the CMD. In addition, photo-z does not work very well at low redshift and is not so useful for picking out cluster galaxies. Therefore, we also use all galaxies with valid photometry to locate the RS in the CMD. We provide more details of our algorithms below, and we use the CModel magnitudes for galaxies. 
In Appendix \ref{sec:app_rs}, we show examples of RS detection and selection. 

First, the archival spec-zs are accessed from NED (Section~\ref{ref:photo_z} and \paperi) 
and matched with the LoVoCCS galaxy photometric catalog. The cluster member galaxies are selected by $|z_{\rm s}-z_{\rm cl}|<0.01$, where $z_{\rm cl}$ is the cluster redshift. Then, we consider both colors $g-r$ and $r-i$ because of their wavelength coverage and depth, and we find that making cuts on two colors instead of one color gives better selection of RS galaxies. We use the $i$-band magnitude as a luminosity proxy because of its depth and small K-correction ($\lesssim0.05$ mag based on model SEDs used in BPZ) which simplifies the comparison between LoVoCCS clusters, and we thus skip the K-correction in the following analysis. 

Next, we detect the RS in the CMD of spectroscopically-selected cluster members and fit the data. We divide the $i$-band magnitude from 13.5 to 19.5 with a step of 0.3 mag,  the $g-r$ color from 0.65 to 1.0 mag with a step of 0.06 mag, and the $r-i$ color from 0.25 to 0.45 with a step of 0.04 mag; each color step approximates to the corresponding RS scatter.  
We then go through the $i$-band magnitude bins, and for each magnitude bin we generate a histogram of the $g-r$ color and find the peak among the color bins; we skip the magnitude bin if the number of galaxies inside is below 10. 
Then, the $i$-band magnitude bins and their corresponding $g-r$ color peaks are fitted by a straight line via \texttt{scipy.optimize.least\_squares}, and we use a loss function \texttt{soft\_l1} to reduce outliers. 
After fitting, we select galaxies that are 0.09 mag around the fitted $g-r$ color of their $i$-band magnitudes, and within those selected galaxies, we repeat the above steps but on the $r-i$ color to make another fit. 
We found that fitting the selected galaxies rather than all galaxies resulted in better RS detection.  

We repeat the same procedure on the whole galaxy photometric catalog and increase the number limit of galaxies per magnitude bin to 15. Then the higher quality of the spectroscopic and photometric fits is selected through visualization and comparison; the former usually  gives a better result when the spec-z data is sufficient. However, only $\sim1/4$ of the cluster sample has enough spec-zs. On the other hand, those clusters provide a validation for the performance of the photometric fit -- the completeness/purity between the RS selections (see the magnitude and color cuts below) based on the spectroscopic fit and the photometric fit is $\gtrsim90\%$. 
In some special cases, we also adjust parameters, e.g., the step, the number limit per bin, or the bright end of the $i$ magnitude, to improve the fit. 
The result shows that in our cluster sample the typical RS fit slopes are $\sim-0.04$ and $\sim-0.02$ in $g-r$ and $r-i$ vs. $i$, respectively, with no obvious dependence on redshift.  
Additionally, the RS seems to be more clear and concentrated (and thus easier to locate and fit) in the CMD of more relaxed clusters, possibly resulting from the cluster evolution.

After we obtain the best RS fit, we make color and magnitude cuts on the whole galaxy photometric catalog. 
We select galaxies which are $<0.09$ mag and $<0.06$ mag around the fitted $g-r$ and $r-i$ colors, respectively; we use those scatter cuts to deal with the RS noise. We also make an $i$ magnitude cut changing with the cluster redshift, which leads to a fixed luminosity limit and simplifies direct comparison among clusters. The fiducial cutoff is $i=20$ for a cluster at redshift $z=0.08$ (approximately the median of our cluster sample and LoVoCCS), and we vary the $i$-band cut based on the cluster luminosity distance (the difference among clusters is $\lesssim 2.5$ mag, much larger than the K-correction). 
This $i$-band limit corresponds to an absolute magnitude limit at $M_{i}\sim -17.8$.  
We set this limit based on the spec-z data depth for the RS fit, and also to remove faint background objects since our clusters are at low redshift. 
We expect that the RS 
contamination from background clusters is low given the color and magnitude cuts.
In addition, our method directly detects the RS in the CMD without any training sets.

We make a further cut on the galaxy sample (after the color and magnitude cuts above) before building the RS density map:  
we use BPZ galaxy type \texttt{t\_b==1} to select a sample of elliptical galaxies. After checking spec-zs and optical images, we find that this cut removes foreground/background galaxies within the color ranges and gives a clean sample of bright cluster member galaxies. 
We note that varying the \texttt{t\_b} cut from 1 (elliptical) to 2 (spiral Sbc) does not strongly affect the RS galaxy distribution morphology especially near the cluster center. 
After the selection, the RS magnitude distribution is close to a Gaussian, and the (apparent) magnitude histogram peak 
(about 3 mag from the magnitude cut; $M_i\sim-21$) 
changes with the cluster redshift. 

Once we have a clean sample of RS galaxies, we compute their 2D distribution by Kernel Density Estimation (KDE) to get the RS density map. 
Similar to the construction of mass maps, we generally consider the field of central $6\times6$ patches (24k$\times$24k pixels), and count the galaxy number in $100\times100$ pixel bins to reduce noise.  We also remove a foreground/background galaxy number density estimated by the density within 2k pixels from the field edge (half of a patch); this density is not high because of the previous selections on the galaxy color and type.  

To make the final RS galaxy distribution, the galaxy count grid is then convolved with a 2D Gaussian with a standard deviation $\sigma$ of 200 kpc (KDE). This smoothing length is chosen to estimate the RS distribution in the cluster and is fixed so that the RS maps of clusters at different redshifts are comparable. 
Note, our RS maps are derived from only  the number density and are not magnitude or flux weighted. 
The resulting RS map for each cluster is presented in Section~\ref{sec:maps}.


\subsection{Triaxiality: Orientation and Centering Analysis}\label{sec:triaxiality}

We study cluster triaxiality by analyzing the orientations and positions of the 2D mass distribution, RS galaxy distribution, and BCG. The BCG and cluster member galaxy distribution can trace the (non-spherical) dark matter distribution and the total mass distribution -- they tend to have similar orientations~\citep[e.g.,][]{Herbonnet19}. For low and medium mass clusters, stacking cluster profiles aligned with the BCG or member galaxy distribution can reveal the cluster triaxiality~\citep[e.g.,][]{Shin2018,Fu2023}. For high mass clusters (like the LoVoCCS sample studied in this work), those types of orientation alignment are less noisier in observations and may be visible in individual cluster fields.  

Measuring the cluster ellipticity is more difficult than measuring the orientation angle on the plane of sky. For example, measuring the ellipticity of the mass distribution can be affected by the shear bias or the convolution in the mass map construction; the same measurement on the galaxy distribution can be affected by the galaxy population; the measurement on the BCG can be affected by the intra-cluster light (ICL). In addition, the ellipticity measurement may also vary with the cluster radius. However, measuring their orientations is less sensitive to these factors noted above. Therefore in this work we only measure the orientation angles.

We also compare the BCG centroid and the map peaks. In relaxed clusters, it has been claimed that the offset between the BCG and the mass center can provide constraints on self-interacting dark matter~\citep[SIDM;][]{Kim2017,Harvey2019}, while in perturbed clusters the offset may result naturally from merging process or projection effects. The mass center can be traced by e.g., the X-ray emission or statistically the lensing mass map.

We summarize our measured angles and centers of the 2D distributions and the BCG of each cluster in Table~\ref{tab:angle_peak},~\ref{tab:angle_peak2} (Appendix~\ref{sec:ap_angle_peak}), and we give details of our measurement method below.

\paragraph{BCG}
To carry out more flexible measurement than the LSP, we first crop a $\sim0.4\times0.4$~Mpc region around the BCG on the optical $r$-band co-added image to reduce neighboring structures e.g., bright stars/galaxies or ICL. Then we use Source Extractor~\citep[SExtractor;][]{Bertin1996} with the detection threshold from $6\sigma$ to $35\sigma$ and a step of $2\sigma$ to detect and measure objects. We then pick the object with the largest area (as the BCG) under each threshold and take its median values of the coordinates (\verb|X_IMAGE| and \verb|Y_IMAGE|) and the angles (\verb|THETA_IMAGE|). As those galaxies have angular sizes much larger than the PSF, we expect the systematics caused by PSF are minimal. 
After visual inspections, we find that this process further reduces the effect of ICL and noise from blended objects but still gives the BCG orientation near the galaxy outer envelope.  
Note if a cluster has multiple bright  central galaxies, we pick the brightest one with the largest area. 

\paragraph{Mass map}
In Section~\ref{sec:maps}, we present the best mass S/N map of each cluster, which is under the aperture that maximizes the S/N of the  cluster-scale structure.  
Starting from that mass map, we first locate the mass map peak within $0.2\deg$ of the BCG mentioned above. Then we consider radial distance cuts of 0.5, 1.0, 2.0~Mpc from that peak on the mass map; they give the properties near the cluster center, within the cluster edge, and within the nearby LSS, respectively. 
We skip the region further out, because the signal is low and the contamination of (projected) neighboring structure is strong there, and it is close to the observation field edge of  clusters at very low redshift. Also, in some cluster fields the map could be ``rounder'' at larger radii, which makes the angle measurement difficult.  At each radial cut, we make small perturbations on the cut from -10\% to 10\% with a step of 2\%, measure an angle in individual cropped mass maps, and take their median to reduce noise. 
The angle is computed by the second moments (similar to the SExtractor algorithm\footnote{\url{https://astromatic.github.io/sextractor/Position.html}}) -- in Eq.~\ref{eq:1st_moment},~\ref{eq:2nd_moment},~\ref{eq:angle}, $i$ is the pixel index, $m_i$ is the mass map pixel value. 

\begin{equation}
    M_\alpha = \Sigma_i m_i \alpha_i / \Sigma_i m_i;~\alpha=x~{\rm or}~y
    \label{eq:1st_moment}
\end{equation}

\begin{equation}
    M_{\alpha\beta} = \Sigma_i m_i \alpha_i \beta_i / \Sigma_i m_i - M_\alpha M_\beta;~\alpha,\beta=x~{\rm or}~y
    \label{eq:2nd_moment}
\end{equation}

\begin{equation}
    \theta = \frac{1}{2}\arctan\left(\frac{2M_{xy}}{M_{xx}-M_{yy}}\right)
    \label{eq:angle}
\end{equation}

\paragraph{RS galaxy map}
The method for measuring the orientation angle and peak of the RS distribution is similar to that used for the mass map above -- we replace the mass map with the RS galaxy map. 

In Section~\ref{sec:compare_angle} and~\ref{sec:compare_peak}, we compare the orientations and centers, respectively, of the BCG, mass and RS distributions of each cluster in our sample and study their statistics. 


\subsection{Data Product Quality Check}\label{sec:quality_check}

We use a ``quality check'' step in the pipeline to examine the quality of our data products by analyzing the statistics of representative quantities. 
The major improvement in our quality check step is evaluating the star-galaxy shape correlation as a function of distance. In \paperi, we use shape correlations between the stars used for the PSF modeling and the galaxies used for the lensing analysis to estimate the PSF modeling error. 
In this work, we improve the calculation of shape correlations by using \verb|TreeCorr|\footnote{\url{https://github.com/rmjarvis/TreeCorr}}~\citep{Jarvis2004}.
\verb|TreeCorr| is implemented in \verb|C++| and uses a tree structure to divide the catalog into small cells. It takes the mean values in the cells that are greatly separated to compute the correlation for that distance bin, instead of using individual pairs of objects in that bin. 
It thus saves the computing resources.  
Therefore, we can reach deeper magnitudes for galaxies ($r<26$), and we find that the star-galaxy shape correlation is $\lesssim10^{-4}$, i.e., the PSF modeling error is $\lesssim10\%$ of the reduced shear. We present more details of assessing the performance of PSF modeling and shape measurement using shape correlations in Discussion (Section~\ref{sec:shape_corr}).


\section{Results}\label{sec:results}

\subsection{Maps}\label{sec:maps}
\begin{figure*}[htbp!]
    \centering
    \includegraphics[width=0.9\textwidth]{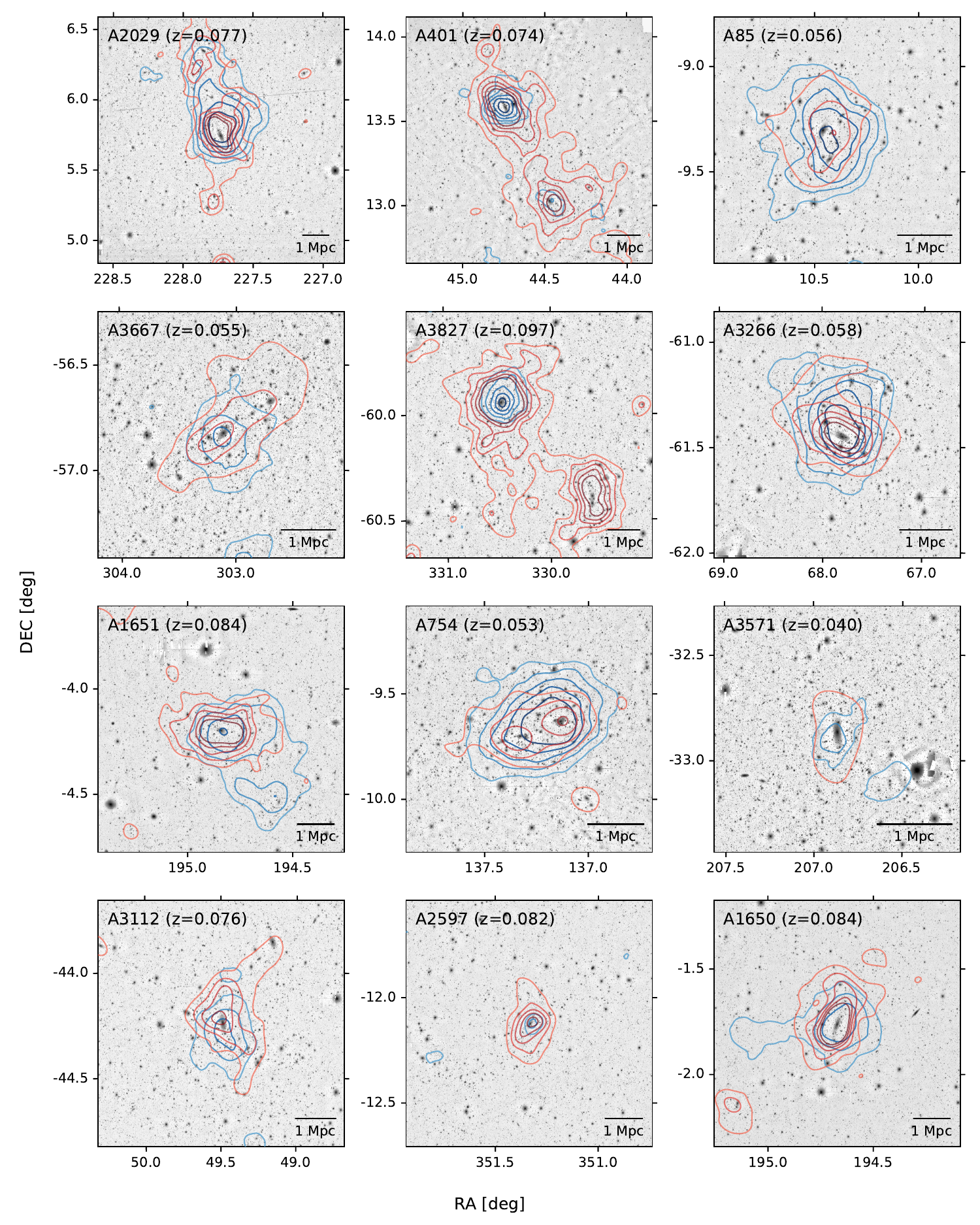}
    \caption{Grid of maps (Part I). In each diagram, the background is the inverted $r$-band co-added image; the blue contours stand for the aperture mass S/N, while the red contours represent the red-sequence galaxy distribution. The mass S/N contours extend from 2 to 6$\sigma$ with a step of 1, and we choose an aperture that maximizes  the S/N of the cluster-scale structure. The galaxy density contours extend from 0.024 to 0.12  per $100\times100$ pixels ($26\farcs3\times26\farcs3$) with a step of 0.024, and we use a smoothing length of 200~kpc ($\sigma$ of a 2D Gaussian) in the Kernel Density Estimation. The redshift is from SIMBAD~\citep{Wenger2000}.  
    The color schemes in all diagrams are the same; a darker color means a higher value. 
    }
    \label{fig:grid1}
\end{figure*}

\begin{figure*}[htbp!]
    \centering
    \includegraphics[width=0.9\textwidth]{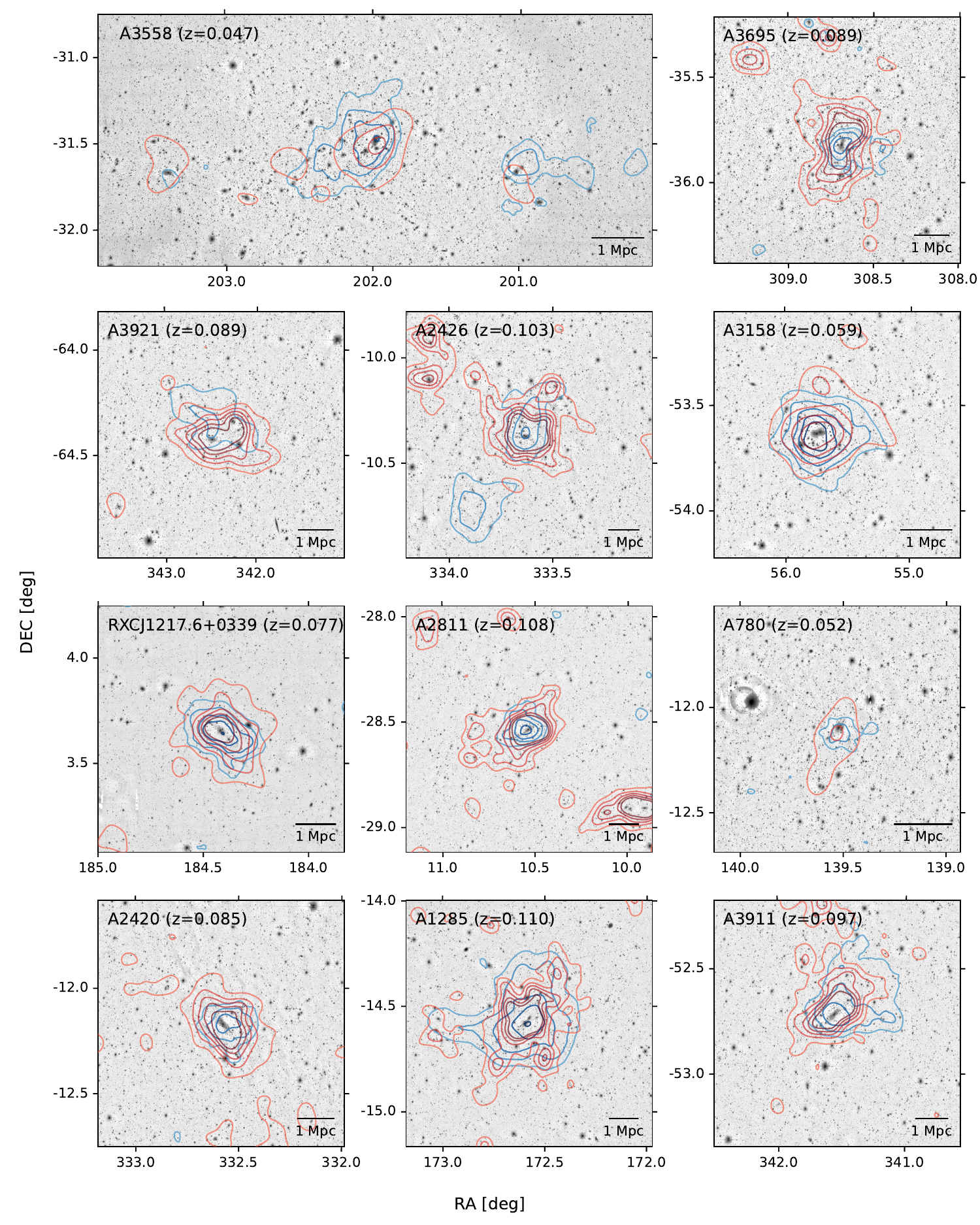}
    \caption{Grid of maps  (Part II). }
    \label{fig:grid2}
\end{figure*}

\begin{figure*}[htbp!]
    \centering
    \includegraphics[width=0.9\textwidth]{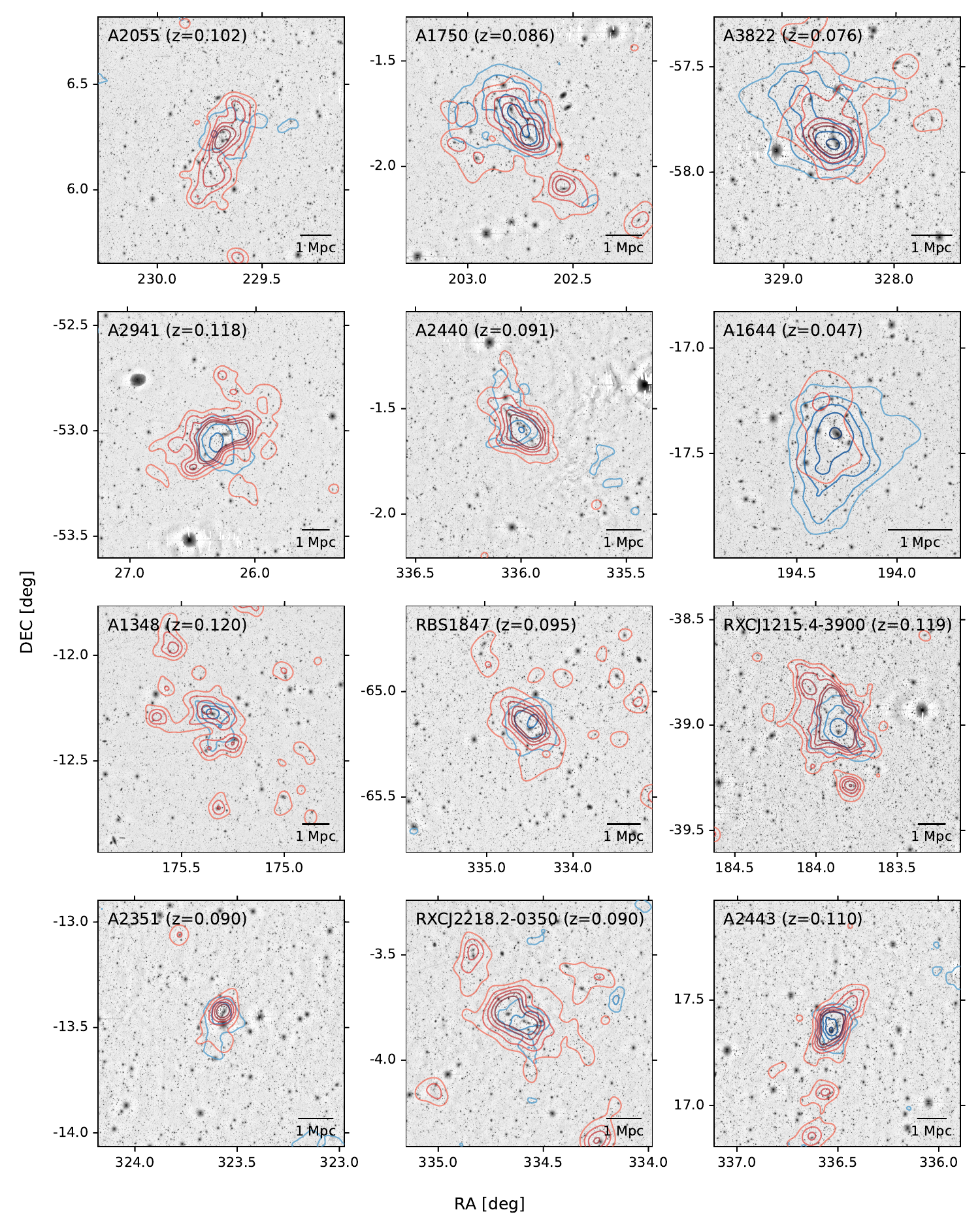}
    \caption{Grid of maps  (Part III). }
    \label{fig:grid3}
\end{figure*}

\begin{figure*}[htbp!]
    \centering
    \includegraphics[width=0.9\textwidth]{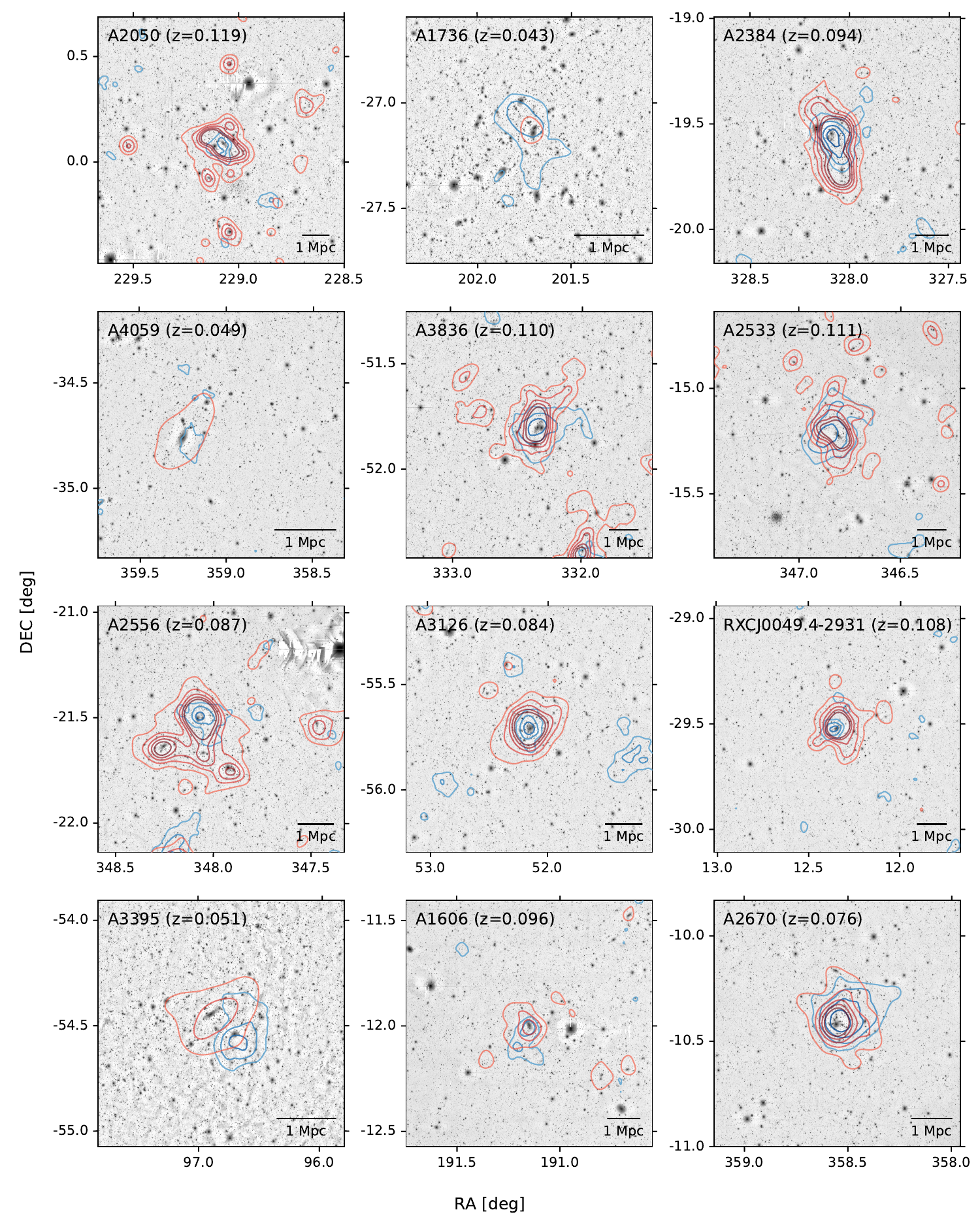}
    \caption{Grid of maps  (Part IV). }
    \label{fig:grid4}
\end{figure*}

\begin{figure*}[htbp!]
    \centering
    \includegraphics[width=0.9\textwidth]{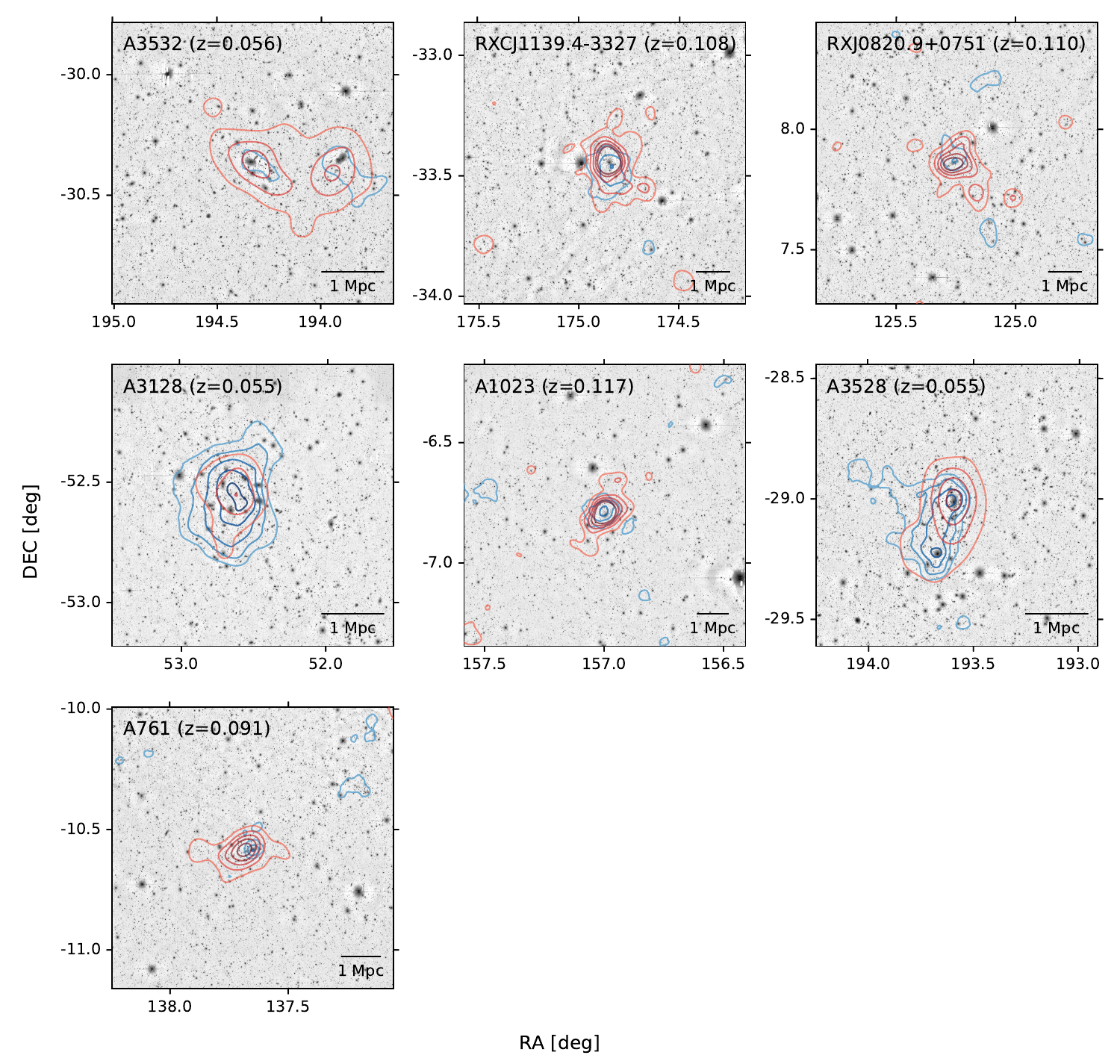}
    \caption{Grid of maps  (Part V). }
    \label{fig:grid5}
\end{figure*}

We present an $r$-band co-added image with the aperture mass S/N map and RS galaxy distribution overlaid in each diagram,  
in Figure~\ref{fig:grid1}, ~\ref{fig:grid2}, ~\ref{fig:grid3}, ~\ref{fig:grid4}, and~\ref{fig:grid5}  -- in total \clndiagram diagrams for \cln clusters sorted by X-ray luminosity; among them 4 diagrams include multiple LoVoCCS clusters (A2029/A2033, A401/A399, A3827/A3825, and A2556/A2554). 
If two LoVoCCS clusters are shown in one diagram, we label it with the one that has the higher X-ray luminosity. 
Each diagram shows the central $4\times4$ patches (16k$\times$16k pixels; $1.2\times1.2$ deg) around the cluster except the cluster pairs above and A3558 (the Shapley supercluster) to show the nearby structure. 
We use an \texttt{arcsinh} scaling to display both faint and bright features on the co-added image, and the local sky over-subtraction effect near bright and large objects is thus visible. Since the over-subtraction mainly affects the neighboring large-area and low-surface-brightness objects, we do not expect it to strongly bias the analysis of weak lensing signal and the detection of red-sequence galaxies -- the lensing sources have small sizes and the selected RS galaxies are bright.  

In the following text, each paragraph is labeled by the LoVoCCS cluster(s) in the corresponding diagram, although other nearby/projected clusters could be included in the field as well. In each field, we compare the optical image with the 2D mass and galaxy distributions to study the origin of clumps in those distributions.  
The term ``aperture size'' refers to the Schirmer filter radius. 
Note, the maximal S/N mentioned in the following text is for the mass map peak within an aperture, and therefore it is \textit{lower} than the detection S/N of the whole cluster. Instead, the detection S/N can be estimated by the error bar of the shear profile or the lensing mass of the cluster (e.g., the S/N of lensing masses presented in \paperi). 
Also, we note that a few clusters have previous lensing mass measurements but no reported lensing mass maps.  \rr1{If a cluster has reported lensing mass measurement but no mass map in the literature, we do not count our map as a ``first'' lensing mass map in the following paragraphs, because we do not know whether previous researchers have built a mass map for that cluster or not. But this indicates that the true number of first mass maps presented in this paper can be larger than what we report. } 
Compared to these previous studies, our dataset has larger wavelength coverage for better determination of galaxy types and redshifts, along with deeper observation and larger FoV for lensing analysis. 

\paragraph{A2029/A2033}
A2029 ($z=0.077$) and A2033 ($z=0.082$) have been widely studied using spectroscopic and lensing measurements in optical windows~\citep[e.g.,][]{Sohn2019,McCleary2020}. 
Compared to the previous study of~\citet{McCleary2020}, this work includes more exposures to improve the depth (by $\sim1$ mag for shape measurements) and uses a different pipeline for image processing and lensing analysis; our results are similar. 
The lensing effect comes from both A2029 and neighboring structures.
Here we use a large aperture (15k pixels in radius) where the S/N of the central peak is maximized ($7.6\sigma$) in the mass map, but the signal of the Southern Infalling Group (SIG) is smoothed out. The lensing signal of SIG reaches a maximum of $1.8\sigma$ at a small aperture (5k pixels), and the RS density clearly shows a concentration of galaxies at SIG. 
The RS map shows a density peak SE of A2033, which is also shown in the member distribution described by~\citet{Sohn2019}. We note that A2033 produces a stronger lensing effect than SIG, but has lower galaxy density than SIG, which is consistent with~\citet{Sohn2019} as well. 
We show the central $6\times6$ patches (24k$\times$24k pixels; $1.8\times1.8$ deg) to include both clusters and nearby structures. 

\paragraph{A401/A399}
Previous studies using X-ray and SZ observations show that A401 ($z=0.074$) and A399 ($z=0.072$) form a pre-merging pair with a gas filament connecting the two~\citep[e.g.,][]{Akamatsu2017,Bonjean2018,Hincks2022}.
Because their redshift difference is small, it is difficult to separate the RS of individual clusters, and thus we study their RS galaxy distributions together.
We find a clear bridge of RS galaxies connecting the two clusters, which is consistent with the gas and previous galaxy analysis~\citep{Bonjean2018}. 
Also, we present the first high-resolution lensing mass map of this cluster pair. We find that A401 has both stronger lensing signal and higher RS galaxy density than A399, which is similar to the gas distribution. Although we see no clear lensing signal from the filament, the mass contours of A399 (and the central galaxies of both clusters) roughly align with the direction connecting the two clusters.  
We use an aperture of 7k pixels in radius that maximizes the S/N peak of A401 ($6.7\sigma$) in the mass map, while the S/N peak of A399 reaches the maximum at 3k pixels ($3.4\sigma$) -- both lensing peaks are well detected despite cirrus and relatively high extinction ($\sim0.3$ mag in $r$ band) in this field. We present the central $5\times5$ patches (20k$\times$20k pixels; $1.5\times1.5$ deg) to show both clusters. 

\paragraph{A85}
A85 is a merging cluster and has been studied via WL by~\citet{McCleary2020} and in \paperi. Here we reanalyze it using the updated pipeline, and we also present its RS galaxy distribution (same for A3921 and A3911 below). The new mass map is consistent with the previous results, and our catalog is $\sim 1$ mag deeper than the analysis of~\citet{McCleary2020}.
We note that A85 has two central BCGs (NE and SW), and that the mass map S/N keeps increasing as the aperture grows because of the substructures (two central clumps around respective BCGs). 
The SW peak has slightly larger S/N than the NE one under small apertures ($<8$k pixels), indicating that it has a smaller physical size than the other, which also corresponds to a smaller BCG. 
We use an aperture (13k pixels; S/N $\sim 6$) before the two peaks merge into one.  
On the other hand, the RS distribution aligns with the orientations of two BCGs (NW-SE).

\paragraph{A3667}
A3667 is \rr1{previously} known as a merging cluster. We find that the RS galaxy distribution closely aligns (SE-NW) with the central galaxy, which also aligns with the gas distribution and the merging direction~\citep{Knopp1996}. The mass distribution  shows alignment with that direction only in the cluster periphery (below $2\sigma$) and has a large offset from that direction near the cluster center. 
Here we use an aperture radius of 14k pixels when  the mass S/N peak reaches a maximum ($4.4\sigma$). 
Our mass map is similar to the early result of~\citet{Joffre2000} for the cluster central region. 

\paragraph{A3827/A3825}
We present A3827 ($z=0.097$) and A3825 ($z=0.074$) in the same diagram because of their proximity on the plane of sky. However, they probably \rr1{are not interacting} because of their large redshift difference ($\Delta z \sim 0.02$), while their RS colors could be quite similar ($\Delta(g-r)\sim0.04$, close to the scatter of RS). We estimate the color based on the elliptical galaxy SED template \texttt{El\_B2004a} used in BPZ.  In the diagram we use an aperture of 7k pixels when the A3827 S/N peak reaches a maximum ($6.5\sigma$). A3825 S/N peak reaches its maximum ($2.4\sigma$) at a small aperture of 3k pixels and is smoothed out in larger apertures. The central galaxies of A3825 \rr1{are aligned} with its galaxy distribution. The central strong lensing feature \rr1{in} A3827 has been used to test the nature of dark matter~\rr1{\citep[e.g.,][]{Massey2018,Lin2023}}. We find a \rr1{$25''$} offset between the WL S/N peak and the BCG, which is below the mass map resolution -- we will further study this in the next section.

\paragraph{A3266}
A3266 is a merging cluster. 
Our result shows a strong correlation between the BCG orientation and the galaxy distribution (NE-SW); the mass distribution \rr1{is tilted} more towards the N-S direction, which is likely caused by a north group (instead of background clusters after visual inspection). The orientations are consistent with the merging direction determined by galaxy motion and gas~\citep{Dehghan2017,Sanders2022}. 
We present the first lensing map of this cluster -- the mass S/N peak reaches the maximum at an aperture radius of 13k pixels ($7.6\sigma$).

\paragraph{A1651}
A1651 exhibits similar morphology between the central galaxy, the galaxy distribution, and the large-scale mass distribution. The mass contours extend to the west. The SW lensing peak comes from a background cluster ($z\sim0.2$). The central mass S/N peak reaches the maximum at 12k pixels ($5.1\sigma$). \citet{Sifon2015} showed that A1651 is a relaxed cluster using spectroscopic measurements. This indicates that even nearly relaxed clusters can have elongated shapes. 

\paragraph{A754}
A754 is a merging cluster. Both the galaxy density and the mass distribution peak at the BCG, and they both align with the BCG orientation (SE-NW). Also, the galaxy density shows another peak at an SE group inside the cluster. In the mass map, a corresponding small clump is only shown in small apertures, which indicates that the substructure has a small characteristic size. The mass map peak gradually shifts to SE from NW as the aperture increases because of the SE subclump. Our mass map is similar to the early result from Subaru for the cluster central region~\citep{Okabe2008}. In the diagram we use a large aperture (14k pixels) that  maximizes the S/N ($7.7\sigma$).

\paragraph{A3571}
A3571 shows alignment between the BCG, the mass distribution, and the RS galaxy distribution (N-S). These components also align with the gas distribution of this relaxed cluster~\citep{Nevalainen2000}. The galaxy density value is relatively small because of the low redshift and the wide angular range. The mass NW tip likely comes from a group of member galaxies. Though we use the LSST pipeline flags to remove objects affected by saturation, here we further mask out objects within $0.25$ deg of the SW bright star (i Cen.) to reduce artifacts, and we find that the cluster mass map does not change greatly after the cut; the residual SW clump could be caused by some background structure. We use an aperture of 11k pixels when the peak S/N reaches the maximum ($3.9\sigma$). This is the first reported WL map of the low-z cluster A3571. It is also part of the Shapley supercluster -- we will present more members of this supercluster later in the text (A3558/A3532/A3528).

\paragraph{A3112}
In A3112, the mass distribution and the galaxy distribution roughly align with the BCG along the N-S direction, and both of their peaks have small offsets from the BCG. The mass peak has an offset towards south, which could be caused by a cluster sub-clump or a background cluster. 
The NW lobe in the RS distribution at large radii could come from a group of member galaxies; it is extending towards A3109 (NW 0.5 deg from A3112) and thus could be influenced by filaments inside the Horologium supercluster. 
We use an aperture of 11k pixels to reach the maximum $4.4\sigma$ in the mass map. This is the first high-resolution lensing map of A3112.

\paragraph{A2597}
A2597 shows weaker lensing signal compared to the clusters that have similar X-ray luminosities. We choose the aperture at 5k pixels to show a S/N maximum ($3.3\sigma$). The galaxy distribution aligns with the BCG at small radii (SE-NW) and tilts more to the N-S direction at large radii due to a group of member galaxies. 

\paragraph{A1650}
A1650 (SDSS-C41041) shows orientation alignment between the BCG, the mass distribution and the RS galaxy distribution. The east mass clump could come from a background cluster/group. 
For the mass map, we use an aperture of 14k pixels that maximizes the peak S/N  ($4.9\sigma$).

\paragraph{A3558}
A3558 lies \rr1{at} the center of the Shapley supercluster. In the diagram we present A3558 (center; $z=0.047$) and its neighboring clusters A3562 (east; $z=0.049$) and A3556 (west; $z=0.048$). We find alignment between the central galaxy, the mass distribution, and the galaxy distribution of  A3558. Both the mass distribution and the galaxy distribution peak at the BCG of A3558 and extend SE towards A3562 -- in between we notice low mass S/N detection ($\sim1\sigma$) and a low-density stream of bright RS galaxies ($\sim0.015$ per $100\times100$ pixels; $\sim25$ per Mpc$^2$), which indicates filaments connecting A3558, A3562, and the central groups between the two. On the other hand, we find no clear signal between A3558 and A3556 in both mass and galaxy distributions. Here we use an aperture of 12k pixels that maximizes the S/N  ($5.0\sigma$), and we show the central $11\times5$ patches (44k$\times$20k pixels; $3.2\times1.5$ deg) to include all three clusters. The mass and galaxy distributions are consistent with the gas distribution~\citep{Planck2014} and are similar to the previous results of ~\citet{Higuchi2020}, and our depth is $\sim1.5$ mag deeper.

\paragraph{A3695} 
We find alignment in orientations between the BCG, the galaxy distribution, and the mass distribution of A3695 at small radii, though these distributions show distorted shapes at large radii, which could be caused by merging. The central galaxy has two peaks (showing a ``dumbbell'' shape) along its major axis, indicating a possible merging direction (NW-SE). We use the mass map that has the highest S/N  peak ($4.9\sigma$) at an aperture of 6k pixels.

\paragraph{A3921}
A3921 is a merging cluster and has been studied via WL in~\paperi. Here we use a higher photo-z cut and obtain a similar mass map. We adopt an aperture of 11k pixels that maximizes the S/N  ($3.3\sigma$). Also, we include an RS map, which has two peaks -- one corresponds to the BCG (SE) and the other corresponds to a group (NW). These peaks also match the main and secondary components of the merging cluster. 

\paragraph{A2426}
A2426 shows alignment between the orientations of the BCG and the RS galaxy distribution (E-W), while the mass distribution is along the N-S direction. There may be an infalling group in the NW corner of the cluster, which would explain the mass distribution elongation. 
The mass concentration in the SE of the diagram is likely caused by a background cluster/group.  
The RS clumps at the NE corner of the diagram may come from a nearby/foreground group. 
For the mass map, we use an aperture of 10k pixels when the peak S/N reaches a maximum ($4.2\sigma$). 

\paragraph{A3158}
A3158 is a merging cluster \rr1{as} suggested by \rr1{its} X-ray emission and member galaxy \rr1{motions}~\citep{Whelan2022}. Similarly, the central galaxy shows two strong peaks, which also align with the galaxy and gas distributions (NW-SE). The mass distribution tilts more to the N-S direction near the cluster center. This is the first high-resolution WL map of A3158 -- here we use an aperture of 11k pixels that maximizes the S/N  ($6.7\sigma$). A3158 is in the Horologium supercluster close to A3128 (presented later; 2.2 deg away). After combining their catalogs, we find no clear sign of filaments between the two clusters except some isolated groups. 

\paragraph{RXCJ1217.6+0339}
The mass and galaxy distributions and the central galaxy of RXCJ1217.6+0339 (RBS1092/ZwCl1215+0400) show orientation alignment (NE-SW), and the offsets between their peaks are small. We use an aperture of 7k pixels that maximizes the peak S/N  ($6.0\sigma$). 

\paragraph{A2811} 
In A2811 the mass distribution orientation weakly aligns with the central galaxy (NW-SE). The galaxy distribution roughly aligns with the E-W direction at small radii and align with the central galaxy at large radii. The peak S/N reaches the maximum ($5.3\sigma$) at an aperture of 6k pixels as shown in the mass map. 
The galaxy overdensity at the SW corner of the diagram comes from A2804 but without clear lensing signal, which is likely caused by its low mass inferred from the X-ray emission~\citep{Obayashi1998,Sato2010}. 

\paragraph{A780}
A780 (Hydra A) shows alignment between the orientations of the BCG and the galaxy distribution (NW-SE); the RS galaxies extend more towards SE due to an infalling group (around LEDA87445; $\sim1$ Mpc SE of the cluster center), which is consistent with previous studies on the X-ray and spectroscopic data~\citep{Girardi2022}. 
The mass distribution is approximately along the E-W direction at small radii. Here we use an aperture of 6k pixels that reaches an S/N maximum of $4.0\sigma$.  

\paragraph{A2420}
The BCG and the galaxy distribution at small radii have similar orientations (NE-SW); at large radii the galaxy distribution is more disturbed due to nearby groups. 
The mass distribution extends in the E-W direction generally; we use an aperture of 7k pixels where the S/N reaches the maximum of $4.9\sigma$. 

\paragraph{A1285}
The BCG(s) of A1285 is  an interacting pair aligning with the NW-SE direction. The mass and RS galaxy distributions also generally align in that direction. 
Both the mass and RS maps indicate a nearby group to the east of the cluster, and there could be a few infalling groups on the periphery of the cluster suggested by the galaxy distribution. 
For the mass map, we use an aperture of 12k pixels when the peak S/N maximized at $6.1\sigma$. After mass fitting, we note that A1285 seems to have a higher mass compared to clusters that have similar X-ray luminosities. 

\paragraph{A3911}
A3911 has been analyzed via WL in~\paperi. We further include the RS galaxy distribution in this paper, which aligns with the BCG along the NW-SE direction. The BCG shows signs of interaction  and has two cores. The lensing mass map is similar to the result in~\paperi, after a higher photo-z cut on the source galaxies implemented in this work. Here we use an aperture of 10k pixels that maximizes the peak S/N  at $4.9\sigma$. The galaxy group at the north edge of the diagram may be related to a nearby cluster A3915 ($z=0.096$; $\sim0.7$ deg from A3911).

\paragraph{A2055}
A2055 has no sign of a giant BCG but a group of bright ellipticals concentrating at the cluster center and several RS galaxies spreading along the NW-SE direction.
The mass map shows a complex shape surrounding the central galaxies, and the peak has a clear offset from those central galaxies, which might indicate that the cluster is undergoing merger activity. We use an aperture of 8k pixels -- the mass S/N map peak reaches a maximum of $3.1\sigma$, which is much lower than the clusters that have similar X-ray luminosities and redshifts (and leads to a low lensing mass). 

\paragraph{A1750} 
A1750 is a merging cluster that has three subclusters located at the north (A1750N), center (A1750C), and south (A1750S).
The galaxy distribution shows peaks located at the BCG of each subcluster. 
The RS galaxy density has the highest peak at A1750C, which is also true for the X-ray surface brightness~\citep{Bulbul2016}. 
Note there is another galaxy density peak at the SW corner of the diagram corresponding to a nearby cluster -- SDSS-C41191 ($z=0.087$; likely low-mass). 
For the mass map, under a small aperture ($\leq5$k pixels) the peak is at A1750C ($\leq4.5\sigma$), but under a large aperture ($\geq6$k pixels) the peak is at A1750N. This suggests that A1750C has a smaller characteristic size than A1750N, but we also note a background cluster XCLASS468 ($z=0.56$) $\sim3$ arcmin SW of A1750N, which could partially contribute to the lensing signal around A1750N.
In the diagram we use an aperture of 10k pixels when the S/N is at a maximum of $5.4\sigma$ to show the cluster-scale structure. 
A1750S has no clear lensing signal (maximum $0.4\sigma$ at a small aperture of 4k pixels), and it produces weak X-ray emission as well~\citep{Bulbul2016}. 
The clump to the east of A1750 could come from a background compact group SDSS-CGB7175.  
Our results are similar to the previous Subaru study~\citep{Okabe2008}, but we use a larger FoV and thus are able to show the structure around A1750.

\paragraph{A3822}
A3822 has two close BCGs and a few bright galaxies at the cluster center. 
There is another BCG in the NE, and the mass and RS galaxy distributions extend towards that direction~\citep[similar phenomenon has been found in the X-ray emission;][]{Lakhchaura2014}, indicating a possible cluster merger.  
The central BCGs align along the NE-SW direction as well. However, near the cluster center the orientations  of mass and RS distributions align more with the E-W direction. 
In addition, there are signs of groups in the NW connecting A3822 with a neighboring  cluster A3806 ($z=0.075$; $\sim1.1$ deg from A3822).   
For the mass map, we use an aperture of 12k pixels where the S/N peak reaches the maximum ($6.3\sigma$). This is the first reported lensing result of A3822.

\paragraph{A2941}
The BCG(s) of A2941  has multiple cores showing signs of merging. The RS distribution aligns with the BCG (E-W). The mass distribution near the center is along the N-S direction and tilts more towards the NE-SW direction at large radii.   %
For the mass map we use an aperture of 9k pixels where the peak S/N reaches the maximum  ($4.4\sigma$). This is the first published lensing result of A2941. 

\paragraph{A2440}
A2440 is a merging cluster that shows a chain of bright elliptical galaxies along the NE-SW direction without clear sign of a single BCG. Nonetheless, the orientations of those bright ellipticals seem to align with the RS galaxy distribution, which is consistent with previous studies~\citep{Maurogordato2011} and we use a larger FoV.
The mass S/N map shows a complex shape -- at small radii, the mass distribution is roughly  along the NW-SE direction. 
We tested further masking a circular region with a radius of 0.05 $\deg$ around the northern bright star (HD212427; within the mass contour) and found no significant change on the mass map. A mass clump near the west edge of the diagram could come from a background cluster (RMJ222223.6-014423.7). Here we use an aperture of 7k pixels where the peak S/N reaches a maximum of $4.1\sigma$.

\paragraph{A1644}
A1644 is a merging cluster. 
The RS galaxy distribution shows two peaks -- a strong peak (A1644N) in the north corresponding to a group, and a weaker peak (A1644S) near the BCG. 
The mass distribution peaks at the BCG and extends to the south. 
In the mass map we use an aperture of 13k pixels where the peak S/N reaches the maximum ($6.2\sigma$). We note that under small apertures there is a small mass peak ($\sim2\sigma$ at an aperture of 3k pixels) near the northern group, but it could be caused by shape noise as well. The RS galaxy distribution approximately aligns with the mass distribution (the N-S direction). 
Our results are similar to the ones of~\citet{Monteiro-Oliveira2020}, and we use deeper data and a different pipeline. 

\paragraph{A1348}
A1348 shows alignment between the orientations of the BCG, the mass and galaxy distributions at small radii; both distributions extend to the south at large radii. The RS density peak shows a small offset from the BCG. 
The galaxy distribution has a complex shape at large radii, which could partially be caused by the infalling of neighboring groups, or projection effects of the nearby structure along the line of sight due to the red sequence scatter; a more detailed study requires sufficient spectroscopic redshift measurements.   
A foreground bright elliptical galaxy to the SW (2MASXJ11405929-1223536; $z=0.074$) and its neighboring galaxies could contaminate the RS distribution because of similar colors. 
For the mass map we use an aperture of 6k pixels that gives the maximal S/N ($4.8\sigma$).

\paragraph{RBS1847}
In RBS1847 (2MAXIJ2216-647) the central galaxy orientation is quite different from the RS and mass distributions. The single BCG is generally along the E-W direction. The mass distribution is along the NW-SE direction (at small radii), while the galaxy distribution is along the NE-SW direction.
The misalignment between the three might be related to the cluster dynamical history. 
We use an aperture of 10k pixels where the S/N peak reaches the maximum ($4.2\sigma$). This is the first lensing mass map of RBS1847.

\paragraph{RXCJ1215.4-3900}
In RXCJ1215.4-3900 the BCG mostly aligns with the E-W direction and is slightly tilted NE-SW. 
However the cluster shows alignment between the mass and galaxy distributions at large radii along the N-S direction, and the RS distribution shows a southern group that could be infalling. 
We use an aperture of 9k pixels where the S/N reaches the maximum ($4.8\sigma$).
This is the first lensing mass map of RXCJ1215.4-3900.

\paragraph{A2351}
A2351 exhibits alignment between the orientations of the BCG, the mass and galaxy distributions (NW-SE), especially near the cluster center. The BCG has a dumbbell shape showing signs of interaction. We tested the effect of adding masks (radius $0.05\deg$) around the central bright stars (V* AS Cap. and HD205131) near the BCG, but this did not change the mass map greatly. 
Here we use an aperture of 8k pixels where the S/N reaches a maximum of $4.0\sigma$.
This is the first lensing mass map of A2351.

\paragraph{RXCJ2218.2-0350}
RXCJ2218.2-0350 is a merging cluster.
Near the cluster center there is a stream of bright ellipticals, and their orientations are generally aligned with the galaxy and mass distributions (NE-SW). 
RXCJ2218.2-0350 can be divided into two subclusters at NE and SW respectively -- X-ray~\citep{Gu2019} and RS peaks are both at the NE subcluster, but the mass map peak is at the SW subcluster, which may have been affected by merging. 
A few nearby groups/clusters (with low masses) are also shown in the RS galaxy distribution.
We show the mass map with an aperture of 5k pixels when the peak S/N is maximized at $3.8\sigma$. 
This is the first mass map of RXCJ2218.2-0350.

\paragraph{A2443}
A2443 is a merging cluster. We find that the galaxy distribution is along the NW-SE direction, which is consistent with the gas~\citep{Clarke2013} and previous studies~\citep{Golovich2019}.
The mass distribution  is oriented NS and peaks at the BCG, while the RS density peak shifts to the north.  
We use an aperture of 5k pixels that maximizes the S/N peak ($6.2\sigma$) of the mass map in the diagram.

\paragraph{A2050}
A2050 shows alignment between the orientations of the BCG, the galaxy distribution, and (at small radii) the mass distribution along the NE-SW direction, and the distributions stretch more to the south at large radii.  
We show the mass map at an aperture of 4k pixels that  maximizes the S/N peak at $4.0\sigma$. 

\paragraph{A1736}
A1736 has multiple BCGs (e.g., IC4252, ESO509-8, ESO509-9; separated by $\sim0.7$~Mpc). The central galaxy ESO509-9 has a close and bright companion LEDA47075, and several bright ellipticals are scattered $\sim0.5 - 1$~Mpc from the cluster center, suggesting that A1736 is undergoing a merger.  
The RS distribution  aligns with the central BCG orientation (N-S) at small radii, while the mass distribution shows the alignment approximately at large radii. 
The mass distribution is along the NE-SW direction at small radii, and the mass peak lies to the NE ($\sim2'\sim0.1$~Mpc) of the central galaxy. 
In the mass map we use an aperture of 11k pixels that  maximizes the S/N peak at $3.9\sigma$. This is the first reported WL mass map of A1736.

\paragraph{A2384}
A2384 is a merging cluster. RXCJ2152.2-1942/A2384(B) is in the SW ($\sim1$ Mpc) of A2384. The BCGs of both (sub-)clusters generally align with the orientations of their combined mass and galaxy distributions (NE-SW) and the gas as well~\citep{Parekh2020}. 
Each cluster has a RS density peak -- A2384 has higher density, and the density drops and rises along the line from A2384 to A2384(B), indicating the filaments connecting the two clusters (consistent with ~\citealt{Maurogordato2011}). 
The mass map peak is close to the BCG of A2384.   
In the diagram we use an aperture of 7k pixels that maximizes the S/N peak of the mass map at $5.7\sigma$. 

\paragraph{A4059} 
A4059 has low (angular) surface density in mass and galaxy number due to its low redshift; the lensing signal  is further decreased by its high critical surface density at low redshift, which leads to low aperture mass S/N.
Despite that, A4059 shows alignment between the orientations of the BCG, the RS galaxy distribution, and roughly the mass distribution (NW-SE), and the gas distribution also has a similar orientation at large radii~\citep{Bartolini2022}. The mass peak is slightly shifted to the SW of the BCG, which could be affected by shape noise because of the low lensing signal, but we note that there is a concentration of gas in the SW of the BCG as well. 
In the diagram we show a mass map using an aperture of 5k pixels that maximizes  the peak S/N at $2.8\sigma$.
This is the first WL map of A4059.

\paragraph{A3836}
In A3836, the RS distribution  aligns with the BCG orientation (N-S), while the mass distribution extends along the E-W direction. 
In the mass map we use an aperture of 8k pixels when the S/N peak reaches a maximum  ($4.9\sigma$). This is the first reported lensing map of A3836. 
Near the diagram SW corner, the mass and RS peaks may correspond to a nearby cluster Ser148-4.

\paragraph{A2533} 
A2533 exhibits alignment between the orientations of the BCG and the RS galaxy distribution at small radii (NE-SW); at large radii the galaxy distribution is more tilted to the N-S direction. The mass distribution is along the NW-SE direction; the peak is slightly shifted to the east. We use an aperture of 8k pixels that  maximizes the S/N peak at $4.6\sigma$.  
This is the first lensing mass map of A2533.

\paragraph{A2556/A2554}
In the diagram we present A2556 (east; $z=0.087$) and A2554 (north; $z=0.109$), together with A2550 (middle-south; $z=0.123$), which all show concentrations of RS galaxies. 
The lensing signal and concentration of RS galaxies at the bottom of the diagram come from A2555 (south; $z=0.111$). 
Similar to A3827/A3825, A2554/A2556 likely have no interaction because of their redshift difference ($\Delta z \sim 0.02$), but their RS colors are similar ($\Delta(g-r)\sim0.04$), and A2556 has no significant lensing signal (maximum $\sim2.8\sigma$ at an aperture of 3k pixels; smaller at larger apertures). 
Likewise, A2550 does not show clear lensing signal (maximum $\sim2.5\sigma$ at an aperture of 3k pixels), and its color difference compared to A2556 is also small ($\Delta(g-r)\sim0.06$).  
A2554 shows alignment between the orientations of the BCG and the RS distribution (N-S), but the mass distribution align with the E-W direction generally. The RS distribution also aligns with the gas distribution, and the X-ray and spectroscopic observations indicate that A2554 is a perturbed system~\citep{Erdim2019}.
For A2556 we find the directions of the BCG and the RS distribution approximately align.
In the diagram we use an aperture of 6k pixels where the peak S/N of A2554 reaches the maximum ($4.5\sigma$), but A2556 and A2550 are at $\sim1.7\sigma$, indicating their small characteristic sizes.

\paragraph{A3126}
In A3126, both the galaxy and mass distributions show circular shapes at small radii, and the BCG is nearly circular as well (but shows signs of merging with a member galaxy at NW). 
At large radii, the galaxy distribution stretches along the NW-SE direction, while the mass distribution extends in the N-S direction. 
On the mass map, we also notice clumps in the north, SE, and SW of A3126, and they are shown in the DES Y3 mass map as well~\citep{Jeffrey2021}; these clumps could come from nearby/background groups. 
We use an aperture of 6k pixels where the peak S/N reaches a maximum at $4.7\sigma$. 

\paragraph{RXCJ0049.4-2931}
RXCJ0049.4-2931 (ACOS84) shows orientation alignment between the the BCG and the mass distribution (SE-NW), while the RS distribution is generally along the N-S direction. In the mass map we use an aperture of 6k pixels when the S/N peak reaches a maximum at $4.2\sigma$.
This is the first lensing mass map of RXCJ0049.4-2931. 

\paragraph{A3395}
A3395 is a merging cluster and has two components (A3395N and A3395S), each with a BCG (ESO161-8 and LEDA19057, respectively).   
The lensing peak is close to the southern BCG, and A3395S produces stronger X-ray emission than A3395N~\citep[e.g.,][]{Reiprich2021}. The RS peak is near the northern BCG, which is also brighter and larger than the southern BCG. The RS distribution and the northern BCG present similar orientations. 
We use an aperture of 10k pixels for the mass map, where the S/N peak reaches the maximum ($4.4\sigma$). This is the first reported WL map of A3395.

\paragraph{A1606}
A1606 has been studied by~\citet{McCleary2020} via WL. Here we re-analyze it using  deeper data ($\sim2$ mag) and present the RS map as well. 
The orientations of the RS distribution and the BCG are roughly aligned.  
In the mass map we use an aperture of 6k pixels where the S/N peak reach the maximum $3.9\sigma$.

\paragraph{A2670}
In A2670, the BCG and the mass and galaxy distributions at small radii are nearly circular.  At large radii the mass distributions is roughly along the NW-SE direction and the RS distribution is generally along the N-S direction. 
In the diagram we show the mass map using an aperture of 10k pixels that maximizes  the peak S/N at $5.8\sigma$. 

\paragraph{A3532}
A3532 has a close companion A3530 that is below our X-ray luminosity cut (west of A3532 in the diagram). 
The two clusters may just start to interact~\citep{Lakhchaura2013}. 
We note that under small apertures A3532 has a higher aperture mass S/N than A3530, while under large apertures it is the opposite. Therefore, we use an aperture of 8k pixels so that both peaks have close S/N ($\sim3$) to show the structure in both clusters. This is the first WL mass map of A3532. 
A3532 has higher RS density than A3530, and both clusters show orientation alignment between the BCG and the RS distribution. A3532 also shows alignment between the BCG and the mass distribution. 
The BCGs of both clusters have dumbbell shapes showing signs of merging.

\paragraph{RXCJ1139.4-3327}
In RXCJ1139.4-3327, we find that the central galaxy has two cores, which are both nearly circular near the center, but their merged galaxy envelope extending to the ICL is along the N-S direction. Interestingly, both the lensing mass and RS galaxies are generally distributed along the N-S direction as well. The distribution peaks are slightly shifted from the BCG, which could be physical or caused by noise. In the mass map we use an aperture of 8k pixels when the peak S/N reaches the maximum at $4.0\sigma$. This is the first lensing mass map of RXCJ1139.4-3327. 

\paragraph{RXJ0820.9+0751}
In RXJ0820.9+0751 (RXJ0821.0+0752), we find that the BCG and both the mass and galaxy distributions at small radii are along the NW-SE direction. 
At large radii, both distributions are more disturbed -- there is a group of RS galaxies to the south of the cluster ($\sim1$ Mpc within the lowest contour of the cluster), which could be an infalling group. Another RS density peak to the SW of the cluster ($\sim0.3$ deg) could come from a foreground group.
For the mass map we use an aperture of 8k pixels where the peak S/N reaches a maximum at $3.1\sigma$. This is the first reported WL mass map of RXJ0820.9+0751. 

\paragraph{A3128}
A3128 is a merging cluster in the Holograms supercluster (see also the paragraph for A3158 above). 
There are a few bright ellipticals at the cluster center showing different orientations, and in the NE of the cluster ($\sim0.5$ Mpc) there is a massive background cluster ACT-CLJ0330-5227/DESJ0330-5228  ($z=0.44$; $M_{\rm 200c}\sim10^{15}M_\odot$; \citealt{McCleary2015,Nord2016}) which can contribute to the lensing signal, and thus we mask the background sources within $r_{\rm 200c}$ of the background cluster.
The final mass map has two central peaks at NE and SW respectively (separated by $\sim4'\sim0.3$ Mpc), which is similar to the previous lensing result of~\citet{McCleary2015}, and we use deeper data (by $\sim 1$ mag in $u,g,z$ bands) and include $i,Y$ bands for photo-z measurements. 
We find that at large radii the mass and galaxy distributions are both approximately aligned with the N-S direction. 
Here we use an aperture of 11k pixels where the mass map S/N peak reaches the maximum at $6.2\sigma$.
Interestingly, the two mass map peaks roughly match the X-ray emission peaks of A3128 (after removing the background cluster emission;~\citealt{Werner2007}). 

\paragraph{A1023}
In A1023, the BCG orientation generally aligns with the galaxy distribution (NW-SE); the mass distribution aligns with that direction only at small radii, and is more circular at large radii. 
The mass peak at the NW corner of the diagram comes from a background cluster A1013. 
We use an aperture of 5k pixels that maximizes  the peak S/N at $4.4\sigma$.
This is the first mass map of A1023.

\paragraph{A3528}
A3528 is a merging cluster that has two BCGs separated by $\sim1$ Mpc at NW and SE, corresponding to two subclusters A3528N and A3528S. 
The RS galaxies gather near the northern BCG (ESO443-4) and their distribution aligns with the orientation of the northern BCG at small radii (the N-S direction); at large radii the galaxy distribution extends to SE and spreads out around the southern BCG (ESO443-7). 
The gas distribution orientation also aligns with the BCGs~\citep{Gastaldello2003}. 
The mass map shows two central peaks corresponding to the two BCGs under small apertures. 
When the aperture expands, the mass S/N map peak reaches a local maximum of $5.0\sigma$ at an aperture of 4k pixels; after that, the peak S/N keeps increasing as the aperture expands 
because of the large separation between the two central clumps, and the corresponding northern and southern peaks start to merge into one central peak at an aperture of 8k pixels. 
In the diagram we thus use an aperture of 7k pixels to show the two peaks. 
In addition, the NE clump in the mass map probably comes from a background cluster SPT-CLJ1256-2851 ($z=0.36$), but we expect that its contribution to the A3528 central lensing signal is small because of the large angular separation between the two clusters.
The mass clump at the bottom of the diagram could come from nearby/background groups.   
This is the first reported lensing mass distribution of A3528. 

\paragraph{A761}
In A761, the mass map peak is close to the BCG (PMNJ0910-1034), but the RS distribution is slightly offset to the east (its peak is close to another bright central galaxy LEDA976301). We use an aperture of 4k pixels for the mass map, where the S/N peak reaches the maximum ($4.1\sigma$). This is the first WL map of A761. The RS distribution shows alignment with the BCG orientation.

\subsection{Orientation Alignment between the Mass and Galaxy Distributions and the Brightest Cluster Galaxy}\label{sec:compare_angle}

\begin{figure*}[htbp!]
    \centering
    \plotone{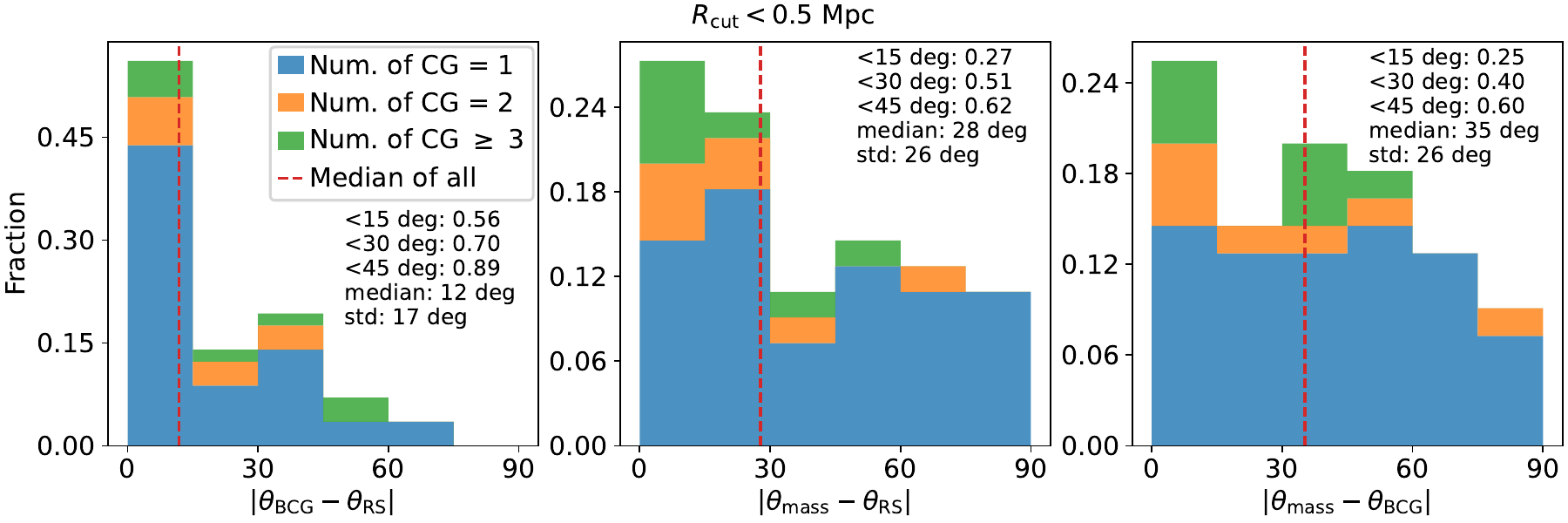}
    \plotone{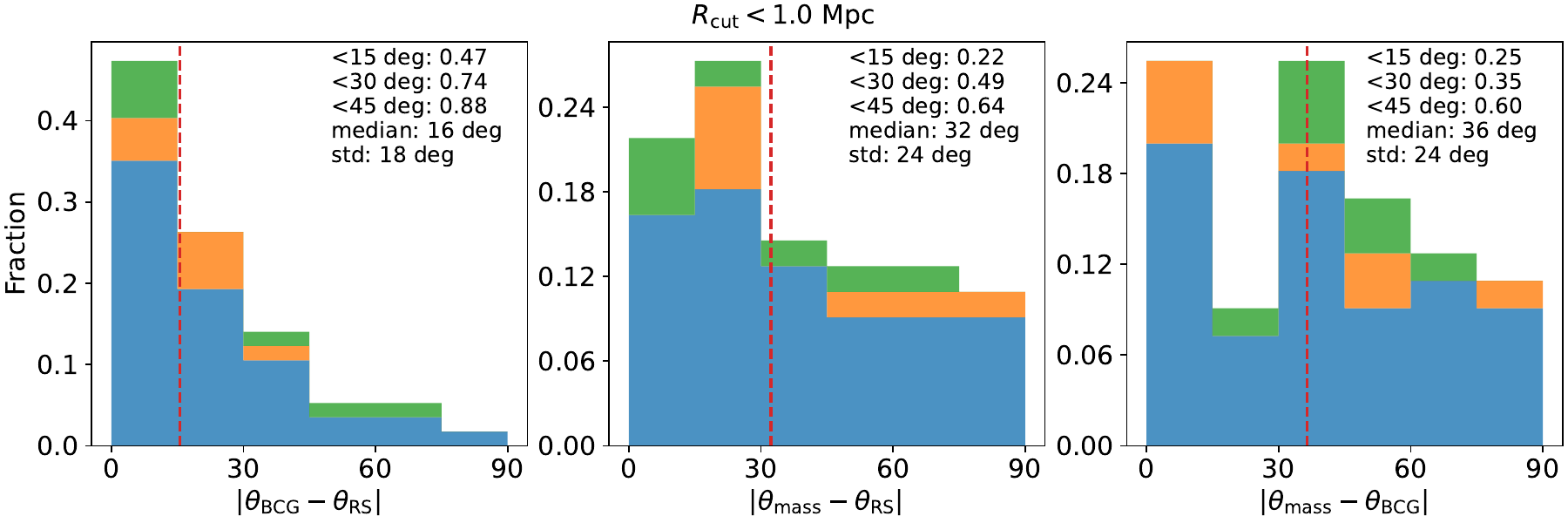}
    \plotone{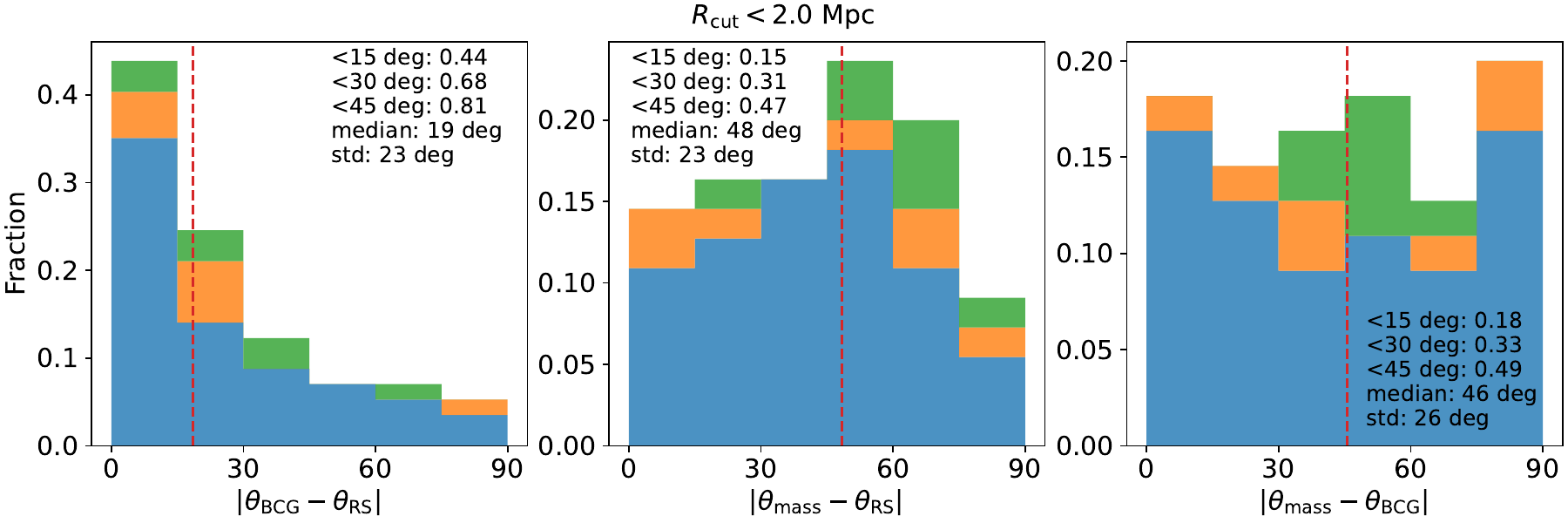}
    \caption{Orientation alignment between the BCG, the RS galaxy distribution, and the lensing mass distribution in each cluster of our sample. The histograms are stacked on top of each other. We consider the distributions within radial distance cuts of 0.5, 1.0, 2.0~Mpc, respectively. The Y-axis shows the fraction in the sample (the total fraction is 1). We give the fractions within 15, 30, and 45 deg in the text box, together with the median (the vertical dashed line) and the standard deviation of all clusters.  
    The angles are in degrees. 
    }
    \label{fig:alignment}
\end{figure*}

The figures in Section~\ref{sec:maps} show striking similarities between the orientations of the mass/RS distributions and the BCG, especially between the RS distribution and the BCG, in individual clusters. 
Therefore, we use the method described  in Section~\ref{sec:triaxiality} to quantify their orientation angles, and then we study the difference between those angles as a function of cluster dynamical state and radial distance range (Figure~\ref{fig:alignment}). 
When computing the  difference angle, we use its supplementary angle if it is larger than $90\deg$.  
Because it is difficult to determine the cluster dynamical state without enough spec-z data, we instead use the number of central galaxies (Number of CG = 1, 2, or $\geq3$) as a proxy -- clusters with more bright central galaxies tend to be more perturbed~\citep{Edwards2012,Mann2012,Furnell2018}. 
We require that the central galaxies being counted are sufficiently larger and  brighter than other member galaxies and clearly separate from neighboring large member galaxies. 
In the future, we will also seek to use the X-ray/SZ information to determine the cluster dynamical state.

In Figure~\ref{fig:alignment}, we find that, statistically, the RS galaxy distribution aligns with the BCG orientation throughout the entire cluster region and the alignment even extends to LSS, while the mass distribution and the RS galaxy distribution show alignment between their orientations only inside the cluster with a larger offset angle, and the mass distribution and the BCG align only near the cluster center. This is similar to the results of~\citet{Oguri2010} and~\citet{Chu2021}, but their samples are in different redshift ranges than ours and are studied using different instruments and methodologies. 
Also, we note that all types of alignment seem not to strongly depend on the cluster redshift and the dynamical state, however this observation could be limited by the cluster sample size. 
The ``significance'' of the orientation alignment can be estimated by Monte Carlo tests or analytically calculated by the cumulative function of a binomial distribution (Appendix~\ref{sec:angle_significance}). 
We find that the probability of more than half of the cluster sample having the angle difference below 30 deg is $\sim 0.5\%$, which is significantly small, assuming the angle difference is uniformly distributed between 0 and 90 deg. 
It indicates that only random fluctuation can not account for the large number of cases where the angle between the  RS distribution and the BCG/mass distribution (within 1 Mpc) is below 30 deg. 

The strong alignment between the orientations of the BCG and the RS distribution may result from the long-term dynamical process that those low-z massive clusters have undergone. The RS galaxies we have selected are relatively bright, and therefore they are older and have remained  in the cluster longer  than other members. The central galaxy and the RS galaxy distribution tend to align because of the gravitational potential~\citep{Donahue2016} or more specifically the filamentary accretion and the tidal field~\citep{Huang2016}.  
The cluster dwarf members and blue members can also affect the galaxy distribution morphology, but their membership is difficult to determine without spectroscopic data.  
For lensing mass distributions, shape noise affects their orientations randomly. Thus, the effect of shape noise is statistically reduced in a large cluster sample (like the sample in this work) for deriving the median misalignment, and the larger sample size helps bolster confidence in our results -- we will revisit this analysis for the whole LoVoCCS sample. 
Additionally, it is worth pointing out that previous studies mostly focus on the clusters at higher redshift ($z>0.1$) which generally experience shorter evolution time than this sample.

\subsection{Distribution of Center Offset}\label{sec:compare_peak}

The figures in Section~\ref{sec:maps} also show a rough consistency between the peaks of the maps and the BCG position. 
Here we compare those positions by studying their distances and relative position angles. 
The positions are measured by the method in Section~\ref{sec:triaxiality}, and we convert their angular separations into physical ones to  simplify comparison.   

\begin{figure*}[htbp!]
    \centering
    \plotone{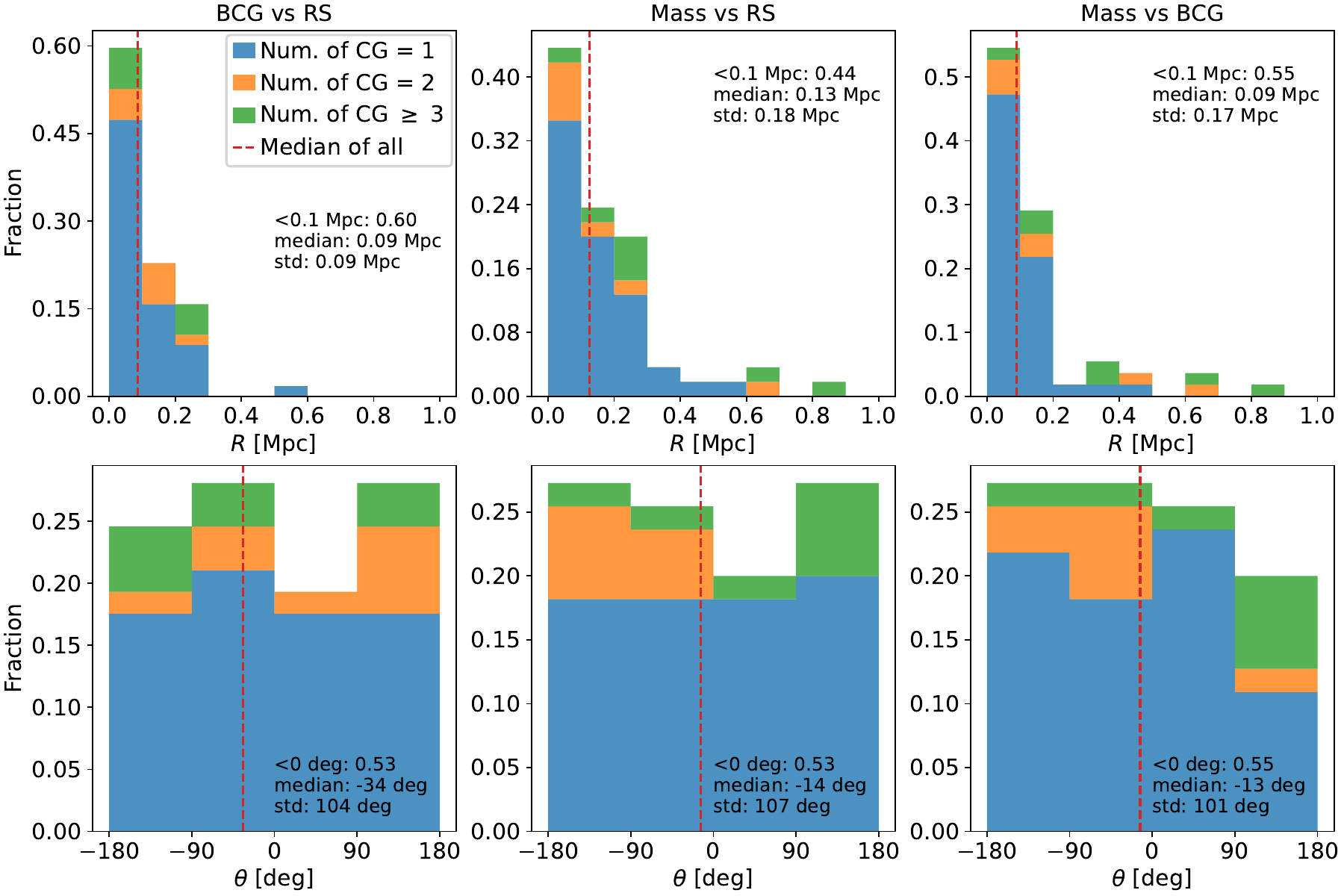}
    \caption{Offsets between the centers of the BCG, the RS distribution, and the mass distribution in each cluster of our sample. The \textit{top} panel shows the distance distribution, while the \textit{bottom} panel shows the (relative azimuth) angle distribution. 
    Similar to Figure~\ref{fig:alignment}, the histograms are stacked for visualization, and the vertical dashed line gives the median of all clusters in the sample. For reference, the standard deviation of a uniform distribution is $\sim29\%$ of the range (here $\sim 104$ deg). 
    }
    \label{fig:offset}
\end{figure*}

The result is presented in Figure~\ref{fig:offset}. 
We find that most positions are aligned (median $\sim0.1$~Mpc), but perturbed clusters tend to have larger separations. Also, the position difference has no clear dependence on the position angle suggesting no significant systematics. 
The distance distribution is similar to previous results~\citep[e.g.,][]{Oguri2010,vonderLinden2014,Donahue2016}. 

We note that mass maps could be affected by background clusters, and RS maps could be affected by photometry noise or nearby groups/LSS. However, when we combine the results from a sample of clusters (as in this work), we can treat the locations of interlopers as random, so that they contribute to noise without systematic bias at the peak location.

As the offset between the mass center and the BCG in relaxed clusters  (the BCG ``wobbling'') may indicate SIDM, 
we further discuss the result  in Section~\ref{sec:mass_bcg_offset}.


\section{Discussion}\label{sec:discussion}

\subsection{Shape Correlation, PSF modeling, and Shape Measurement}\label{sec:shape_corr}
As mentioned in Section~\ref{sec:quality_check}, we use \texttt{TreeCorr} to compute shape correlations -- 
the cross-correlation between the shapes of stars and galaxies, and the auto-correlation of stellar shapes. We use those correlations to estimate the PSF modeling bias in the lensing shear measurements.  The details are given below. 

We consider the correlations in  Eq.~\ref{eq:correlation} and~\ref{eq:correlation2}, where the superscripts $\alpha,\beta$ are star (s) or galaxy (g), the subscripts $i,j$ are 1 or 2,  $e_i(\textbf{R})$ is $i$th component of the ellipticity/shape of an object at coordinate $\textbf{R}$, $\textbf{r}$ is a vector with length $r$ connecting two objects, and $\hat{\textbf{r}}$ is the unit vector of $\textbf{r}$. 

\begin{equation}
    C_{ij}^{\alpha\beta}(r)=\langle e_{i}^{\alpha}(\textbf{R})\cdot e_{j}^{\beta}(\textbf{R+r})\rangle_{\textbf{R},\hat{\textbf{r}}}
    \label{eq:correlation}
\end{equation}

\begin{equation}
    C^{\alpha\beta}(r) 
    = C_{11}^{\alpha\beta}(r)+C_{22}^{\alpha\beta}(r)
    \label{eq:correlation2}
\end{equation}

To compute the correlations, we use the stars for PSF modeling in the central $6\times6$ patches, and we use their SDSS shapes derived from second moments without PSF correction.   
For galaxies, we use the high-quality selection from  lensing analysis and utilize their PSF-corrected HSM shapes (Section~\ref{sec:mass_map}; $r$-band depth $\sim26$ mag). 
The measured galaxy shape can be described as a sum of three elements (in the first order expansion) -- the intrinsic shape (shape noise), the WL induced distortion (proportional to shear), and the PSF residual (the PSF ``leakage''). After cross-correlating the galaxy shapes with the star shapes (equivalent to the PSF), we expect that the contributions from shape noise and lensing in the galaxy shapes should be averaged out, as they are random to PSF. If we assume that the PSF residual is proportional to the PSF, then the proportionality factor can be estimated  by the ratio between the galaxy-star and star-star correlations ($C^{\rm sg}/C^{\rm ss}$).
Here, we consider the contributions from the 1-1 and 2-2 component correlations; our test shows that the contributions from the 1-2 and 2-1 correlations are smaller.
Next, we can obtain an estimate  of the PSF shape magnitude from the star-star correlation ($\sqrt{ C^{\rm ss} }$), and thereby estimate the magnitude of the PSF residual by $C^{\rm sg}/\sqrt{C^{\rm ss}}$. 
Note, that term must be divided by 2 when compared to the (reduced) shear $g$ since $g_i\sim\langle e_i \rangle/(2\mathcal{R})$ in WL, and the typical shear responsivity $\mathcal{R}$ is $\sim1/1.2$ (\paperi) and can be treated as a secondary factor here. 
Similar methods have been used in the studies of~\citet{Jarvis2016} and~\citet{Mandelbaum2018a}.

\begin{figure*}[htb]
    \centering
    \plotone{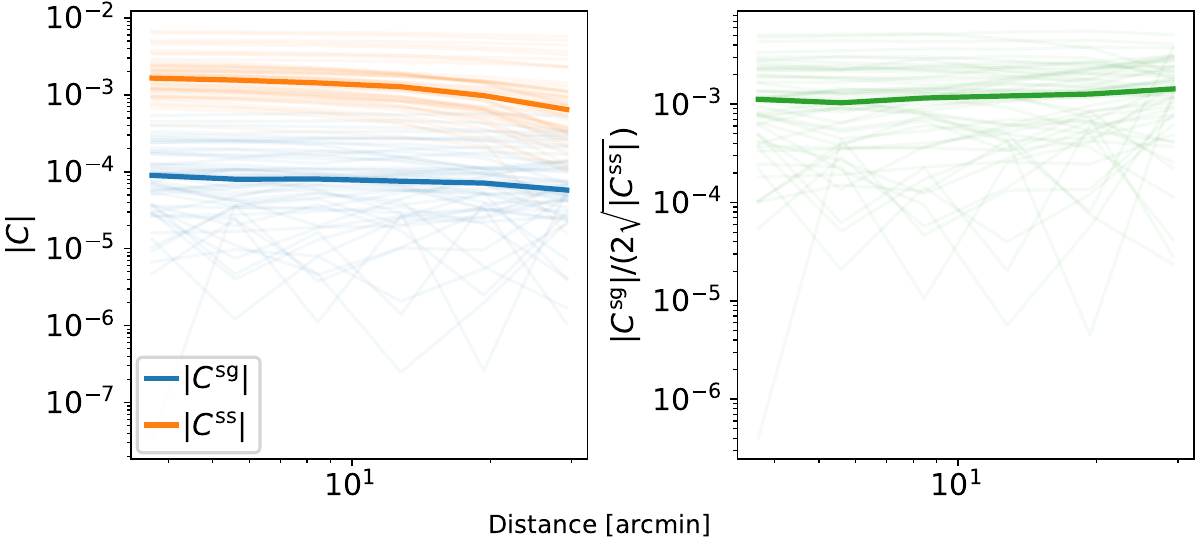}
    \caption{\textit{Left}: Star-galaxy shape correlations  ($C^{\rm sg}$) and star-star correlations ($C^{\rm ss}$) in absolute values. 
    \textit{Right}: The effect of PSF modeling bias on the shear measurement. 
    In both diagrams, the values in individual cluster fields are shown with low opacity, and their median values in each distance bin are shown in bold solid lines. 
    }
    \label{fig:shape_correlation}
\end{figure*}

Figure~\ref{fig:shape_correlation} shows that the effect of PSF modeling bias on the shear measurement -- the  $C^{\rm sg}/(2\sqrt{C^{\rm ss}})$ ratio -- is at the level of $\sim10^{-3}$. 
We tested and found that using the calibrated per-object shear estimate instead can produce a similar result. 
Since the typical shear value in the cluster region that we are interested in is $\sim0.02$ (at $\sim 0.5 r_{\rm 200c}$ for representative NFW halos; Section~\ref{sec:mass_bcg_offset}), we conclude that the PSF modeling bias on shear measurements is $\lesssim10\%$  (at percent level on average). 
We also note that the clusters with larger ratio values usually have larger masses, and thus their lensing shear signals are also larger, which compensates for the higher biases.

\subsection{Mass Map}

Some LoVoCCS clusters are covered by DES, and
we compare our mass maps with the recent DES Y3 WL mass maps~\citep{Jeffrey2021}. %
We select the Wiener map from the DES Y3 maps, which has the best performance among the methods used in that study. The Wiener map maximizes the convergence posterior using a Gaussian prior and gives better reconstruction of mass distribution for LSS than clusters.  
While the DES Y3 maps used sources from different redshift bins, we use the full redshift range that includes all galaxies at different redshifts, since our clusters are at very low redshift $z\lesssim0.1$. 
Although the DES Y3 maps are posterior convergence maps instead of aperture mass maps (as in this work), we expect the mass distribution morphologies described by the two types of maps to be consistent (Section~\ref{ref:photo_z},~\ref{sec:mass_map}). 
After comparison, we find that in those cluster fields our maps are similar to the DES Y3 maps, and that our maps have higher resolution resulting from depth.


\subsection{Mass-BCG Center Offset}\label{sec:mass_bcg_offset}

Following Section~\ref{sec:compare_peak}, here we further discuss the centroid offset between the lensing mass distribution and the BCG in relaxed clusters as a possible tracer of SIDM. 
We find that  clusters with only one bright central galaxy show a median offset $\sim0.09$~Mpc ($\sim1$~arcmin) between the centroids of the 2D mass distribution and the BCG. 
However, this offset may be limited by the catalog-binning size, and some of those clusters may still be perturbed. Also, shape noise may contribute to this offset~\citep{McCleary2020}.  

\begin{figure*}[htbp!]
    \centering
    \plottwo{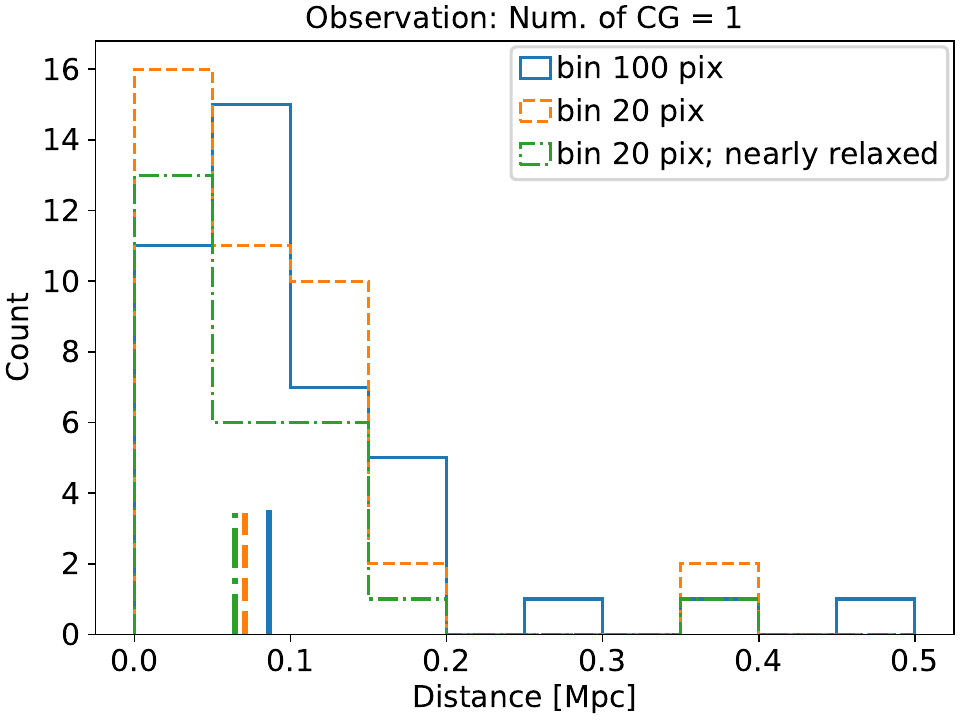}{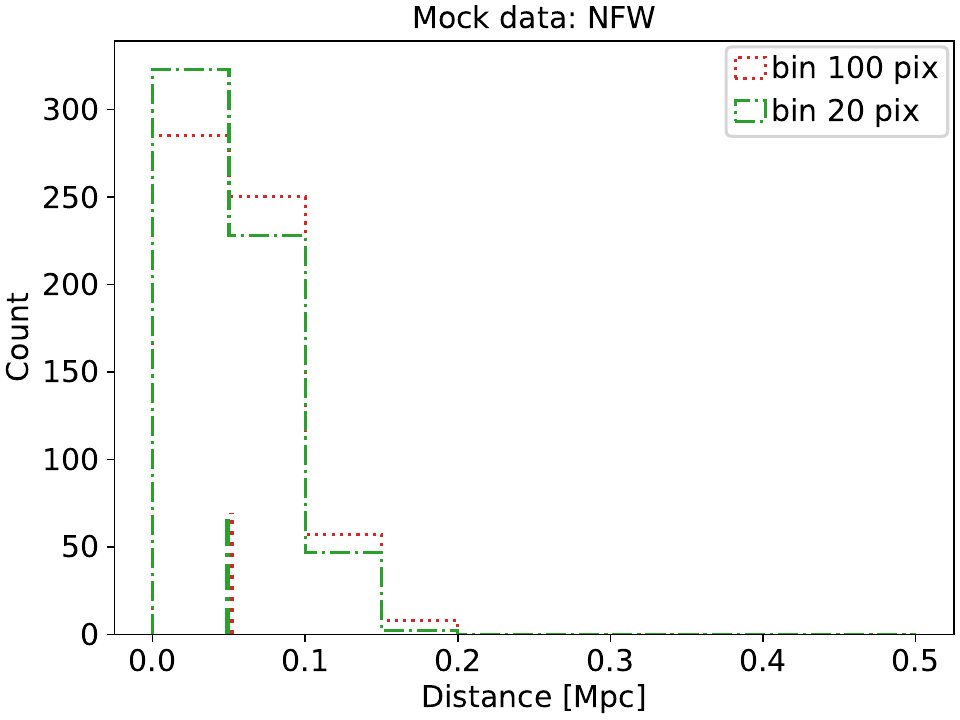}
    \caption{\textit{Left}: Histogram of mass-BCG center distance from observation. The blue solid curve corresponds to the distribution in Figure~\ref{fig:offset}. 
    The orange dashed curve shows the distribution of the same clusters under the 20-pixel binning. 
    The green dash-dot curve shows the distribution of selected clusters with the 20-pixel binning. 
    The vertical  lines at the bottom gives the median of each distribution. 
    \textit{Right}: Histogram of the distance between the NFW halo center and the detected lensing peak derived from the mock data (600 runs). 
    }
    
    \label{fig:mass-BCG_distance}
\end{figure*}

To test those effects, we first reduce the binning size to $20\times20$ pixels and rebuild the mass maps in higher resolution. 
Since the smaller binning leads to much longer computational time, we only produce maps in the central 
$\sim5-10\%$ 
region near the original peak and ensure the new central peak is within the window. This reanalysis indicates that the bin size reduction does not significantly change the trend between the aperture radius and the peak S/N. 
Additionally, we remove the clusters that have reported mergers (Section~\ref{sec:maps}) or have clear signs of interaction at the BCG (e.g., a dumbbell shape with double cores) by visual inspection. 
Then, we remake the  offset histogram and compare that with the previous result -- the left panel of Figure~\ref{fig:mass-BCG_distance} shows that the offset drops (median $\sim64$~kpc) when we use smaller bins and more relaxed clusters. 

Next, we study the contribution from shape noise by building mock shape catalogs distorted by a typical cluster in our sample with representative lensing sources. 
We consider a spherical NFW halo at redshift $z=0.08$ with mass $M_{\rm 200c}=7\times10^{14}M_{\odot}$, and we set the source galaxies at redshift $z=0.7$ with per-component shape dispersion 0.4 and number density 7 per arcmin$^2$  
-- those are approximately the median values in our final catalogs for lensing analysis. 
Then, we bin the catalog by 100 pixels and 20 pixels, make  mass maps respectively, and measure the distance between the mass map central peak and the NFW halo center. The distributions of their offsets derived from mock catalogs are shown in the right panel of Figure~\ref{fig:mass-BCG_distance}. 
For the 100-pixel binning, the median values in the real (nearly relaxed) and mock observations are $\sim90$~kpc and  $\sim52$~kpc, respectively. 
Taking the square root of the difference between their squares (the intrinsic offset and the shape noise are independent), we conclude that the intrinsic/net offset is $\lesssim73$~kpc. Similarly, under the 20-pixel binning, the median values in the real and mock observations are $\sim64$~kpc and  $\sim49$~kpc, respectively, and we estimate a net offset $\lesssim41$~kpc. 
Note, the true value could still be smaller since some clusters might not be fully relaxed, the sample size is limited, and there could be other contributing noise factors. 
Nonetheless, our result sets an upper limit of the intrinsic offset, which is still larger than the common values in SIDM tests~\citep[$\lesssim10$~kpc, e.g.,][]{Harvey2019}. 
In the future, we will revisit this problem through the entire LoVoCCS sample, utilize the gas centroid (obtained from high-resolution X-ray observations) as a tracer of the mass peak, and use more complete spectroscopic data to determine the dynamical state. We will further test the effect of shape noise using mock catalogs derived from cosmological simulations~\citep[e.g.,][]{Dietrich2012b}. 
Another interesting follow-up study might examine the relationship between the offset and the BCG orientation.

\section{Summary}\label{sec:summary}
In this work, we analyzed the complete DECam observations of \cln LoVoCCS clusters spanning a wide range of dynamical states, redshifts, and X-ray luminosities and produced their lensing mass and red-sequence galaxy maps. The depth of observation is unprecedented for the majority of our clusters, and for about half of the sample, the weak lensing maps presented in this work are the first in the literature.

We present and inspect the mass and galaxy distributions, and compare them with previous studies (Section~\ref{sec:maps}). 
We also study the correlations between their orientation angles and centroids (Section~\ref{sec:compare_angle} and~\ref{sec:compare_peak}). 
Many clusters in our sample exhibit alignment between the orientations and/or centroids of the BCG and mass/galaxy distributions, without strong dependence on the dynamical state, redshift, and mass. 
The orientation alignment is stronger near the cluster center than the outer region, and it is stronger between the BCG and the RS galaxy distribution. 
The BCG and the RS distribution have a median misalignment angle of 19 deg within a radial cut of 2~Mpc; the median center offset is 0.09~Mpc. 
The mass and RS distributions show a median misalignment angle of 32 deg within 1~Mpc; the median center offset is 0.13~Mpc. 
Between the mass distribution and the BCG, the median misalignment angle within 0.5~Mpc is 35 deg, and the median center offset is 0.09~Mpc. 
Our results are comparable to recent studies of different samples~\citep[e.g.,][]{Umetsu2018}. 
\rr1{
Additionally, \citet{Zhou2023} showed that the central galaxy and member galaxy distribution are more closely aligned at lower redshift and higher cluster mass  (i.e., the cluster evolution enhances the alignment), which is consistent with our results -- our sample focuses on nearby massive clusters and show signs of stronger alignment between the central galaxy and galaxy distribution compared to the literature. 
}

We are actively acquiring additional spectroscopic redshifts of both background source galaxies and cluster members. Combined with anticipated contributions from the Dark Energy Spectroscopic Instrument (DESI) survey, future observations are expected to significantly enhance the spec-z completeness for LoVoCCS clusters, which will be instrumental in refining the calibration of photometric redshifts. Improved spec-z measurements will also enable more accurate determination of cluster galaxy membership and dynamical state, sharpening our detection of red sequence and blue cloud galaxies and providing valuable insights into the star formation histories of our cluster sample. 
In parallel, alternative methods for distinguishing red and blue cluster member galaxies are under study, such as the Gaussian mixture model-based Red Dragon algorithm~\citep{Black2022}. The future scope of our work includes broadening the alignment studies to test the intrinsic alignment of cluster member galaxies, and to incorporate gas morphology via X-ray and Sunyaev–Zel'dovich (SZ) observations. We intend to utilize the distributions of mass, galaxies, and gas to explore splashback features in greater detail. Finally, the results of parametric mass fitting for LoVoCCS clusters will be detailed in an upcoming publication. 
Our data products including co-added images and final catalogs will be made available to the research community, and interested parties can obtain the mass and galaxy maps from the corresponding author. 
The LoVoCCS DECam exposures are publicly available on the NOIRLab Astro Data Archive website.\footnote{\url{https://astroarchive.noirlab.edu}}


\section*{Acknowledgments}
\rr1{We thank the anonymous reviewer for constructive comments that have helped improve the paper.}
We thank the LSST Data Management team and the CTIO observation support team.
We thank Xiangchong Li, Chun-Hao To, Yuanyuan Zhang, Thomas Matheson, Camille Avestruz for 
comments and support. 
We thank the members of the Observational Cosmology, Gravitational Lensing and Astrophysics Research Group at Brown University for contributing to the tests of the LoVoCCS \texttt{run\_steps} pipeline on different galaxy clusters, especially previous/current students and group members: Isabel Horst, Taylor Knapp, Derek Waleffe, Andrea Minot, Claire Hawkins, Alexandra Ekstrom, Natalie Rugg, Nicholas Conroy, Miguel Ruiz Granda, Liam Storan, Jessica Washington, Pei-Jun Huang, Francesco Serraino, Richard Dong, Youngik Lee, Dani Romero Mejia, Tim Launders, Vini Amit Rupchandani, Alexander Green. 

I.D. and D.C. are thankful for support from the National Science Foundation (No. AST-2108287; Collaborative Research; LoVoCCS: The Local Volume Complete Cluster Survey). 
G.W. gratefully acknowledges support from the National Science Foundation through grant AST-2205189 and from HST program number GO-16300. M.D., D.T., and A.E. are grateful for support from the National Aeronautics and Space Administration Astrophysics Data Analysis Program (NASA-80NSSC22K0476). 
K.U. acknowledges support from the National Science and Technology Council of Taiwan (grant NSTC 112-2112-M-001-027-MY3) and from the Academia Sinica Investigator Award (AS-IA-112-M04). 
J.S. was supported by the National Research Foundation of Korea (NRF) grant funded by the Korea government (MSIT) (No. RS-2023-00210597). P.N.~acknowledges support from DOE grant \#DE-SC0017660. E.P. is supported by NASA grant 80NSSC23K0747. The work of S.F. and T.L. is supported by NOIRLab, which is managed by the Association of Universities for Research in Astronomy (AURA) under a cooperative agreement with the National Science Foundation. 

This research was conducted using computational resources and services at the Center for Computation and Visualization, Brown University. 

This work has made use of data from the European Space Agency (ESA) mission
{\it Gaia} (\url{https://www.cosmos.esa.int/gaia}), processed by the {\it Gaia}
Data Processing and Analysis Consortium (DPAC,
\url{https://www.cosmos.esa.int/web/gaia/dpac/consortium}). Funding for the DPAC
has been provided by national institutions, in particular the institutions
participating in the {\it Gaia} Multilateral Agreement. 

The Pan-STARRS1 Surveys (PS1) and the PS1 public science archive have been made possible through contributions by the Institute for Astronomy, the University of Hawaii, the Pan-STARRS Project Office, the Max-Planck Society and its participating institutes, the Max Planck Institute for Astronomy, Heidelberg and the Max Planck Institute for Extraterrestrial Physics, Garching, The Johns Hopkins University, Durham University, the University of Edinburgh, the Queen's University Belfast, the Harvard-Smithsonian Center for Astrophysics, the Las Cumbres Observatory Global Telescope Network Incorporated, the National Central University of Taiwan, the Space Telescope Science Institute, the National Aeronautics and Space Administration under Grant No. NNX08AR22G issued through the Planetary Science Division of the NASA Science Mission Directorate, the National Science Foundation Grant No. AST-1238877, the University of Maryland, Eotvos Lorand University (ELTE), the Los Alamos National Laboratory, and the Gordon and Betty Moore Foundation. \rr1{All the {\it PS1} data used in this paper can be found in MAST~\citep{PS1-DB}.}

The national facility capability for SkyMapper has been funded through ARC LIEF grant LE130100104 from the Australian Research Council, awarded to the University of Sydney, the Australian National University, Swinburne University of Technology, the University of Queensland, the University of Western Australia, the University of Melbourne, Curtin University of Technology, Monash University and the Australian Astronomical Observatory. SkyMapper is owned and operated by The Australian National University's Research School of Astronomy and Astrophysics. The survey data were processed and provided by the SkyMapper Team at ANU. The SkyMapper node of the All-Sky Virtual Observatory (ASVO) is hosted at the National Computational Infrastructure (NCI). Development and support of the SkyMapper node of the ASVO has been funded in part by Astronomy Australia Limited (AAL) and the Australian Government through the Commonwealth's Education Investment Fund (EIF) and National Collaborative Research Infrastructure Strategy (NCRIS), particularly the National eResearch Collaboration Tools and Resources (NeCTAR) and the Australian National Data Service Projects (ANDS).

Funding for the Sloan Digital Sky Survey (SDSS) has been provided by the Alfred P. Sloan Foundation, the Participating Institutions, the National Aeronautics and Space Administration, the National Science Foundation, the U.S. Department of Energy, the Japanese Monbukagakusho, and the Max Planck Society. The SDSS Web site is \url{https://www.sdss.org}.
The SDSS is managed by the Astrophysical Research Consortium (ARC) for the Participating Institutions. The Participating Institutions are The University of Chicago, Fermilab, the Institute for Advanced Study, the Japan Participation Group, The Johns Hopkins University, Los Alamos National Laboratory, the Max-Planck-Institute for Astronomy (MPIA), the Max-Planck-Institute for Astrophysics (MPA), New Mexico State University, University of Pittsburgh, Princeton University, the United States Naval Observatory, and the University of Washington. 

The VISTA Hemisphere Survey data products served at Astro Data Lab are based on observations collected at the European Organisation for Astronomical Research in the Southern Hemisphere under ESO programme 179.A-2010, and/or data products created thereof. 
Based on data obtained from the ESO Science Archive Facility with DOI: \url{https://doi.org/10.18727/archive/57}. 
Based on data products created from observations collected at the European Organisation for Astronomical Research in the Southern Hemisphere under ESO programme 179.A-2010 and made use of data from the VISTA Hemisphere survey~\citep{McMahon2013}  with data pipeline processing with the VISTA Data Flow System~\citep{Irwin2004,Lewis2010,Cross2012}. 

The Hyper Suprime-Cam (HSC) collaboration includes the astronomical communities of Japan and Taiwan, and Princeton University. The HSC instrumentation and software were developed by the National Astronomical Observatory of Japan (NAOJ), the Kavli Institute for the Physics and Mathematics of the Universe (Kavli IPMU), the University of Tokyo, the High Energy Accelerator Research Organization (KEK), the Academia Sinica Institute for Astronomy and Astrophysics in Taiwan (ASIAA), and Princeton University. Funding was contributed by the FIRST program from the Japanese Cabinet Office, the Ministry of Education, Culture, Sports, Science and Technology (MEXT), the Japan Society for the Promotion of Science (JSPS), Japan Science and Technology Agency (JST), the Toray Science Foundation, NAOJ, Kavli IPMU, KEK, ASIAA, and Princeton University. 
This paper makes use of software developed for the Large Synoptic Survey Telescope. We thank the LSST Project for making their code available as free software at  \url{https://www.lsst.org/about/dm}. 
This paper is based [in part] on data collected at the Subaru Telescope and retrieved from the HSC data archive system, which is operated by the Subaru Telescope and Astronomy Data Center (ADC) at National Astronomical Observatory of Japan. Data analysis was in part carried out with the cooperation of Center for Computational Astrophysics (CfCA), National Astronomical Observatory of Japan. The Subaru Telescope is honored and grateful for the opportunity of observing the Universe from Maunakea, which has the cultural, historical and natural significance in Hawaii. 

This research has made use of the Spanish Virtual Observatory (\url{https://svo.cab.inta-csic.es}) project funded by MCIN/AEI/10.13039/501100011033/ through grant PID2020-112949GB-I00. 

This research has made use of the NASA/IPAC Extragalactic Database (NED), which is funded by the National Aeronautics and Space Administration and operated by the California Institute of Technology.

This research has made use of the SIMBAD database, operated at CDS, Strasbourg, France.

This research has made use of data, software and/or web tools obtained from the High Energy Astrophysics Science Archive Research Center (HEASARC), a service of the Astrophysics Science Division at NASA/GSFC and of the Smithsonian Astrophysical Observatory's High Energy Astrophysics Division.

This project used data obtained with the Dark Energy Camera (DECam), which was constructed by the Dark Energy Survey (DES) collaboration. This work is based on observations at Cerro Tololo Inter-American Observatory, NSF's NOIRLab (NOIRLab Prop. ID 2019A-0308; PI: I. Dell'Antonio). 
This project used public archival data from the Dark Energy Survey (DES). Funding for the DES Projects has been provided by the U.S. Department of Energy, the U.S. National Science Foundation, the Ministry of Science and Education of Spain, the Science and Technology Facilities Council of the United Kingdom, the Higher Education Funding Council for England, the National Center for Supercomputing Applications at the University of Illinois at Urbana-Champaign, the Kavli Institute of Cosmological Physics at the University of Chicago, the Center for Cosmology and Astro-Particle Physics at the Ohio State University, the Mitchell Institute for Fundamental Physics and Astronomy at Texas A\&M University, Financiadora de Estudos e Projetos, Funda{\c c}{\~a}o Carlos Chagas Filho de Amparo {\`a} Pesquisa do Estado do Rio de Janeiro, Conselho Nacional de Desenvolvimento Cient{\'i}fico e Tecnol{\'o}gico and the Minist{\'e}rio da Ci{\^e}ncia, Tecnologia e Inova{\c c}{\~a}o, the Deutsche Forschungsgemeinschaft, and the Collaborating Institutions in the Dark Energy Survey.
The Collaborating Institutions are Argonne National Laboratory, the University of California at Santa Cruz, the University of Cambridge, Centro de Investigaciones Energ{\'e}ticas, Medioambientales y Tecnol{\'o}gicas-Madrid, the University of Chicago, University College London, the DES-Brazil Consortium, the University of Edinburgh, the Eidgen{\"o}ssische Technische Hochschule (ETH) Z{\"u}rich,  Fermi National Accelerator Laboratory, the University of Illinois at Urbana-Champaign, the Institut de Ci{\`e}ncies de l'Espai (IEEC/CSIC), the Institut de F{\'i}sica d'Altes Energies, Lawrence Berkeley National Laboratory, the Ludwig-Maximilians Universit{\"a}t M{\"u}nchen and the associated Excellence Cluster Universe, the University of Michigan, the National Optical Astronomy Observatory, the University of Nottingham, The Ohio State University, the OzDES Membership Consortium, the University of Pennsylvania, the University of Portsmouth, SLAC National Accelerator Laboratory, Stanford University, the University of Sussex, and Texas A\&M University.
Based on observations at Cerro Tololo Inter-American Observatory, a programme of NOIRLab (NOIRLab Prop. 2012B-0001; PI J. Frieman).

The Legacy Surveys consist of three individual and complementary projects: the Dark Energy Camera Legacy Survey (DECaLS; Proposal ID \#2014B-0404; PIs: David Schlegel and Arjun Dey), the Beijing-Arizona Sky Survey (BASS; NOAO Prop. ID \#2015A-0801; PIs: Zhou Xu and Xiaohui Fan), and the Mayall z-band Legacy Survey (MzLS; Prop. ID \#2016A-0453; PI: Arjun Dey). DECaLS, BASS and MzLS together include data obtained, respectively, at the Blanco telescope, Cerro Tololo Inter-American Observatory, NSF's NOIRLab; the Bok telescope, Steward Observatory, University of Arizona; and the Mayall telescope, Kitt Peak National Observatory, NOIRLab. The Legacy Surveys project is honored to be permitted to conduct astronomical research on Iolkam Du'ag (Kitt Peak), a mountain with particular significance to the Tohono O'odham Nation.
BASS is a key project of the Telescope Access Programme (TAP), which has been funded by the National Astronomical Observatories of China, the Chinese Academy of Sciences (the Strategic Priority Research Programme `The Emergence of Cosmological Structures' Grant \# XDB09000000), and the Special Fund for Astronomy from the Ministry of Finance. The BASS is also supported by the External Cooperation Programme of Chinese Academy of Sciences (Grant \# 114A11KYSB20160057), and Chinese National Natural Science Foundation (Grant \# 11433005).
The Legacy Survey team makes use of data products from the Near-Earth Object Wide-field Infrared Survey Explorer (\textit{NEOWISE}), which is a project of the Jet Propulsion Laboratory/California Institute of Technology. \textit{NEOWISE} is funded by the National Aeronautics and Space Administration.
The Legacy Surveys imaging of the DESI footprint is supported by the Director, Office of Science, Office of High Energy Physics of the U.S. Department of Energy under Contract No. DE-AC02-05CH1123, by the National Energy Research Scientific Computing Center, a DOE Office of Science User Facility under the same contract; and by the U.S. National Science Foundation, Division of Astronomical Sciences under Contract No. AST-0950945 to NOAO.

This research uses services or data provided by the Astro Data Lab at NSF’s NOIRLab. 

This research is based on data obtained from the Astro Data Archive at NSF's NOIRLab. These data are associated with the observing programs listed in Table \ref{tab:archival}. 


%

\vspace{5mm}
\facilities{
CTIO:4m(Blanco/DECam), Astro Data Lab, Astro Data Archive, IRSA
}


\software{
\rr1{Astropy~\citep{Astropy2013,Astropy2018,AstropyCollaboration2022}},
Astroquery~\citep{Ginsburg2019}, 
LSST Science Pipelines~\citep{Juric2017,Bosch2019},
Matplotlib~\citep{Hunter2007}, 
Numpy~\citep{Harris2020}, Scipy~\citep{Virtanen2020},  
Source Extractor~\citep{Bertin1996} 
}



\appendix

\section{Survey Depth}\label{sec:app_depth}

We use Table~\ref{tab:depth},~\ref{tab:depth2} to show the $5\sigma$ depth of the cluster fields ($\sim2\deg$ in diameter) whose maps are presented in Section~\ref{sec:maps}. Those LoVoCCS clusters are observation-time complete.  We consider an annular region from a radial distance of 10 arcmin to 1 deg towards the cluster in each field to reduce the contamination effect of cluster member galaxies, although some fields include scattered cluster pairs/groups (e.g., A401/A399, A3827/A3825, and A2556/A2554); as the background galaxies are much fainter and greatly outnumber the foreground galaxies, we expect the statistical bias caused by member galaxies is minimal.

\begin{table*}[htbp!]
    \centering
    \begin{tabular}{lrrcccccccccc}
\hline
Name & R.A. (deg) & Decl. (deg) &$u_{\rm p5}$ & $g_{\rm p5}$ & $r_{\rm p5}$ & $i_{\rm p5}$ & $z_{\rm p5}$ & $u_{\rm c5}$ & $g_{\rm c5}$ & $r_{\rm c5}$ & $i_{\rm c5}$ & $z_{\rm c5}$ \\
\hline
A2029 & 227.73 & 5.82 & 24.9 & 26.7 & 26.3 & 25.5 & 25.0 & 24.6 & 26.4 & 25.9 & 25.1 & 24.6 \\
A401 & 44.74 & 13.58 & 24.7 & 25.5 & 25.7 & 25.1 & 24.2 & 24.4 & 25.2 & 25.3 & 24.8 & 23.9 \\
A85 & 10.40 & -9.33 & 25.3 & 25.8 & 26.0 & 25.1 & 23.4 & 25.0 & 25.5 & 25.7 & 24.8 & 23.1 \\
A3667 & 303.14 & -56.84 & 25.3 & 26.2 & 26.3 & 25.8 & 25.2 & 25.1 & 25.8 & 25.9 & 25.2 & 24.7 \\
A3266 & 67.80 & -61.41 & 25.3 & 26.3 & 26.1 & 25.6 & 24.8 & 25.0 & 26.0 & 25.8 & 25.2 & 24.5 \\
A1651 & 194.84 & -4.20 & 25.2 & 25.5 & 26.0 & 25.5 & 24.5 & 24.9 & 25.3 & 25.7 & 25.1 & 24.2 \\
A754 & 137.21 & -9.64 & 25.7 & 26.1 & 25.9 & 25.3 & 25.2 & 25.5 & 25.8 & 25.5 & 25.0 & 24.8 \\
A3571 & 206.87 & -32.87 & 25.3 & 25.9 & 26.1 & 25.3 & 24.6 & 25.0 & 25.7 & 25.8 & 25.1 & 24.4 \\
A3112 & 49.47 & -44.24 & 25.4 & 25.5 & 26.3 & 25.4 & 24.4 & 25.1 & 25.3 & 26.0 & 25.1 & 24.1 \\
A2597 & 351.33 & -12.12 & 25.2 & 26.1 & 26.1 & 25.4 & 25.0 & 25.0 & 25.9 & 25.8 & 25.1 & 24.7 \\
A1650 & 194.68 & -1.76 & 25.2 & 26.0 & 25.7 & 25.9 & 25.0 & 24.9 & 25.7 & 25.4 & 25.6 & 24.6 \\
A3558 & 201.98 & -31.49 & 25.3 & 26.3 & 26.0 & 25.3 & 24.6 & 25.0 & 26.0 & 25.7 & 24.9 & 24.3 \\
A3695 & 308.70 & -35.83 & 25.2 & 26.1 & 26.1 & 25.9 & 25.0 & 24.9 & 25.9 & 25.8 & 25.5 & 24.7 \\
A3921 & 342.50 & -64.43 & 25.2 & 25.6 & 26.0 & 25.5 & 24.3 & 25.0 & 25.4 & 25.7 & 25.2 & 24.0 \\
A2426 & 333.61 & -10.37 & 25.4 & 26.0 & 26.1 & 25.6 & 24.5 & 25.2 & 25.7 & 25.8 & 25.3 & 24.2 \\
A3158 & 55.66 & -53.63 & 25.4 & 25.6 & 26.2 & 25.2 & 24.4 & 25.1 & 25.3 & 25.4 & 24.9 & 24.1 \\
RXCJ1217.6+0339 & 184.43 & 3.65 & 25.0 & 24.8 & 25.8 & 26.1 & 23.6 & 24.8 & 24.5 & 25.4 & 25.8 & 23.3 \\
A2811 & 10.54 & -28.54 & 25.6 & 26.2 & 26.1 & 25.5 & 24.9 & 25.3 & 25.9 & 25.8 & 25.1 & 24.6 \\
A780 & 139.62 & -12.26 & 25.5 & 26.0 & 26.0 & 25.3 & 24.4 & 25.2 & 25.8 & 25.7 & 25.0 & 24.0 \\
A2420 & 332.59 & -12.19 & 25.3 & 25.9 & 26.1 & 25.4 & 24.8 & 25.1 & 25.7 & 25.8 & 25.1 & 24.5 \\
A1285 & 172.61 & -14.56 & 25.0 & 25.7 & 26.0 & 25.1 & 24.4 & 24.8 & 25.5 & 25.6 & 24.8 & 24.1 \\
A3911 & 341.59 & -52.74 & 25.5 & 25.9 & 26.2 & 25.6 & 24.3 & 25.3 & 25.7 & 25.8 & 25.3 & 24.0 \\
A2055 & 229.69 & 6.23 & 25.1 & 25.5 & 25.9 & 25.4 & 24.6 & 24.8 & 25.2 & 25.6 & 25.1 & 24.3 \\
A1750 & 202.71 & -1.86 & 25.0 & 26.0 & 26.0 & 25.4 & 24.7 & 24.7 & 25.8 & 25.6 & 25.1 & 24.4 \\
A3822 & 328.53 & -57.85 & 25.2 & 26.2 & 25.9 & 25.5 & 24.6 & 24.9 & 26.0 & 25.6 & 25.2 & 24.3 \\
A2941 & 26.27 & -53.02 & 25.4 & 25.9 & 26.0 & 25.1 & 24.5 & 25.1 & 25.7 & 25.7 & 24.8 & 24.1 \\
A2440 & 335.97 & -1.60 & 24.9 & 25.6 & 25.9 & 25.2 & 24.5 & 24.6 & 25.4 & 25.6 & 24.9 & 24.1 \\
A1644 & 194.30 & -17.41 & 25.1 & 26.2 & 26.1 & 25.3 & 24.6 & 24.8 & 25.9 & 25.8 & 25.1 & 24.3 \\
\hline
    \end{tabular}
    \caption{Cluster field depth  (Part I). We show the $5\sigma$ PSF magnitudes of point sources ($m_{\rm p5}$) and CModel magnitudes of extended sources ($m_{\rm c5}$), in DECam $u,g,r,i,z$ bands of individual cluster fields (sorted by X-ray luminosity/cluster rank) and their median values (Table~\ref{tab:depth2}). 
    These are apparent AB magnitudes without Galactic extinction correction ($\lesssim0.1$ mag in $r$ band). The magnitude zero points have been corrected for the color-terms between the reference catalogs (PS1/SkyMapper/SDSS) and DECam, and here we only consider the magnitude uncertainties  reported in the LSST pipeline forced photometry of co-added images. The residual zero point offsets are $\lesssim0.03$ mag in $g,r,i,z$ when compared with DES (if available) and $\lesssim0.05$ mag in $u$ when compared with model stellar loci in the color-color space. The celestial coordinates come from SIMBAD (ICRS, J2000). 
    }
    \label{tab:depth}
\end{table*}

\begin{table*}[htbp!]
    \centering
    \begin{tabular}{lrrcccccccccc}
\hline
Name & R.A. (deg) & Decl. (deg) &$u_{\rm p5}$ & $g_{\rm p5}$ & $r_{\rm p5}$ & $i_{\rm p5}$ & $z_{\rm p5}$ & $u_{\rm c5}$ & $g_{\rm c5}$ & $r_{\rm c5}$ & $i_{\rm c5}$ & $z_{\rm c5}$ \\
\hline
A1348 & 175.35 & -12.27 & 25.4 & 26.1 & 25.9 & 26.2 & 24.7 & 25.1 & 25.9 & 25.6 & 25.9 & 24.4 \\
RBS1847 & 334.50 & -65.18 & 25.2 & 26.1 & 26.0 & 25.0 & 24.1 & 24.9 & 25.8 & 25.7 & 24.7 & 23.8 \\
RXCJ1215.4-3900 & 183.89 & -38.99 & 25.4 & 25.7 & 25.6 & 25.0 & 24.6 & 25.0 & 25.4 & 25.3 & 24.7 & 24.2 \\
A2351 & 323.60 & -13.39 & 25.3 & 25.8 & 26.1 & 25.2 & 24.3 & 25.0 & 25.6 & 25.8 & 24.9 & 24.0 \\
RXCJ2218.2-0350 & 334.58 & -3.83 & 25.2 & 25.7 & 25.8 & 24.9 & 24.1 & 24.9 & 25.5 & 25.5 & 24.6 & 23.7 \\
A2443 & 336.56 & 17.42 & 25.0 & 26.0 & 26.0 & 25.0 & 24.5 & 24.7 & 25.6 & 25.6 & 24.7 & 24.1 \\
A2050 & 229.08 & 0.06 & 25.3 & 26.3 & 26.2 & 25.7 & 24.8 & 25.0 & 26.0 & 25.9 & 25.3 & 24.5 \\
A1736 & 201.72 & -27.11 & 24.8 & 26.7 & 26.5 & 25.4 & 24.5 & 24.5 & 26.4 & 26.0 & 25.0 & 24.2 \\
A2384 & 328.08 & -19.61 & 25.5 & 25.6 & 25.9 & 25.2 & 25.0 & 25.2 & 25.3 & 25.6 & 24.9 & 24.6 \\
A4059 & 359.17 & -34.67 & 25.4 & 25.8 & 25.9 & 25.3 & 24.3 & 25.1 & 25.5 & 25.6 & 25.0 & 24.1 \\
A3836 & 332.40 & -51.84 & 25.0 & 26.6 & 25.9 & 25.5 & 24.1 & 24.7 & 26.3 & 25.6 & 25.2 & 23.8 \\
A2533 & 346.81 & -15.22 & 25.2 & 26.0 & 25.5 & 25.0 & 24.3 & 25.0 & 25.7 & 25.2 & 24.8 & 24.0 \\
A2556 & 348.26 & -21.63 & 25.3 & 25.8 & 25.7 & 25.3 & 24.7 & 25.0 & 25.5 & 25.4 & 25.0 & 24.3 \\
A3126 & 52.18 & -55.71 & 24.8 & 25.9 & 26.1 & 25.0 & 24.2 & 24.5 & 25.7 & 25.8 & 24.8 & 23.9 \\
RXCJ0049.4-2931 & 12.35 & -29.53 & 25.1 & 25.7 & 26.0 & 25.3 & 24.6 & 24.9 & 25.5 & 25.7 & 25.0 & 24.3 \\
A3395 & 96.88 & -54.40 & 25.4 & 26.2 & 26.3 & 25.7 & 25.1 & 25.0 & 25.9 & 25.9 & 25.2 & 24.6 \\
A1606 & 191.15 & -11.99 & 24.4 & 26.1 & 26.1 & 25.8 & 24.6 & 24.1 & 25.8 & 25.7 & 25.5 & 24.4 \\
A2670 & 358.54 & -10.39 & 24.9 & 24.8 & 25.9 & 25.3 & 24.6 & 24.7 & 24.6 & 25.5 & 25.0 & 24.4 \\
A3532 & 194.33 & -30.37 & 25.4 & 26.3 & 25.9 & 25.0 & 24.3 & 25.1 & 26.0 & 25.6 & 24.8 & 24.0 \\
RXCJ1139.4-3327 & 174.86 & -33.45 & 25.5 & 26.1 & 26.0 & 25.5 & 24.1 & 25.1 & 25.8 & 25.7 & 25.0 & 23.8 \\
RXJ0820.9+0751 & 125.26 & 7.86 & 25.0 & 25.8 & 26.1 & 25.7 & 24.9 & 24.7 & 25.5 & 25.8 & 25.3 & 24.5 \\
A3128 & 52.64 & -52.55 & 25.9 & 26.4 & 26.3 & 25.4 & 24.9 & 25.6 & 26.1 & 25.6 & 25.1 & 24.5 \\
A1023 & 157.00 & -6.80 & 25.2 & 26.1 & 26.0 & 25.1 & 24.2 & 24.9 & 25.7 & 25.7 & 24.7 & 23.8 \\
A3528 & 193.58 & -29.02 & 24.9 & 25.8 & 26.3 & 25.3 & 24.6 & 24.6 & 25.6 & 26.0 & 25.1 & 24.3 \\
A761 & 137.65 & -10.58 & 25.5 & 26.2 & 26.0 & 25.3 & 24.6 & 25.2 & 25.9 & 25.7 & 25.0 & 24.3 \\
A3825 & 329.59 & -60.39 & 25.3 & 26.1 & 26.1 & 25.8 & 24.0 & 25.0 & 25.8 & 25.3 & 25.3 & 23.7 \\
Median & -- & -- & 25.3 & 26.0 & 26.0 & 25.3 & 24.6 & 25.0 & 25.7 & 25.7 & 25.1 & 24.2 \\
\hline
    \end{tabular}
    \caption{Cluster field depth  (Part II). 
    }
    \label{tab:depth2}
\end{table*}

\section{DECam Archival Data}\label{sec:app_archival}
In Table~\ref{tab:archival}, we list the proposal IDs of the public exposures used in our co-addition. We thank the researchers of those programs for their observations.

\begin{table*}[htbp!]
    \centering
\begin{tabular}{llllllll}
\hline
PROPID & PI & PROPID & PI & PROPID & PI & PROPID & PI \\
\hline
2012B-0001 & Frieman & 2014A-0270 & Grillmair & 2016A-0189 & Rest & 2018A-0911 & Forster \\
2012B-0003 & DES SV & 2014A-0306 & Dai & 2016A-0190 & Dey & 2018A-0913 & Zenteno \\
2012B-0569 & Allen & 2014A-0321 & Geha & 2016A-0366 & Bechtol & 2018A-0914 & Makler \\
2012B-3004 & Dell'Antonio & 2014A-0339 & Hargis & 2016A-0384 & McCleary & 2018B-0905 & Oh \\
2012B-3007 & Thorman & 2014A-0386 & Dell'Antonio & 2016A-0386 & Malhotra & 2019A-0205 & Goldstein \\
2012B-9999 & Engineering & 2014A-0390 & Bloom & 2016A-0397 & Zenteno & 2019A-0265 & Finkbeiner \\
2013A-0327 & Rest & 2014A-0412 & Rest & 2016A-0618 & Mackey & 2019A-0272 & Zenteno \\
2013A-0360 & von der Linden & 2014A-0415 & von der Linden & 2016A-0622 & Penny & 2019A-0305 & Drlica-Wagner \\
2013A-0400 & Bloom & 2014A-0608 & Forster & 2016A-0951 & Bechtol & 2019A-0308 & Dell'Antonio \\
2013A-0411 & Nidever & 2014A-0610 & Taylor & 2016B-0124 & Frieman & 2019A-0913 & Carballo-Bello \\
2013A-0455 & Sheppard & 2014A-0613 & Rodriquez & 2016B-0301 & Rest & 2019A-0917 & Lopes \\
2013A-0529 & Rich & 2014A-0621 & Mackey & 2016B-0905 & Jerjen & 2019B-0323 & Zenteno \\
2013A-0611 & Mackey & 2014A-0624 & Jerjen & 2016B-0909 & Navarete & 2019B-0372 & Soares-Santos \\
2013A-0612 & Sheen & 2014B-0146 & Sullivan & 2017A-0260 & Soares-Santos & 2019B-0403 & Martinez-Vazquez \\
2013A-0613 & Munoz & 2014B-0244 & von der Linden & 2017A-0388 & Zenteno & 2019B-1004 & Chaname \\
2013A-0704 & Wood & 2014B-0265 & Dell'Antonio & 2017A-0909 & Cooke & 2019B-1014 & Olivares \\
2013A-0717 & Dell'Antonio & 2014B-0404 & Schlegel & 2017A-0916 & Carballo-Bello & 2020A-0399 & Zenteno \\
2013A-0719 & Saha & 2014B-0608 & Jaffe & 2017A-0925 & Munoz & 2020A-0402 & Vivas \\
2013A-0724 & Allen & 2015A-0062 & French & 2017B-0103 & Barkhouse & 2020A-0908 & Olivares \\
2013A-0737 & Sheppard & 2015A-0110 & De Boer & 2017B-0110 & Frieman & 2020A-0909 & Arevalo \\
2013A-0741 & Schlegel & 2015A-0163 & Grillmair & 2017B-0253 & Carlin & 2020B-0053 & Shen \\
2013A-2101 & Walker & 2015A-0306 & Balbinot & 2017B-0279 & Rest & 2020B-0241 & Zenteno \\
2013A-9999 & Lee & 2015A-0397 & Walker & 2017B-0904 & Lopes & 2020B-0909 & Chaname \\
2013B-0418 & Rest & 2015A-0608 & Forster & 2017B-0907 & Munoz & 2021A-0117 & Kotulla \\
2013B-0421 & Rest & 2015A-0609 & Carballo-Bello & 2017B-0951 & Vivas & 2021A-0149 & Zenteno \\
2013B-0440 & Nidever & 2015A-0616 & Jerjen & 2018A-0215 & Carlin & 2021A-0275 & Rest \\
2013B-0502 & Dell & 2015A-0617 & Nataf & 2018A-0242 & Bechtol & 2021A-0922 & Nilo-Castellon \\
2013B-0531 & Rest & 2015A-0618 & Lidman & 2018A-0276 & Dell'Antonio & 2022A-388025 & Palmese \& Wang \\
2013B-0612 & Chaname & 2015A-0619 & Goncalves & 2018A-0369 & Rest & 2022A-777564 & Zhao \\
2013B-0617 & Mackey & 2015A-0620 & Bonaca & 2018A-0380 & Rest & 2022A-975778 & Kelkar \\
2013B-0627 & Lima Neto & 2015B-0187 & Berger & 2018A-0386 & Zenteno & 2022B-297190 & Palmese \& Wang \\
2014A-0157 & Favia & 2015B-0191 & Rice & 2018A-0907 & Munoz & -- & -- \\
2014A-0256 & Eckert & 2016A-0004 & Bonaca & 2018A-0909 & Puzia & -- & -- \\
\hline
\end{tabular}
    \caption{DECam public datasets used in our co-addition (sorted by Proposal ID). }
    \label{tab:archival}
\end{table*}

\section{Mass Map Example}\label{sec:app_mass}
\begin{figure*}
    \centering
    \includegraphics[width=0.5\textwidth]{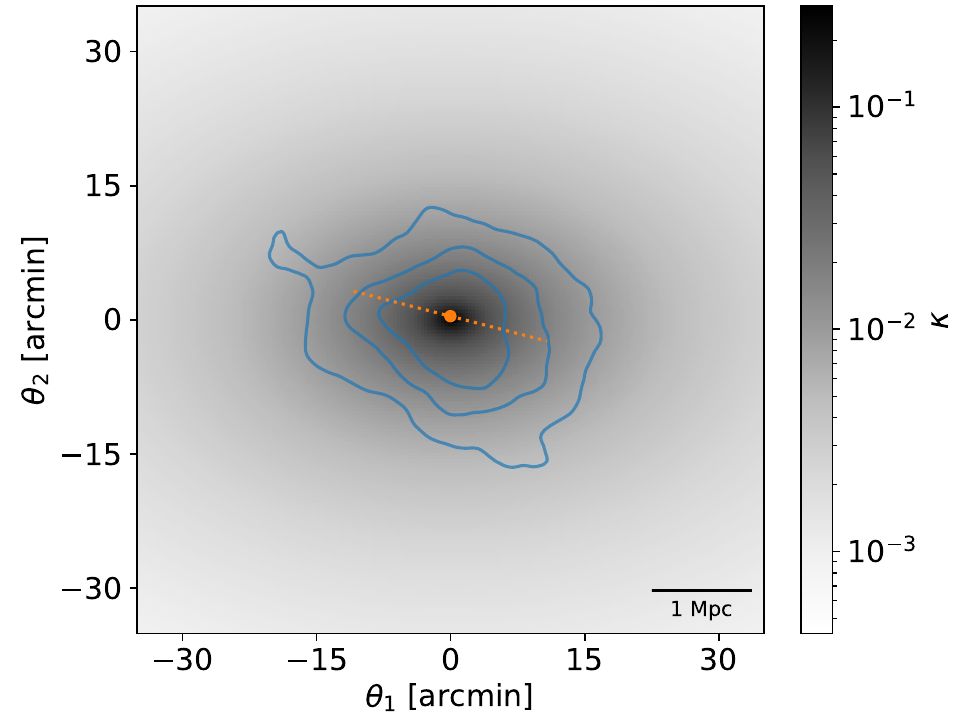}
    \caption{Convergence of an elliptical halo (centered in the diagram and elongated horizontally; gray scale in the background) overlaid with the corresponding aperture mass map (blue contours showing S/N $=2,4,6$). 
    The orange point and the dotted line give the detected lensing peak and the derived orientation angle (within 1 Mpc), respectively. 
    We vary the aperture radius and use the aperture when the peak S/N reaches the maximum (10k pixels $\sim 10 r_{\rm s}$). The diagram spans 16k$\times$16k DECam pixels ($4\times4$ patches; $0\farcs263$ per pixel), which is consistent with the figures in Section~\ref{sec:maps}. }
    \label{fig:el_map}
\end{figure*}

We present an example of using the aperture mass map to detect the orientation (and center) of an elliptical NFW halo. We apply the Schirmer aperture to a mock lensing catalog derived from the study of~\citet{Fu2023} (see also~\citealt{Oguri2010} and~\citealt{Oguri2010b}). The catalog spans 24k$\times$24k DECam pixels ($6\times6$ patches). The halo mass, source and lens redshifts, shape noise, and source density are set to be the median values in our LoVoCCS observations (see also Section~\ref{sec:mass_bcg_offset}). 
In Figure~\ref{fig:el_map}, we show the convergence of the elliptical halo ($\kappa({\theta_1,\theta_2})=\kappa_{\rm NFW}(\sqrt{q\theta_1^2+\theta_2^2/q});~q=2/3$) overlaid with the lensing mass S/N map. The orientations of the halo and the map are generally consistent. 

\section{Red Sequence Examples}\label{sec:app_rs}
\begin{figure*}[htbp!]
    \centering
    \includegraphics[width=0.32\textwidth]{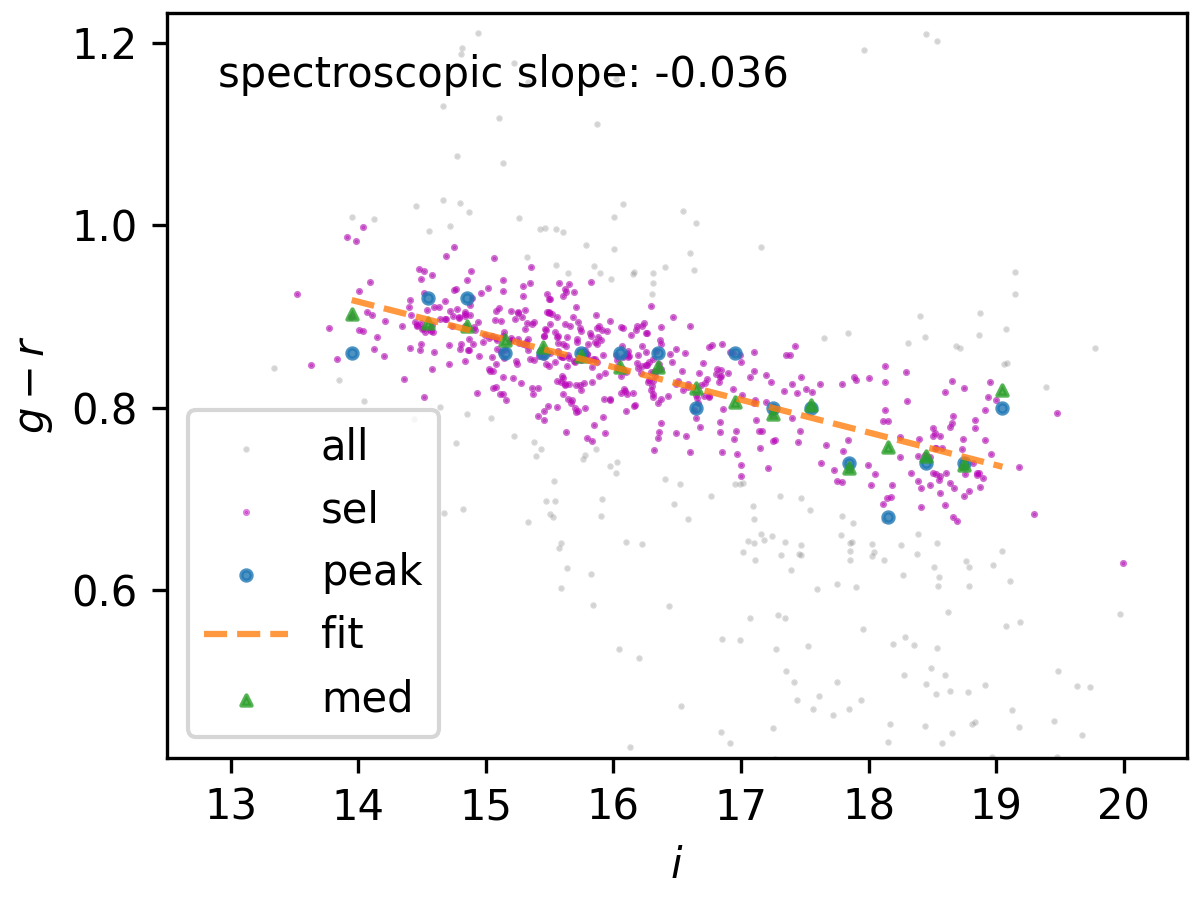}
    \includegraphics[width=0.32\textwidth]{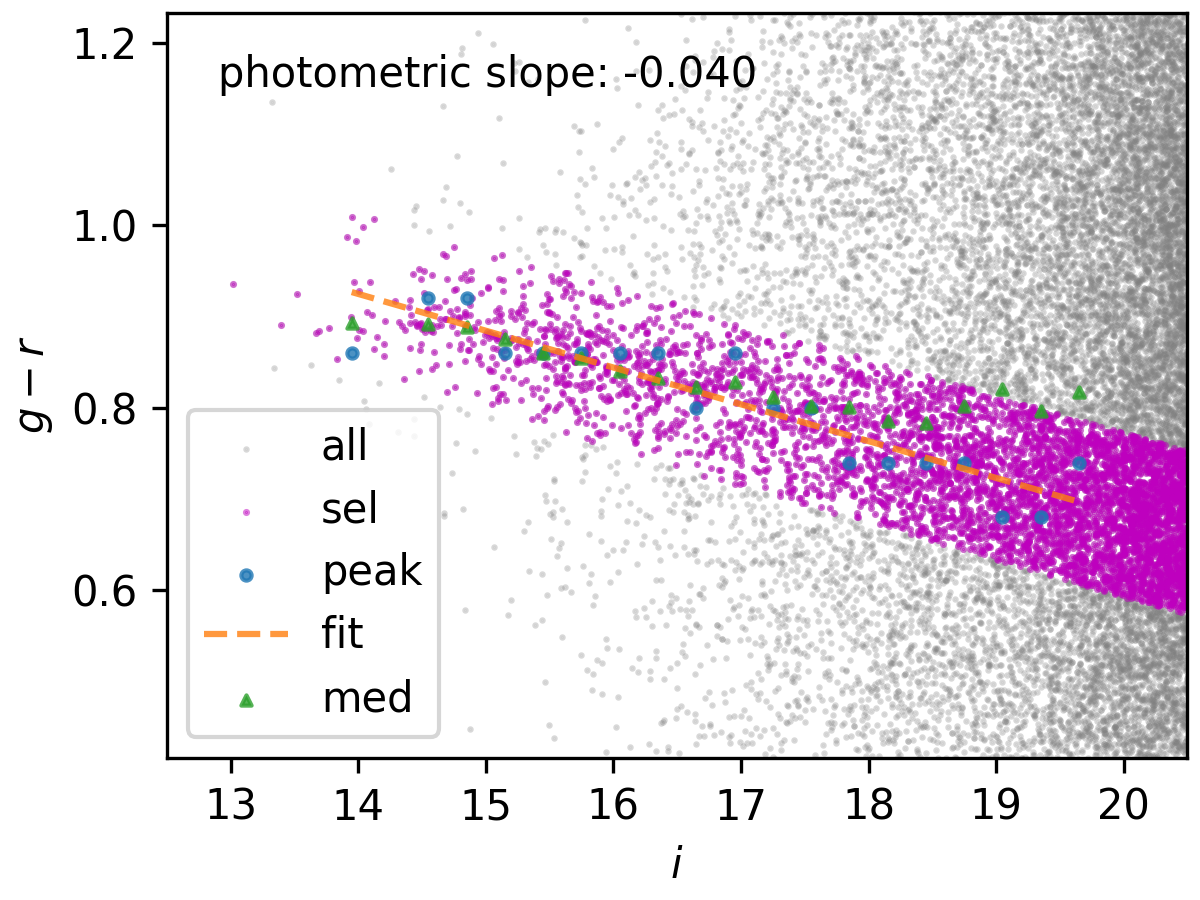}
    \includegraphics[width=0.32\textwidth]{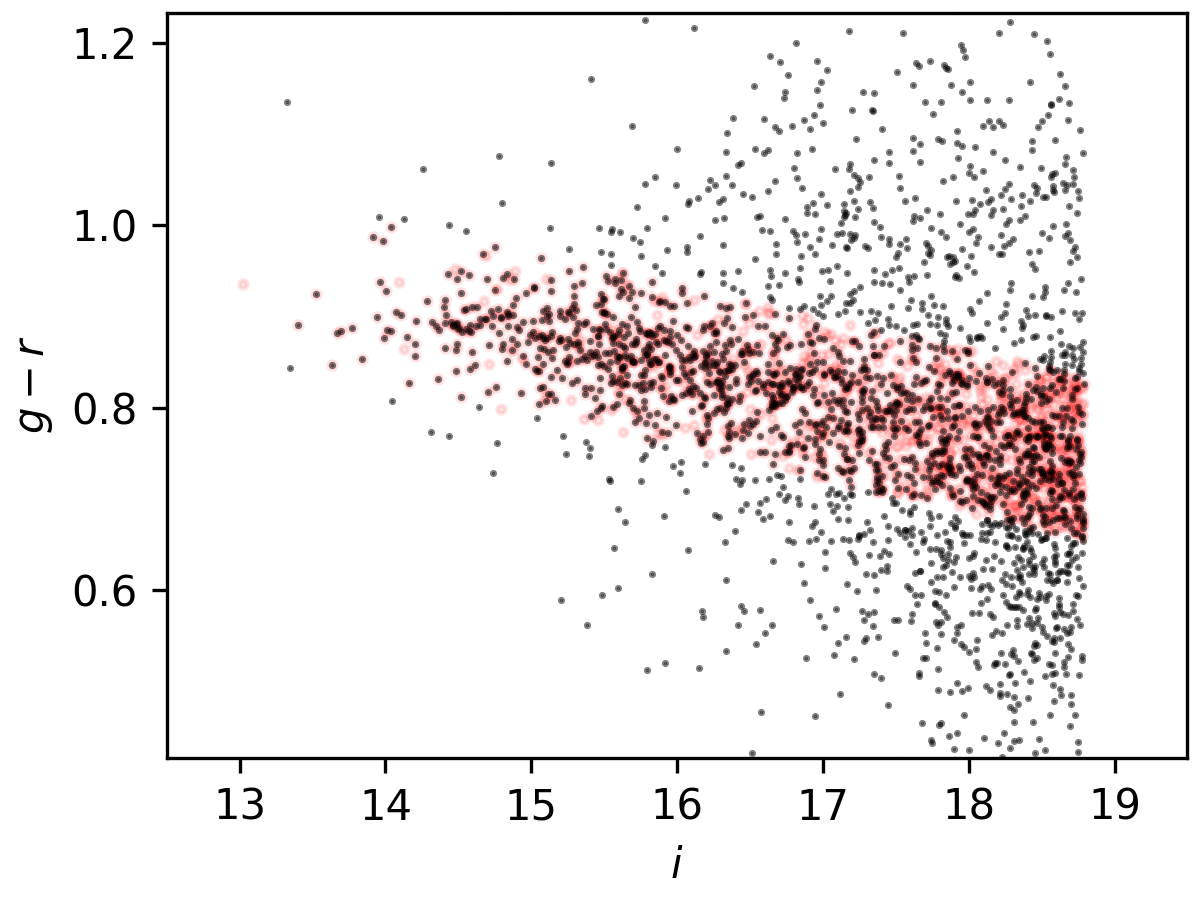}
    \includegraphics[width=0.32\textwidth]{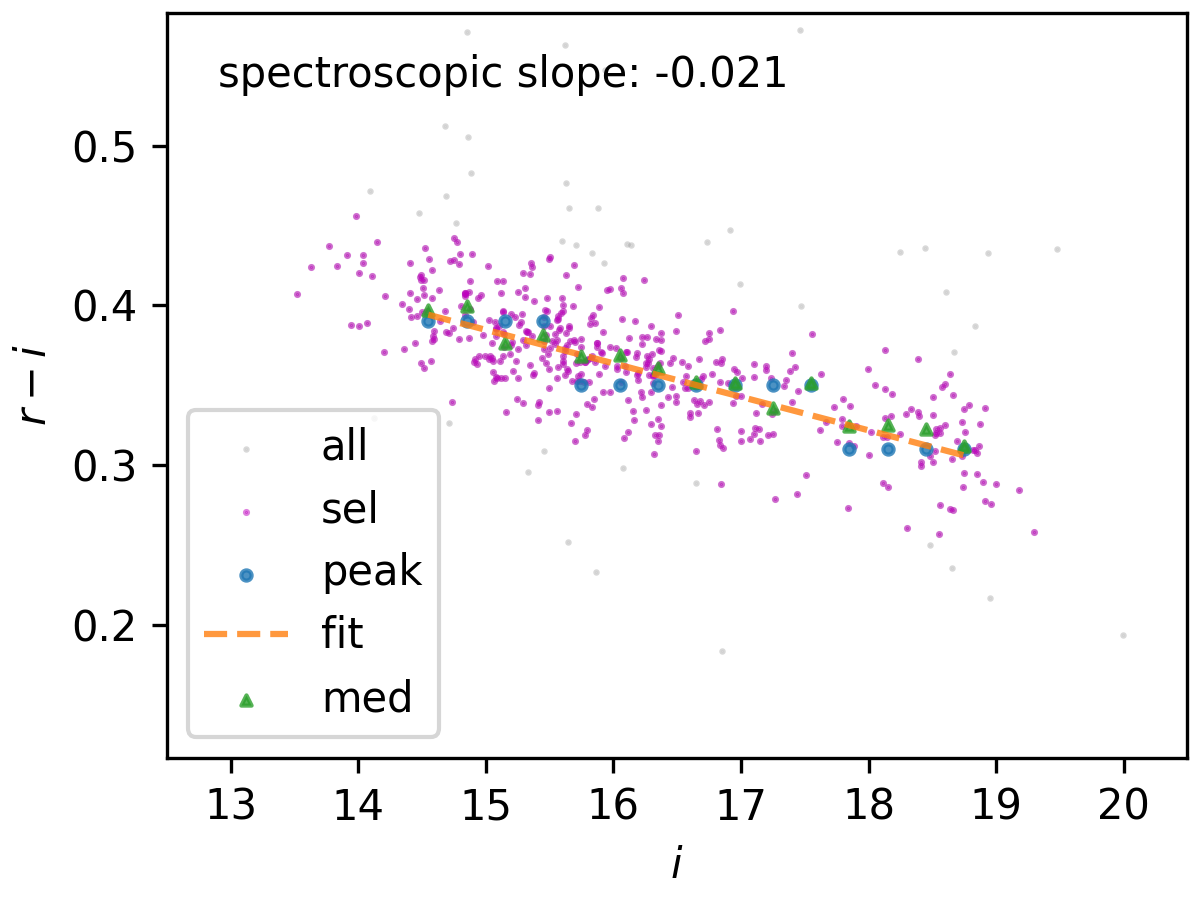}
    \includegraphics[width=0.32\textwidth]{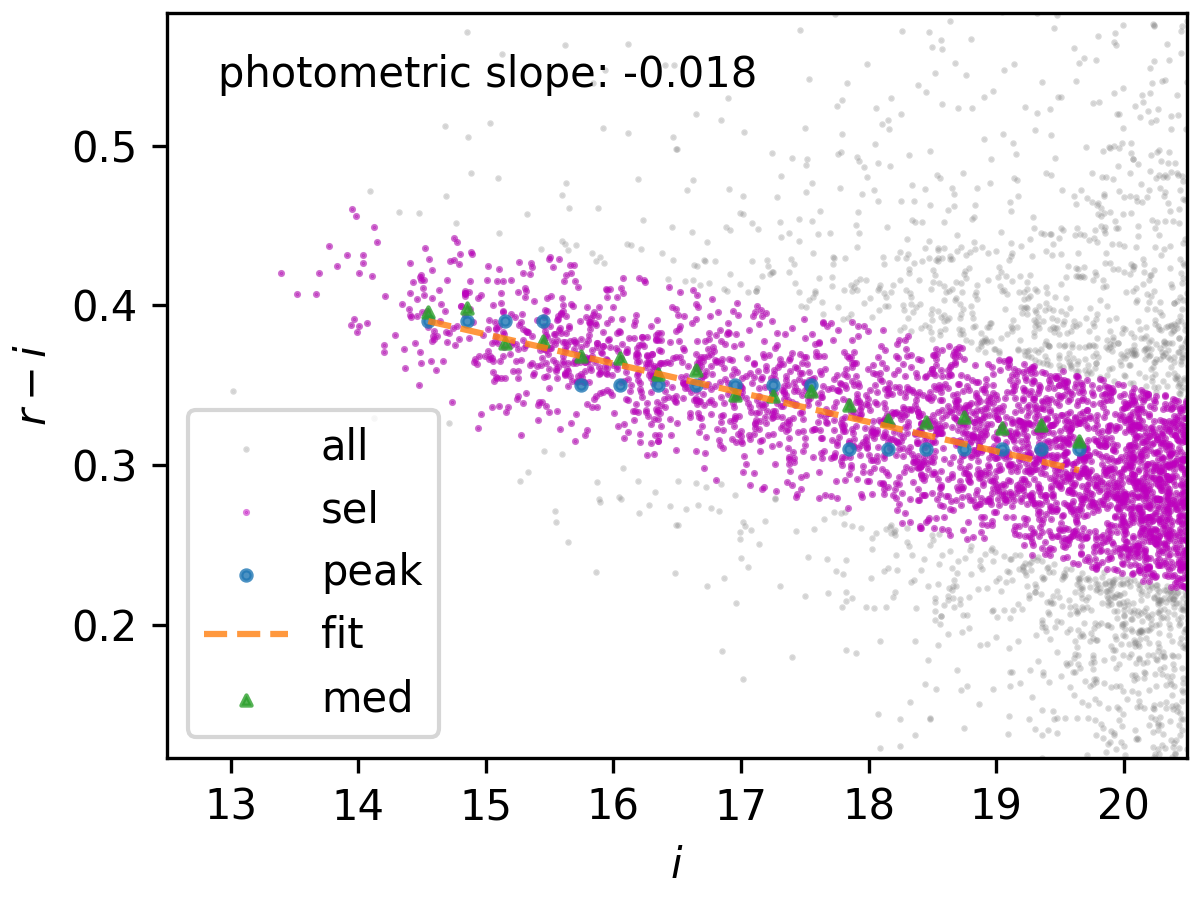}
    \includegraphics[width=0.32\textwidth]{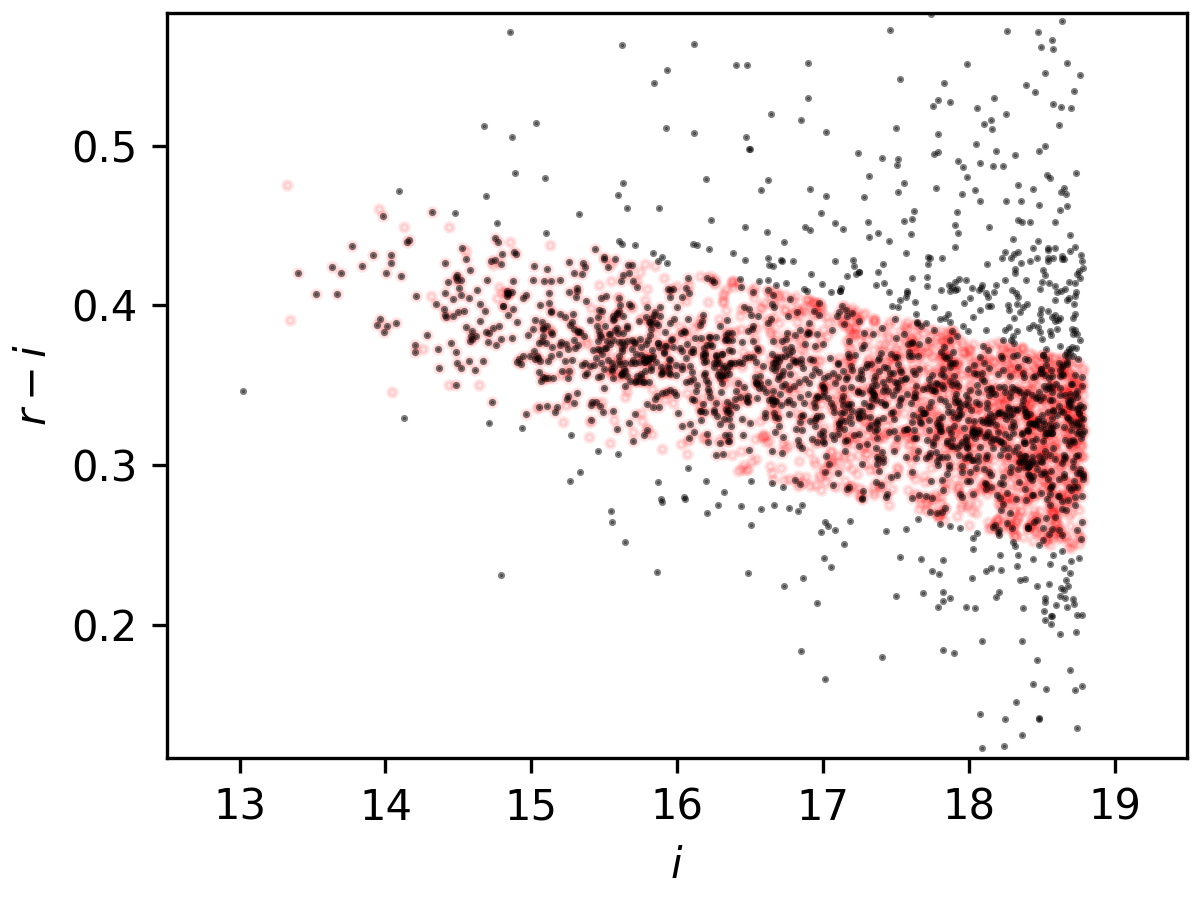}
    \caption{Red sequence examples in A3558 ($z=0.047$). 
    \textit{Top left}: The gray points are all member galaxies selected by spec-z. The blue points are the histogram peaks of those gray points in individual $i$ magnitude bins (Section~\ref{sec:RS}), while the green triangles are the median values in those bins (consistent with blue points generally). The orange dash line is the linear fit of those blue points. The magenta points are the galaxies selected by the scatter cut around the fitted line. 
    \textit{Bottom left}: The gray points are member galaxies selected by both spec-z and $g-r$ (i.e., the galaxies denoted by magenta points in the \textit{top left} diagram) -- the RS is more clear. Other markers have similar meanings to the ones in the \textit{top left} diagram. 
    \textit{Top middle}: Now the gray points are all galaxies in the catalog with valid photometry. Other markers have similar meanings to the \textit{top left} diagram. Note the green median points have large biases at faint magnitudes, but the blue peak points are consistent with the trend at the bright end. 
    \textit{Bottom middle}: The gray points correspond to the magenta ones in the \textit{top middle} diagram. Other markers have similar meanings. Again, the RS is more clear than its top counterpart because of the first selection in $g-r$. 
    \textit{Top right}: We start from all galaxies. The red circles show the galaxies selected by the $g-r$ fit (the spectroscopic one in this case), while the black points are the galaxies selected by the $r-i$ fit.  
    Here an extra $i$-magnitude cut is also included to fix the luminosity limit. The overlap between red and black points is the RS galaxy sample.   
    \textit{Bottom right}: Similar to the \textit{top right} diagram -- the red circles are selected by the $r-i$ fit, while the black circles are selected by the $g-r$ fit. Those diagrams indicate that a single-color selection can include foreground/background galaxies, but a two-color selection can give a much cleaner sample of RS galaxies. 
    }
    \label{fig:rs1}
\end{figure*}

\begin{figure*}[htbp!]
    \centering
    \includegraphics[width=0.32\textwidth]{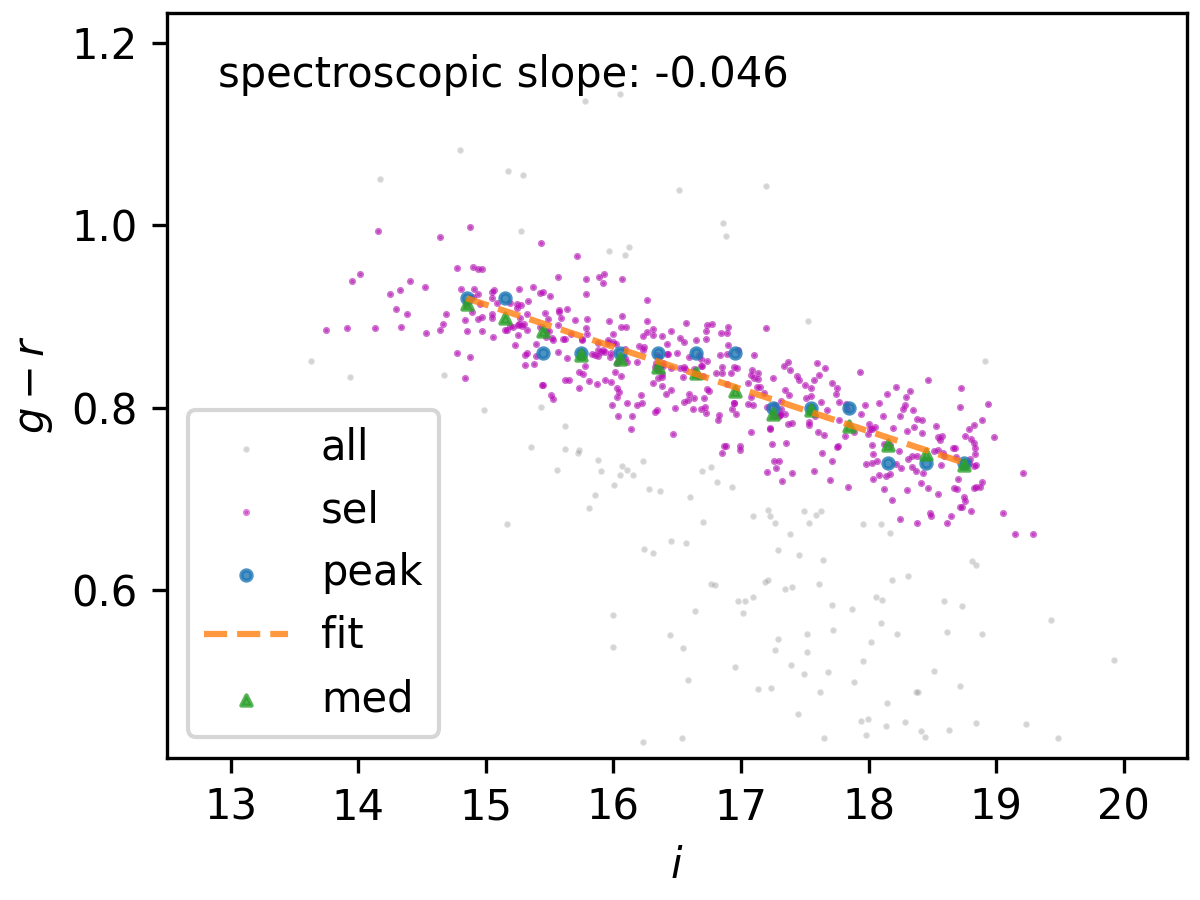}
    \includegraphics[width=0.32\textwidth]{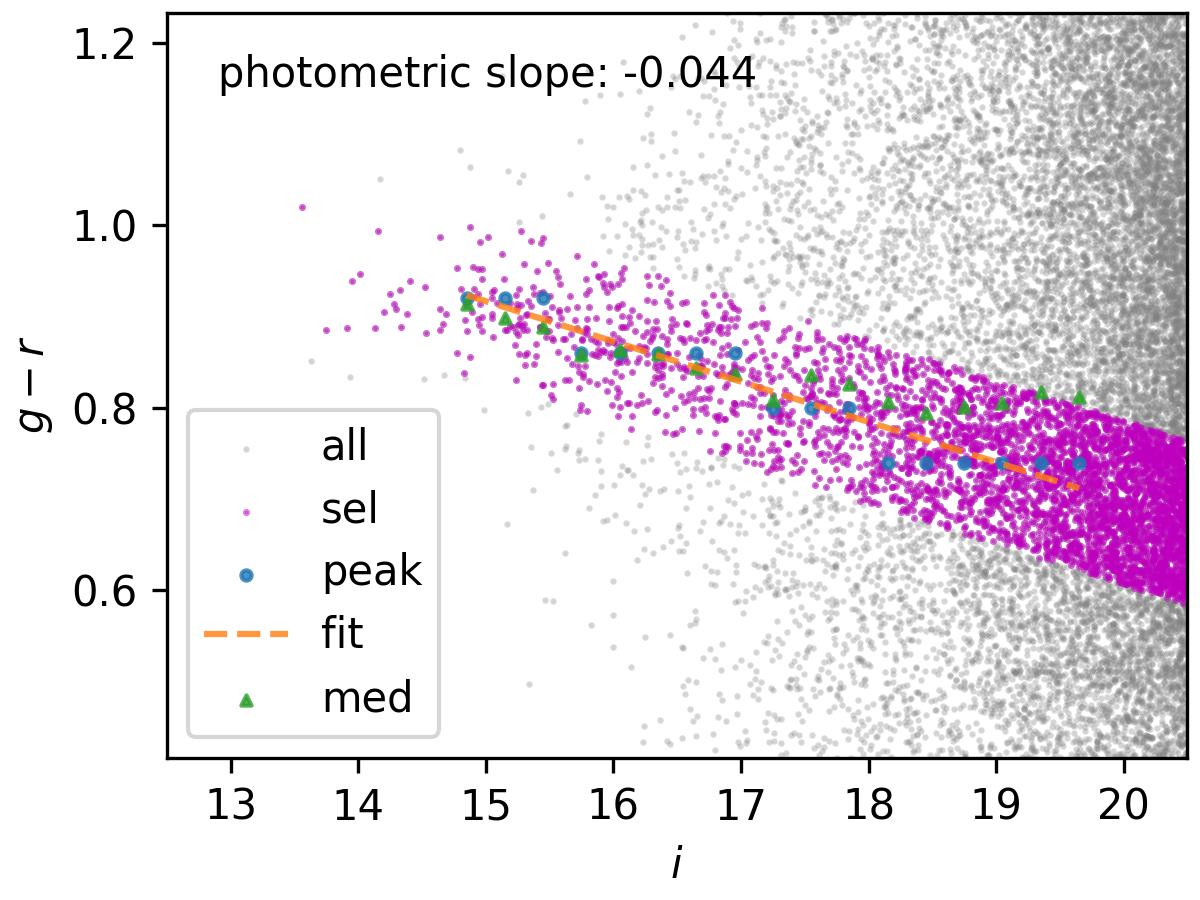}
    \includegraphics[width=0.32\textwidth]{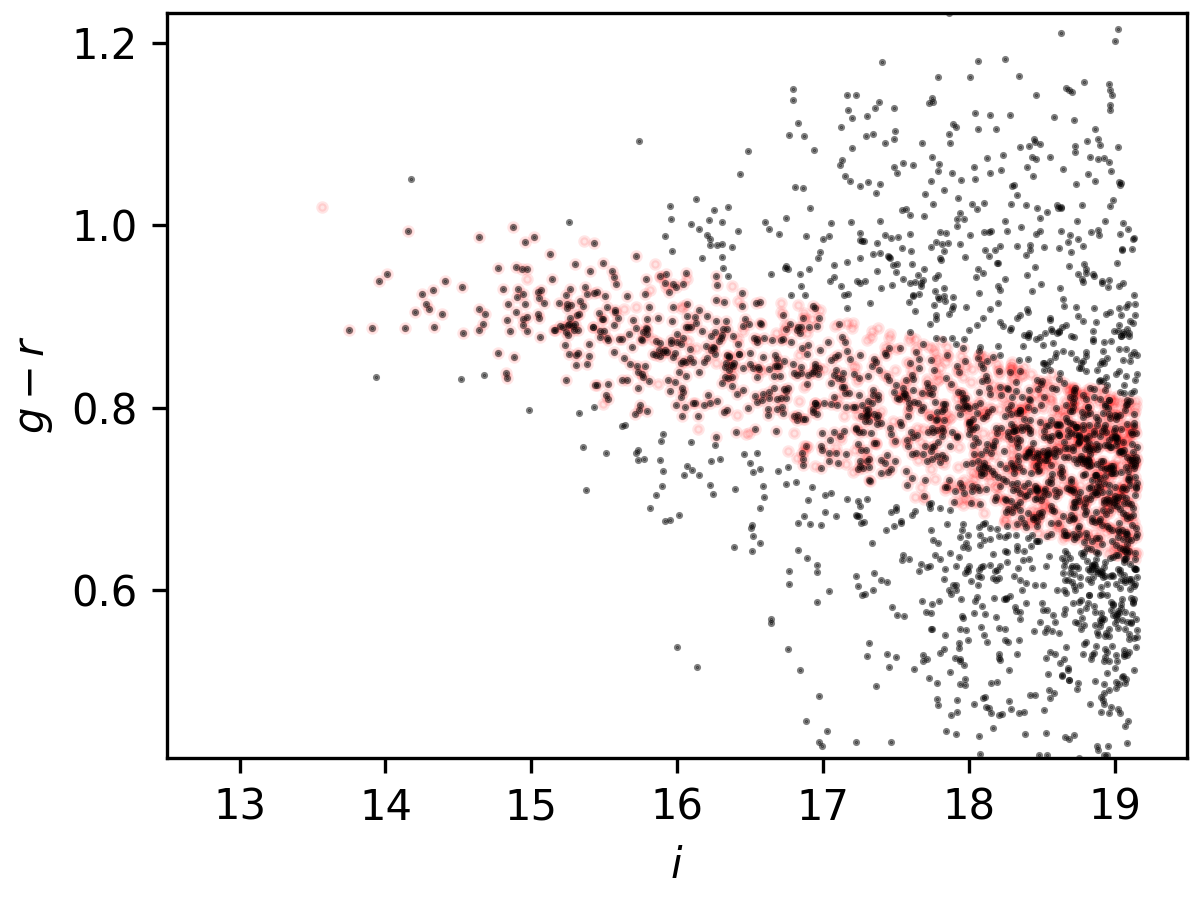}
    \includegraphics[width=0.32\textwidth]{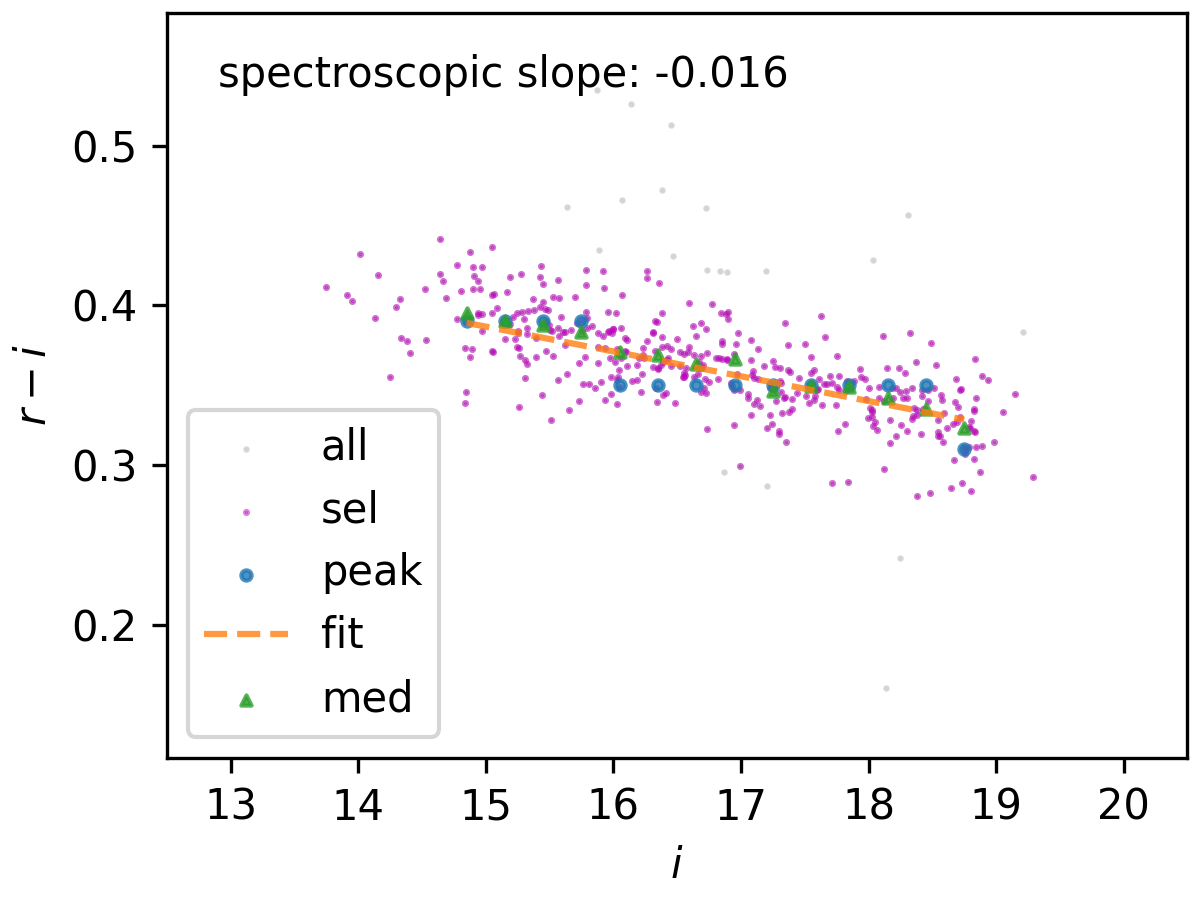}
    \includegraphics[width=0.32\textwidth]{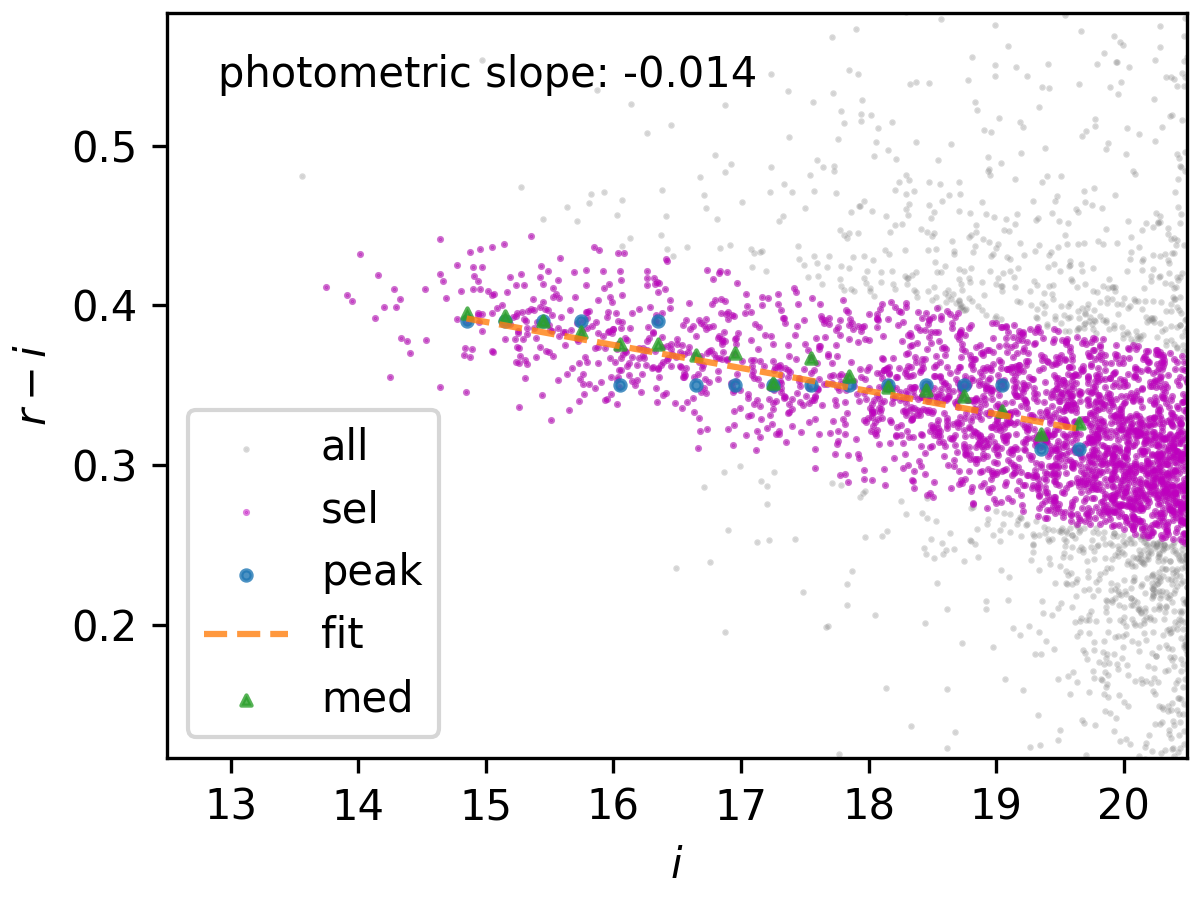}
    \includegraphics[width=0.32\textwidth]{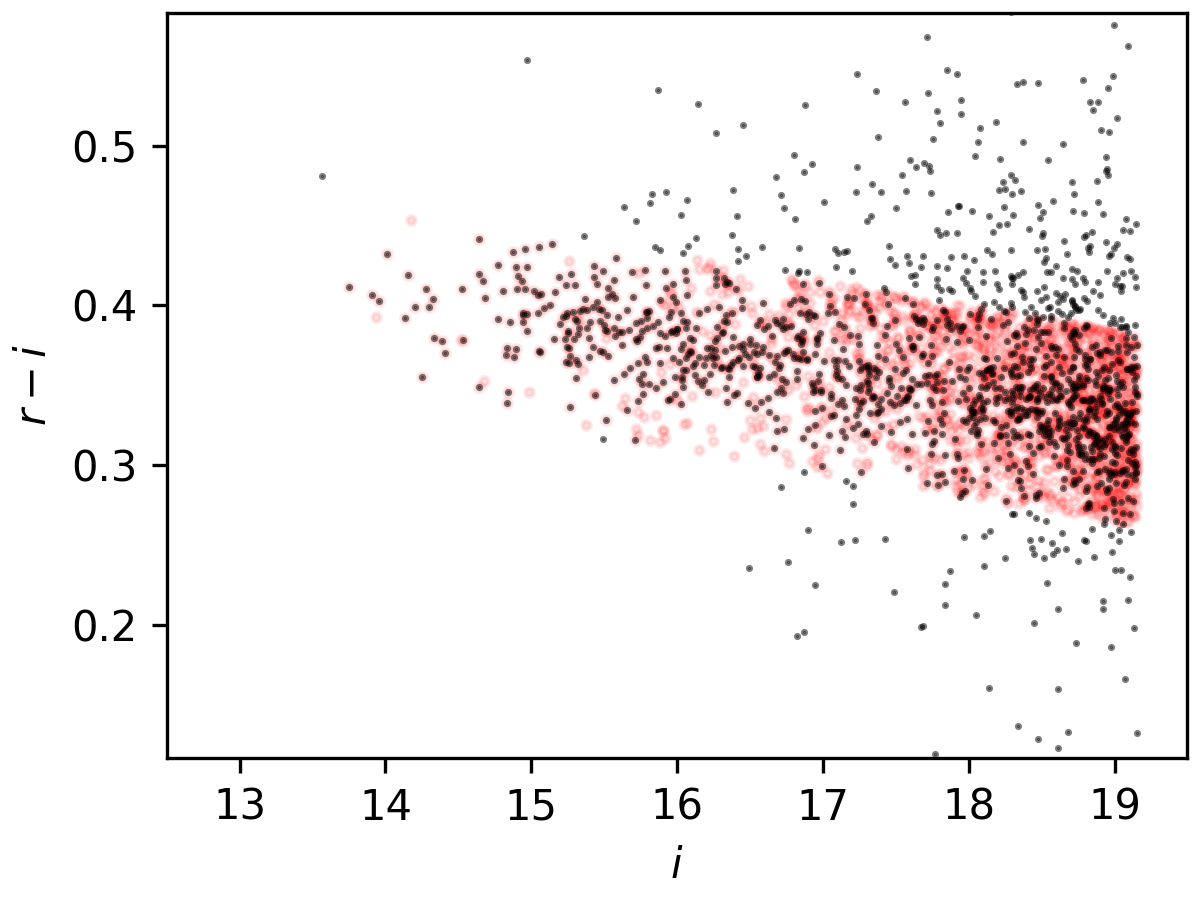}
    \caption{Counterparts of the red sequence examples in A3667 ($z=0.055$). They are similar to the ones for A3558 (Figure~\ref{fig:rs1}). }
    \label{fig:rs2}
\end{figure*}

We present the examples of detecting and selecting the red sequence galaxies in the color--magnitude diagrams in Figure~\ref{fig:rs1} (A3558) and Figure~\ref{fig:rs2} (A3667). Those two clusters have enough spec-z data to show RS in CMD, and thus we use the RS linear fit parameters detected from the spec-z selected galaxies instead of all galaxies (though their fitting results are close). 
The algorithm was presented in Section~\ref{sec:RS}.

\section{Orientations and Centers}\label{sec:ap_angle_peak}

Table~\ref{tab:angle_peak},~\ref{tab:angle_peak2} show the measured orientation angles and centers of the BCG, the RS and mass distributions (with different radial cuts) of individual clusters (and their types/CG counts). 

\begin{figure*}
    \centering
    \plotone{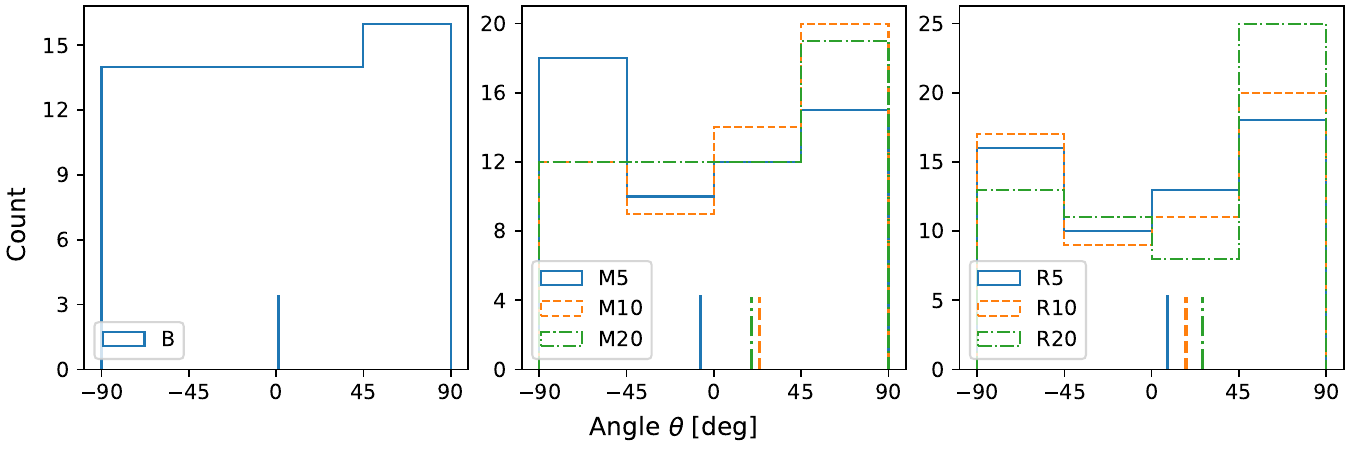}
    \caption{Angle distributions. The labels have the same meanings as the ones in Table~\ref{tab:angle_peak},~\ref{tab:angle_peak2}. The vertical lines give the respective median values. }
    \label{fig:angle_distribution}
\end{figure*}

We also examine the distributions of those angles (Figure~\ref{fig:angle_distribution}). 
The distributions are generally
uniform.  
Large radial cuts may lead to large noise in the angle measurement, because at large radii the number density of member galaxies and the mass density are low.

\begin{table*}[htbp!]
    \centering
    \begin{tabular}{lrrrrrrrrrrrrrrr}
\hline
Name & $z$ & $\alpha_{\rm B}$ & $\delta_{\rm B}$ & $\alpha_{\rm M}$ & \rr1{$\delta_{\rm M}$} & $\alpha_{\rm R}$ & $\delta_{\rm R}$ & $\theta_{\rm B}$ & $\theta_{\rm M5}$ & $\theta_{\rm M10}$ & $\theta_{\rm M20}$ & $\theta_{\rm R5}$ & $\theta_{\rm R10}$ & $\theta_{\rm R20}$ & $n_{\rm CG}$ \\
\hline
A2029 & 0.08 & 227.73 & 5.74 & 227.73 & 5.78 & 227.74 & 5.77 & -63 & 85 & 84 & -70 & -74 & -74 & -84 & a \\
A401 & 0.07 & 44.74 & 13.58 & 44.74 & 13.58 & 44.74 & 13.58 & -63 & -7 & 1 & -25 & -61 & -48 & -58 & a \\
A85 & 0.06 & 10.46 & -9.30 & 10.45 & -9.33 & 10.41 & -9.31 & 59 & -70 & -84 & -41 & 85 & 72 & 75 & b \\
A3667 & 0.06 & 303.11 & -56.83 & 303.14 & -56.85 & 303.15 & -56.84 & 44 & -69 & 80 & -84 & 38 & 44 & 48 & a \\
A3827 & 0.10 & 330.47 & -59.95 & 330.46 & -59.94 & 330.48 & -59.94 & 72 & 78 & 69 & 57 & 73 & 64 & 83 & a \\
A3266 & 0.06 & 67.81 & -61.45 & 67.80 & -61.43 & 67.82 & -61.45 & -10 & -74 & 80 & 86 & -25 & -24 & -44 & a \\
A1651 & 0.08 & 194.84 & -4.20 & 194.83 & -4.20 & 194.85 & -4.20 & -13 & 21 & 23 & -66 & -6 & -2 & -10 & a \\
A754 & 0.05 & 137.13 & -9.63 & 137.14 & -9.65 & 137.12 & -9.63 & 22 & 29 & 35 & 51 & 19 & 21 & 8 & a \\
A3571 & 0.04 & 206.87 & -32.86 & 206.89 & -32.91 & 206.85 & -32.88 & -85 & 65 & 62 & 18 & 84 & 88 & 86 & a \\
A3112 & 0.08 & 49.49 & -44.24 & 49.49 & -44.27 & 49.50 & -44.22 & -84 & -75 & -87 & 82 & -53 & -68 & 84 & a \\
A399 & 0.07 & 44.47 & 13.03 & 44.46 & 13.03 & 44.44 & 13.02 & -43 & -33 & -10 & -10 & -64 & -34 & -40 & a \\
A2597 & 0.08 & 351.33 & -12.12 & 351.32 & -12.13 & 351.31 & -12.11 & 50 & 68 & -67 & -66 & 52 & 81 & 76 & a \\
A1650 & 0.08 & 194.67 & -1.76 & 194.67 & -1.78 & 194.68 & -1.77 & 71 & 43 & 54 & 11 & 70 & 65 & 67 & a \\
A3558 & 0.05 & 201.99 & -31.50 & 201.98 & -31.48 & 201.97 & -31.52 & 74 & 70 & 67 & 54 & 47 & 45 & 11 & a \\
A3695 & 0.09 & 308.69 & -35.82 & 308.69 & -35.83 & 308.69 & -35.78 & 63 & 42 & 3 & -12 & 54 & 80 & 82 & a \\
A3921 & 0.09 & 342.49 & -64.43 & 342.48 & -64.35 & 342.44 & -64.43 & -5 & -76 & -39 & -23 & 0 & 5 & -11 & a \\
A2426 & 0.10 & 333.63 & -10.37 & 333.63 & -10.35 & 333.64 & -10.37 & -5 & 77 & 82 & 68 & -14 & -11 & 76 & a \\
A3158 & 0.06 & 55.72 & -53.63 & 55.75 & -53.65 & 55.75 & -53.64 & 3 & -77 & -22 & 44 & 13 & 33 & 56 & a \\
RXCJ1217.6+0339 & 0.08 & 184.42 & 3.66 & 184.40 & 3.64 & 184.43 & 3.67 & -82 & -29 & -32 & 21 & -22 & -45 & -56 & a \\
A2811 & 0.11 & 10.54 & -28.54 & 10.54 & -28.54 & 10.54 & -28.54 & 41 & -7 & 50 & 53 & 6 & 35 & 31 & a \\
A780 & 0.05 & 139.52 & -12.10 & 139.52 & -12.12 & 139.53 & -12.12 & 59 & 3 & 33 & 49 & 68 & 68 & 59 & a \\
A2420 & 0.08 & 332.58 & -12.17 & 332.57 & -12.16 & 332.56 & -12.19 & -42 & 16 & 13 & -63 & -64 & -54 & -44 & a \\
A1285 & 0.11 & 172.60 & -14.58 & 172.58 & -14.59 & 172.59 & -14.58 & 32 & 47 & 44 & 52 & 31 & 62 & 61 & b \\
A3911 & 0.10 & 341.56 & -52.72 & 341.54 & -52.72 & 341.52 & -52.69 & 46 & -25 & 3 & -4 & 40 & 55 & 55 & a \\
A2055 & 0.10 & 229.69 & 6.23 & 229.73 & 6.26 & 229.69 & 6.25 & 65 & 57 & -76 & -62 & 58 & 66 & 68 & c \\
A1750 & 0.09 & 202.71 & -1.86 & 202.78 & -1.73 & 202.71 & -1.87 & -11 & -58 & -59 & 20 & -46 & -56 & -50 & c \\
A3822 & 0.08 & 328.52 & -57.87 & 328.55 & -57.87 & 328.55 & -57.85 & -64 & -23 & 63 & 39 & -26 & -46 & -51 & b \\
A2941 & 0.12 & 26.24 & -53.02 & 26.31 & -53.06 & 26.28 & -53.02 & 3 & 85 & -78 & 33 & 14 & 23 & 30 & b \\
A2440 & 0.09 & 335.99 & -1.58 & 336.00 & -1.60 & 335.97 & -1.62 & -34 & 3 & 25 & 19 & -44 & -41 & -47 & c \\
\hline
    \end{tabular}
    \caption{Cluster centers and orientation angles  (Part I).  We present the measurements of the BCG (B), the lensing mass distribution (M), and the RS galaxy distribution (R) of each LoVoCCS cluster studied in this work (sorted by X-ray luminosity/cluster rank). The redshift ($z$) is from SIMBAD. The coordinates ($\alpha/\delta$ for R.A./Decl. in ICRS) and angles ($\theta$)  are in degrees. The angles are counterclockwise from the X-axis, with the X-axis pointing to the west and the Y-axis pointing to the north. We present the angles within 0.5~Mpc, 1.0~Mpc, 2.0~Mpc from the central peak with tags 5, 10, 20, respectively. The last column ($n_{\rm CG}$) gives the cluster type based on the number of bright central galaxies; 1, 2, $\geq$3 correspond to type a, b, c (43, 8, 7 clusters, respectively). 
    We skip the noisy mass/RS distribution results of some clusters. 
    }
    \label{tab:angle_peak}
\end{table*}
\begin{table*}[htbp!]
    \centering
    \begin{tabular}{lrrrrrrrrrrrrrrr}
\hline
Name & $z$ & $\alpha_{\rm B}$ & $\delta_{\rm B}$ & $\alpha_{\rm M}$ & $\delta_{\rm M}$ & $\alpha_{\rm R}$ & $\delta_{\rm R}$ & $\theta_{\rm B}$ & $\theta_{\rm M5}$ & $\theta_{\rm M10}$ & $\theta_{\rm M20}$ & $\theta_{\rm R5}$ & $\theta_{\rm R10}$ & $\theta_{\rm R20}$ & $n_{\rm CG}$ \\
\hline
A1644 & 0.05 & 194.30 & -17.41 & 194.30 & -17.41 & 194.38 & -17.25 & -44 & -80 & 83 & 52 & 75 & 89 & 76 & a \\
A1348 & 0.12 & 175.35 & -12.28 & 175.35 & -12.27 & 175.38 & -12.26 & -10 & -12 & -20 & 82 & -19 & -14 & -36 & a \\
RBS1847 & 0.10 & 334.50 & -65.18 & 334.46 & -65.14 & 334.50 & -65.17 & -7 & 62 & -84 & -17 & -46 & -46 & -69 & a \\
RXCJ1215.4-3900 & 0.12 & 183.86 & -39.03 & 183.87 & -39.01 & 183.83 & -39.05 & -33 & 88 & -70 & -21 & -41 & -50 & -73 & a \\
A2351 & 0.09 & 323.57 & -13.43 & 323.58 & -13.42 & 323.57 & -13.42 & 62 & -69 & -87 & -82 & 56 & 71 & 74 & a \\
RXCJ2218.2-0350 & 0.09 & 334.67 & -3.78 & 334.56 & -3.80 & 334.67 & -3.77 & 2 & -9 & -42 & 46 & -51 & -32 & -37 & c \\
A2443 & 0.11 & 336.53 & 17.36 & 336.54 & 17.36 & 336.52 & 17.39 & -55 & 87 & 85 & 80 & 74 & 63 & 61 & c \\
A2050 & 0.12 & 229.07 & 0.09 & 229.08 & 0.08 & 229.07 & 0.08 & -49 & -80 & 67 & 48 & -38 & -34 & -45 & a \\
A1736 & 0.04 & 201.70 & -27.14 & 201.71 & -27.11 & 201.71 & -27.13 & 77 & -59 & -38 & -29 & -77 & 85 & -85 & c \\
A2384 & 0.09 & 328.09 & -19.55 & 328.09 & -19.56 & 328.08 & -19.55 & -61 & -56 & -56 & -26 & 85 & -78 & -82 & b \\
A4059 & 0.05 & 359.25 & -34.76 & 359.21 & -34.79 & 359.25 & -34.74 & 62 & 83 & -72 & 25 & 53 & 54 & 74 & a \\
A3836 & 0.11 & 332.34 & -51.81 & 332.34 & -51.80 & 332.35 & -51.81 & 20 & 29 & 23 & 26 & 81 & 74 & 55 & a \\
A2533 & 0.11 & 346.81 & -15.22 & 346.86 & -15.23 & 346.81 & -15.21 & -48 & 44 & 52 & 76 & -49 & -67 & -72 & a \\
A2556 & 0.09 & 348.26 & -21.63 & -- & -- & 348.26 & -21.65 & 42 & -- & -- & -- & 10 & 8 & 24 & a \\
A3126 & 0.08 & 52.15 & -55.71 & 52.15 & -55.71 & 52.15 & -55.71 & -66 & 85 & 87 & -75 & 76 & 45 & 76 & a \\
RXCJ0049.4-2931 & 0.11 & 12.35 & -29.52 & 12.36 & -29.52 & 12.34 & -29.52 & 35 & 29 & 28 & 24 & 75 & -83 & 67 & a \\
A2554 & 0.11 & 348.08 & -21.50 & 348.08 & -21.49 & 348.07 & -21.49 & -85 & -12 & -5 & 2 & -73 & -72 & 47 & a \\
A2033 & 0.08 & 227.86 & 6.35 & -- & -- & -- & -- & -36 & -- & -- & -- & -- & -- & -- & a \\
A3395 & 0.05 & 96.90 & -54.45 & 96.67 & -54.59 & 96.85 & -54.45 & 34 & 45 & 82 & -74 & 35 & 28 & 45 & b \\
A1606 & 0.10 & 191.15 & -11.99 & 191.15 & -12.01 & 191.15 & -12.01 & -80 & 64 & 48 & 75 & -68 & 56 & 26 & a \\
A2670 & 0.08 & 358.56 & -10.42 & 358.54 & -10.41 & 358.56 & -10.41 & 30 & -62 & 19 & 54 & 85 & -65 & -63 & a \\
A3532 & 0.06 & 194.34 & -30.36 & 194.33 & -30.36 & 194.32 & -30.37 & -8 & -27 & 3 & 58 & -40 & -26 & -13 & a \\
RXCJ1139.4-3327 & 0.11 & 174.85 & -33.45 & 174.83 & -33.46 & 174.85 & -33.44 & 77 & -70 & 85 & -38 & -85 & -84 & 78 & a \\
RXJ0820.9+0751 & 0.11 & 125.26 & 7.86 & 125.26 & 7.87 & 125.27 & 7.86 & 15 & -56 & 87 & -74 & 8 & 45 & -39 & a \\
A3128 & 0.06 & 52.66 & -52.62 & 52.64 & -52.53 & 52.62 & -52.56 & -81 & -80 & 78 & 50 & -86 & -87 & -64 & c \\
A1023 & 0.12 & 156.99 & -6.80 & 157.00 & -6.79 & 156.99 & -6.79 & 49 & 27 & -20 & -32 & 44 & 53 & 75 & a \\
A3528 & 0.06 & 193.59 & -29.01 & 193.61 & -29.00 & 193.59 & -29.00 & 89 & -86 & -76 & -87 & -89 & 77 & 75 & b \\
A761 & 0.09 & 137.65 & -10.58 & 137.65 & -10.59 & 137.68 & -10.59 & 1 & -56 & 32 & -48 & 35 & 18 & 1 & a \\
A3825 & 0.07 & 329.61 & -60.39 & -- & -- & 329.62 & -60.36 & 80 & -- & -- & -- & -78 & -77 & -7 & b \\
\hline
    \end{tabular}
    \caption{Cluster centers and orientation angles   (Part II). 
    }
    \label{tab:angle_peak2}
\end{table*}

\section{Significance of Orientation Alignment}\label{sec:angle_significance}

Consider a set of angles $\{ \theta_i \},~i=1,2,...,N$. 
Each angle independently obeys a uniform distribution $\theta_i \sim U(0,\Phi)$. 
The probability of an angle $\theta_i$ being smaller than $\phi$ is $\phi/\Phi$. 
The probability of $n$ angles being smaller than  $\phi$ (and the others being larger than $\phi$) is ${N \choose n}(\phi/\Phi)^n (1-\phi/\Phi)^{N-n}$. 
Then the probability that more than $m$ angles have their values smaller than $\phi$ is $\sum_{i=m+1}^N {N \choose i}(\phi/\Phi)^i (1-\phi/\Phi)^{N-i}=1-F(m;N,\phi/\Phi)$, where $F$ is the cumulative distribution function of a binomial distribution with $m$ times of successes in $N$ independent tests and success probability $\phi/\Phi$. 

This result can be extended to the case where $\theta_i$ is between $\alpha$ and $\beta$. If $\alpha<\beta<\Phi$ and $\psi=\beta-\alpha$, the probability that more than $m$ angles have values between $\alpha$ and $\beta$ is similarly $1-F(m;N,\psi/\Phi)$.  

\begin{figure*}
    \centering
    \includegraphics[width=0.5\textwidth]{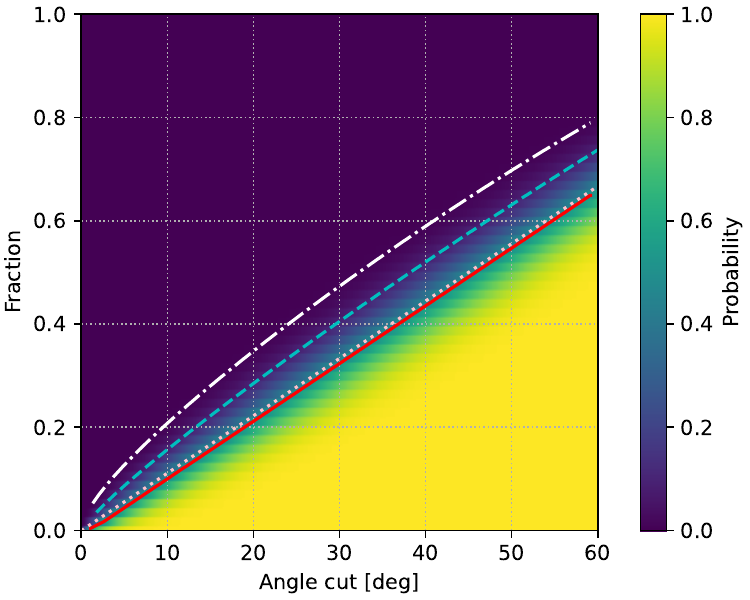}
    \caption{Probability of more than a fraction of the clusters have angles smaller than an angle cut. The red solid line, the cyan dashed line, and the white dash-dotted line correspond to probabilities of 0.5, 0.1, and 0.01, respectively. The pink dotted line gives the angle cut divided by 90 deg for reference. }
    \label{fig:angle}
\end{figure*}

Let $\theta_i$ be the angle between two orientations (e.g., BCG and RS distribution) and $\Phi$ be the angle upper limit (90 deg). 
Let $N$ be the number of clusters (e.g., 57 in our analysis after removing the noisy cluster). 
If we assume the angle is randomly distributed between 0 and 90 deg, then the probability that more than a fraction of $f$ of those clusters have the angles smaller than $\phi$ is $1-F(57f;57,\phi/90)$. Figure~\ref{fig:angle} gives this probability as a function of the fraction $f$ and the angle cut $\phi$. 
The probability drops quickly as the angle cut drops or the fraction increases. 
For example, the probability that more than 1/3 of the clusters have their angles smaller than 30 deg is $1-F(57/3;57,30/90)\sim0.4$, the probability that more than half of the cluster sample have the angles below 30 deg is $\sim5\times10^{-3}$, and  the probability that more than half of the sample have the angles below 15 deg is $\sim3\times10^{-9}$. 



\bibliography{main}{}
\bibliographystyle{aasjournal}



\end{document}